\documentclass[preprint,5p,twocolumn]{elsarticle}

\usepackage[fleqn]{amsmath}
\usepackage[latin1]{inputenc}
\usepackage[english]{babel}
\usepackage{amssymb,amsmath}
\usepackage{mathrsfs}
\usepackage{bm}
\usepackage[colorinlistoftodos]{todonotes}
\usepackage{enumitem}
\usepackage{multirow}
\usepackage{booktabs}
\usepackage{wrapfig}
\usepackage{cuted}
\usepackage{caption}
\usepackage[normalem]{ulem}

\newcommand{\vvv}{\textsf{v}}
\newcommand{\www}{\textsf{w}}

\newcommand{\HH}{\mathscr H}
\newcommand{\FF}{\mathscr F}
\newcommand{\R}{\mathscr R}

\newcommand{\FFF}{\mathcal F}
\newcommand{\RRR}{\mathcal R}

\newcommand{\K}{\mathcal K}

\newcommand{\C}{\mathcal C}

\newcommand{\tond}[1]{{\left(#1\right)}}
\newcommand{\quadr}[1]{{\left[#1\right]}}
\newcommand{\inter}[1]{{\left\langle#1\right\rangle}}
\newcommand{\graff}[1]{\{#1\}}

\newcommand{\derp}[2]{{\frac{\partial #1}{\partial #2}}}

\newcommand{\norm}[1]{\left\|#1\right\|}

\renewcommand{\leq}{\leqslant}
\renewcommand{\geq}{\geqslant}

\newcommand*\circled[1]{\raisebox{.5pt}{\textcircled{\raisebox{-.9pt} {#1}}}}

\newcommand{\pa}{\partial}

\newcommand{\bc}{\begin{center}}
\newcommand{\ec}{\end{center}}

\def\CC{\mathbb{C}}
\def\RR{\mathbb{R}}
\def\ZZ{\mathbb{Z}}

\def\Id{\mathbb I}

\def\eps{\epsilon}
\def\vphi{\varphi}
\def\im{i}

\def\ph{\varphi}

\newtheorem{theorem}{Theorem}[section]
\newtheorem{theorem*}{Theorem}
\newtheorem{lemma}{Lemma}[section]

\newtheorem{proposition}{Proposition}[section]

\newtheorem{remark}{Remark}[section]

\newenvironment{proof}[0]{\noindent{\bf Proof: }}{$\square$\par\medskip}

\begin{document}
\small
\begin{frontmatter}
\title{ On the nonexistence of degenerate phase-shift multibreathers\\
  in a zigzag Klein-Gordon model}

\author[1]{T. Penati}
\author[2]{V.  Koukouloyannis}
\author[1]{M. Sansottera} 
\author[3]{P.G. Kevrekidis}
\author[1]{S. Paleari}

\address[1]{Department of Mathematics ``F.Enriques'', Milano University, via
  Saldini 50, Milano, Italy, 20133}

\address[2]{Department of Mathematics, Statistics and Physics, College of Arts
  and Sciences, Qatar University, P.O. Box 2713, Doha, Qatar}

\address[3]{Department of Mathematics and Statistics, University of
  Massachusetts, Amherst, MA 01003-4515, USA}

\begin{abstract}
In this work, we study the existence of low amplitude four-site
phase-shift multibreathers for small values of the coupling $\epsilon$
in Klein-Gordon (KG) chains with interactions longer than the
classical nearest-neighbour ones. In the proper parameter regimes, the
considered lattices bear connections to models beyond one spatial
dimension, namely the so-called zigzag lattice, as well as the
two-dimensional square lattice. We examine initially the persistence
conditions of the system, in order to seek for vortex-like
waveforms. Although this approach provides useful insights, due to the
degeneracy of these solutions, it does not allow us to determine if
they constitute true solutions of our system.  In order to overcome
this obstacle, we follow a different route.  In the case of the zigzag
configuration, by means of a Lyapunov-Schmidt decomposition, we are
able to establish that the bifurcation equation for our model can be
considered, in the small energy and small coupling regime, as a
perturbation of a corresponding non-local discrete nonlinear
Schr\"odinger (NL-dNLS) equation. There, nonexistence results of
degenerate phase-shift discrete solitons can be demonstrated by
exploiting the expansion of a suitable density current of the NL-dNLS,
obtained in recent literature.  Finally, briefly considering a
one-dimensional model bearing similarities to
the square lattice, we conclude that the above strategy
is not efficient for the proof of the existence or nonexistence of
vortices due to the higher degeneracy of this configuration.
  
\end{abstract}

\end{frontmatter}

\section{Introduction}
\label{s:1}

The study of nonlinear dynamical lattices of Klein-Gordon, as well as
Fermi-Pasta-Ulam and related types has received considerable attention
over the past two decades due to the intense interest in waveforms
which are exponentially localized in space and periodic in time,
namely the so-called discrete
breathers~\cite{KivsharDBReview,FlachPR2008}. These states have been
recognized as emerging rather generically in systems that combine
discreteness and nonlinearity. Relevant experimental examples abound
and range from Josephson junction
arrays~\cite{TriasPRL2000,BinderPRL2000} to electrical transmission
lines \cite{EnglishPRE2008}, from micro-mechanical cantilever arrays
\cite{SatoPRL2003,SatoC2003} to coupled torsion
pendula~\cite{EnglishPRL2009}, and from coupled antiferromagnetic
layers \cite{SchwarzPRL1999} to granular
crystals~\cite{BoechlerPRL2010,jinkyu}, to name just a few examples.

Most of these studies concern fundamental localized states, and most
of them are predominantly in simpler, more controllable
one-dimensional settings~\cite{FlachPR2008}. However,
optical~\cite{moti}, atomic~\cite{emergent} and other settings suggest
an interest in exploring higher-dimensional settings. In the latter,
novel structures (such as discrete vortices, also referred to as
phase-shift multibreathers) emerge~\cite{aubry,archilla} and
occasional surprises arise, such as the existence of energy thresholds
for breather existence~\cite{kladko} or the potential of higher charge
vortices to be more stable than their lower charge counterparts under
appropriate conditions~\cite{vassilis_kour}. It has been argued that
(as will also be discussed further below) suitable adaptations of
beyond-nearest-neighbor interactions~\cite{KouKCR13} and the so-called
zigzag~\cite{efrem} chains share some of the intriguing features of
higher-dimensional settings, while remaining effectively
one-dimensional in their formulation. For this reason, the latter will
represent the starting point for our study in what follows.

More specifically, in this work, we are interested in Klein-Gordon
(KG) models with range of interactions beyond nearest-neighbour, with
Hamiltonian
\begin{equation*}
  \HH = \sum_{j\in\ZZ} \quadr{\frac12y^2_j + V(x_j)} 
  +\sum_{j\in\ZZ} \sum_{h=1}^{r} \eps_h\frac{(x_{j+h}-x_j)^2}2
  \ ,
\end{equation*}
where $V(x_j) = \frac12x_j^2+\frac14x_j^4$. This Hamiltonian describes
an infinite chain of anharmonic oscillators with linear interactions
between them up to $r$ neighbours and vanishing boundary conditions at
infinity $\lim_{n\to\pm\infty} x_n = \lim_{n\to\pm\infty} y_n = 0$,
which are automatically satisfied since we set
$\ell^2(\RR)\times\ell^2(\RR)$ as the phase space of the
system. {We will denote by $E$ the energy of the system,
  i.e. the (conserved along the dynamics) value of the Hamiltonian.}

In what follows we will limit our analysis to range of interaction $r=3$. By
considering $\eps_j=k_j\eps$, with $k_1=1$, the above Hamiltonian becomes
\begin{equation}
  \begin{aligned}
  \label{ham_long}
  \HH &= \HH_0+\epsilon\HH_1=\\
  &=\sum_{j\in\ZZ} \left({\frac{y_j^2}{2} +V(x_j)}\right) \\ &\quad+\frac{\eps}2 \sum_{j\in\ZZ}
  \quadr{(x_{j}-x_{j+1})^2
  +k_2(x_{j}-x_{j+2})^2+
  k_3(x_{j}-x_{j+3})^2}\, .
  \end{aligned}
\end{equation}

We are interested in the existence, in the small coupling limit (i.e., for values
of the coupling close to the anti-continuum limit~\cite{macaub} of
$\epsilon\rightarrow0$), of multibreather solutions. These constitute a class of
periodic orbits whose energy is spatially localized on few oscillators (or
sites). More precisely, in this paper we focus on solutions localized on four
adjacent oscillators (namely with indices $j\in S=\graff{1,2,3,4}$), for a
reason that will be clear in a while. If, in the uncoupled case $\eps=0$, they
are given the same energy (or action), any orbit is periodic (having the
oscillators moving with the same frequency), irrespectively of the phase
differences between them, forming in this way a completely resonant four-dimensional
torus. Our investigation can thus be seen to fall within the general question of
the perturbation of low-dimensional resonant tori in Hamiltonian dynamics.

When we consider only nearest neighbours interactions in
\eqref{ham_long}, i.e., setting $k_2=k_3=0$, it is well known that
only multibreathers with {\it standard phase-differences} ($\varphi=0$
or $\pi$) between adjacent oscillators survive the breaking of the
resonant torus \cite{Kou13}.  If next-to-nearest
(or longer range) neighbour interactions are added, other solutions
with non-standard phase differences may survive: these are called {\it
  phase-shift} multibreathers (see e.g.~\cite{KouK09,PelS12}). The emergence of phase-shift
multibreathers in both one-dimensional KG and dNLS models with
interactions longer than these of the nearest-neighbours interactions,
have been investigated in some recent literature
\cite{KouKCR13,Kev09,ChoCMK11}. This issue partially overlaps with the
study of vortex structures in two-dimensional lattices, like in
\cite{PelKF05b,archilla,KouKLKF10}. Indeed, a suitable long-range
interaction in a one-dimensional lattice allows to reproduce the local
interactions involved in a two-dimensional vortex, for example in a
hexagonal or square lattice, thus providing an emulation of the
two-dimensional object by a one-dimensional one at leading order in
the coupling perturbation parameter $\eps$; such an approximation
clearly fails at higher orders, due to the differences in terms of
lattice shape and interaction among sites.

A special case of a two-dimensional lattice is the
so-called zigzag lattice~\cite{efrem}.  This lattice consists of just two
oscillator chains which are connected as shown in Fig.~\ref{fig:ZZ}.
In this case, we can easily see that {\it vortex} solutions of Fig.~\ref{fig:ZZ}
correspond to {\it four-site} multibreathers in the system of
Fig.~\ref{fig:ZZ_1D}.  The zigzag system is described by a
Hamiltonian
\begin{equation}
  \label{e.KG.zz}
  \begin{aligned}
    \HH_{110} &= \sum_{j\in\ZZ} \left({\frac12 y^2_j + V(x_j)}\right)\\
    &\quad+\frac\eps2\sum_{j\in\ZZ}\quadr{(x_{j+1}-x_j)^2+(x_{j+2}-x_j)^2}\ ,
  \end{aligned}
\end{equation}
that corresponds to a Hamiltonian~\eqref{ham_long} with $k_2=1$ and
  $k_3=0$.  Indeed, the subscript of $\HH$ refers to the values of the coupling
  constants $k$ (including $k_1$ which is always 1 in our notation).

\begin{strip}
\begin{minipage}{\linewidth}
  \makebox[\linewidth]{
    \includegraphics[width=0.5\linewidth]{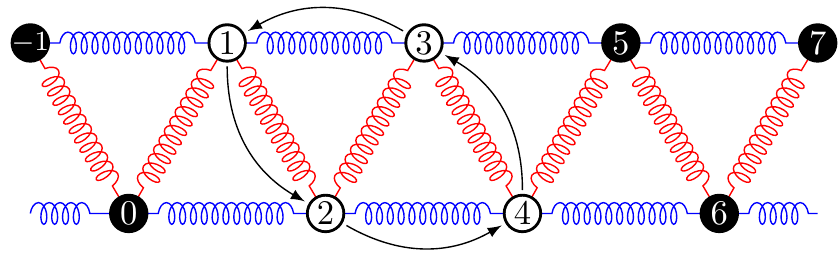}}
    \captionof{figure}{The two-dimensional zigzag model: all the interactions are nearest neighbor ones
      with the same strength. The indexing indicates the energy flow of the vortex
      solutions. Color online.}\label{fig:ZZ}
\end{minipage}

\begin{minipage}{\linewidth}
  \makebox[\linewidth]{
    \includegraphics[width=\linewidth]{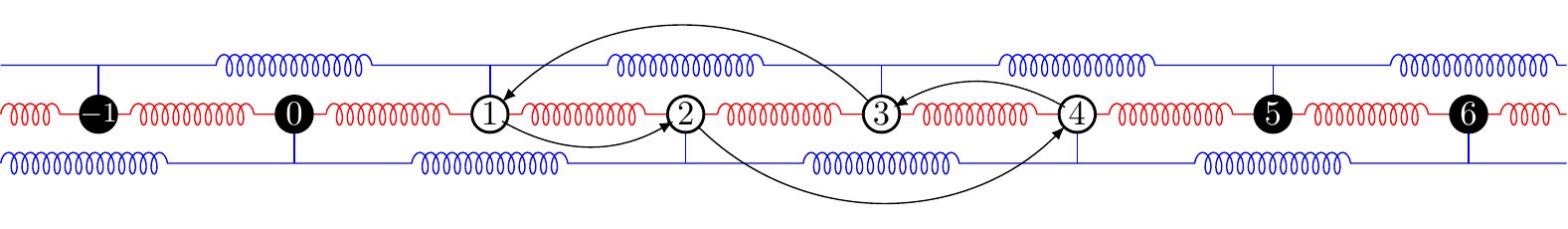}}
  \captionof{figure}{The corresponding one-dimensional zigzag
      model. The numbers
    indicate the correspondance to the two-dimensional
    zigzag model. Color online.}
  \label{fig:ZZ_1D}
\end{minipage}
\end{strip}

Both in the one-dimensional and in the two-dimensional case, the
existence of multibreathers is typically performed via implicit
function theorem arguments, which rely on the non-degeneracy of some
linearized equation. This is the case, for example, of the classical
result in~\cite{AhnMS02}, where true multibreather solutions are
obtained from approximate solutions which correspond to critical
points of an averaged (effective) Hamiltonian: in this context, an
approximate solution has to satisfy some {\sl persistence conditions}
(see e.g.~\cite{KouKCR13}) which select admissible candidates of
phase-differences for a possible continuation. The same analytical
tool, i.e.,~the implicit function theorem, can be used also in a
different scheme: approaching the original problem with a
Lyapunov-Schmidt decomposition (with the torus being resonant), it is
used to solve the Range equation, and then the use of some symmetry,
like time-reversibility, can remove the Kernel directions (see
\cite{PelS12}). However, in some degenerate cases, the candidate
solutions we acquire from the persistence conditions do not correspond
to true solutions of our systems. In such cases, a deeper analysis is
required which typically involves higher order terms of the
bifurcation (kernel) equation.

By studying the persistence conditions for the zigzag system
\eqref{e.KG.zz}, we realize that the candidate vortex solutions
      {of Fig.1 are not isolated, but} appear as two
      one-parameter families within the three-dimensional manifold of
      phase-differences. These two families intersect in {
        what we call {\it symmetric vortex} configuration, since it
        features} the standard vortex phase differences $\Phi^{({\rm
          sv})}\equiv\pmb{\vphi}=\pm(\pi/2, \pi, -\pi/2)$\footnote{The
        reason that the $\Phi^{({\rm sv})}$ configuration is the one
        with $\pm(\pi/2, \pi, -\pi/2)$ and not the $\pm(\pi/2, \pi/2,
        \pi/2)$ as one could have expected, is that, as we can see
        from Fig.\ref{fig:ZZ} the vortex-flow is
        $1\rightarrow2\rightarrow4\rightarrow3\rightarrow1$ while the
        phase differences are calculated using consecutive
        oscillators.}, where $\pmb{\vphi}\equiv(\varphi_1, \vphi_2,
      \vphi_3)$, see~\eqref{e.phi}. On the other hand, we will call
      all the other solutions of these two families, with
      $\pmb{\vphi}\neq\Phi^{({\rm sv})}$ as {\it
        asymmetric vortices}. Let us note here that these families
      also include some of the standard ($\varphi_i \in \{0, \pi\}$)
      multibreather solutions in addition to the isolated standard
      solutions of the persistence conditions.

Due to the degeneracy, which manifests itself into the presence of
families of candidate solutions, and even more in their intersection
points, we attempt to complement our analysis by performing a
numerical investigation of the persistence conditions of the full
problem \eqref{ham_long} in the neighbourhood of the values
$(k_2,k_3)=(1,0)$, which correspond to the zigzag configuration. In
this study, we realize first of all that there exist
families\footnote{Since now we consider the persistence conditions of
  the full problem \eqref{ham_long}, the families are considered in
  the $(k_2, k_3$)-space.} of solutions which are non-degenerate and
consequently easily continued to real solutions. In addition, there is
a solution family at $k_2=1$ for all values of $k_3$.  As $k_3 \to 0$,
we observe that some of the non-degenerate families geometrically
converge also to the $k_2=1$ family increasing in this way the
degeneracy and for $k_3=0$ they become the two vortex families of
solutions. Thus, it is difficult to get a definitive answer on the
existence of true vortex solutions in the case of the Hamiltonian
$\HH_{110}$ only by the study of the persistence conditions.

To get a complete description of the continuation we thus follow a
different route, exploiting the corresponding dNLS model
\begin{equation}
  \begin{aligned}
  \label{e.NLdNLS.zz}
  {H_{110}} &= \sum_j|\psi_j|^2 + \frac38\sum_j|\psi_j|^4\\ &\quad+
  \frac{\eps}2\sum_j \quadr{|\psi_{j+1}-\psi_j|^2 +
    |\psi_{j+2}-\psi_j|^2} \ ,
  \end{aligned}
\end{equation}
as a bridge to $\HH_{110}$. Indeed the former can be shown to be a
good approximation of the latter in the energy regime $E\ll 1$ and for
couplings $\eps\ll\sqrt{E}$ (see, e.g., \cite{BamPP10,PelPP16} or
Subsection \ref{ss:approax}). Moreover, although the system ${H_{\rm
    110}}$ shares the same degeneracy as the original KG model, we are
able to more straightforwardly derive the nonexistence of any
phase-shift discrete soliton of ${H_{\rm 110}}$ following the scheme
of~\cite{PenSPKK16} by exploiting the expansion of an invariant
quantity, i.e., the {\it Density Current}.  Since this efficient
nonexistence strategy is based on some minimal smoothness assumption
with respect to $\eps$, to get nonexistence assuming only continuity,
we also expand the bifurcation equation at leading orders showing that
this ${H_{\rm 110}}$ case is less degenerate than the one studied
in~\cite{PenSPKK16}. This weaker degeneracy allows to deduce
nonexistence of the continuation by verifying a sufficient condition
on the linearized bifurcation equation. Since this sufficient
condition is robust under small perturbation, we are then able to
transfer the nonexistence result of ${H_{110}}$ to the original system
$\HH_{110}$, showing the {\sl nonexistence} of any vortex solution
(symmetric or asymmetric) for the degenerate model \eqref{e.KG.zz}, in
the prescribed regime of the two main parameters $E$ and $\eps$. We
may claim that the nonexistence result itself, in the presence of
degeneracy, and the above mentioned indirect strategy here applied,
represents the key point of the paper.

In order to state such a result (which can be found in
Section~\ref{s:cont} in a slightly more technical formulation), we
introduce the four-dimensional resonant torus filled by periodic
orbits, belonging to the possible solutions of $\HH_{110}$ for
$\eps=0$
\begin{equation}
  \label{e.u0}
  \bar u_j(\tau)=
  \begin{cases}
    0\ ,                &j\not\in S \\
    x(\tau+\theta_j)\ , &j    \in S
  \end{cases}\ ,
\end{equation}
where $ S=\{1,2,3,4\}$ and $x(\tau)$ is a nonlinear oscillation of
\begin{equation}
  \label{e.nonlin.osc}
  \gamma^2 x'' + x + x^3=0\ ,\qquad x(0)=\rho\ ,
\end{equation}
where $\tau:=\gamma t$ is the rescaled time induced by the frequency
$\gamma$ associated to the (small) amplitude $\rho$ of the
oscillation, and $\varphi_j$ are phase differences between the above
mentioned (which are also called as ``central'') successive
oscillators with
\begin{equation}
\vphi_j := \theta_{j+1} - \theta_j\ ,\quad j\in S^*=\{1,2,3\}. 
\label{e.phi}
\end{equation}
We have then
\begin{theorem}
\label{t.nonexistence}
For $\eps$ small enough ($\eps\not=0$), the only four-site unperturbed
solutions \eqref{e.u0} that can be continued, at fixed frequency
$\gamma$, to solutions $u_j(\rho,\eps,\tau)$ of \eqref{e.KG.zz},
correspond to $\varphi_j\in\graff{0,\pi}$.
\end{theorem}

Moreover, in all the cases which appear to be non-degenerate, the
``dNLS approximation'' strategy, allows to { derive any
  existence result for \eqref{ham_long} from the existence result for
  the corresponding} dNLS model
\begin{equation}
  \begin{aligned}
    \label{e.NLdNLS}
          {H} &= \sum_j|\psi_j|^2 + \frac38\sum_j|\psi_j|^4
          \\ &\quad+\frac{\eps}2\sum_j\quadr{ |\psi_{j+1}-\psi_j|^2 +
            k_2|\psi_{j+2}-\psi_j|^2+k_3|\psi_{j+3}-\psi_j|^2} \ ,
  \end{aligned}
\end{equation}
and provides explicit (though not sharp) estimates of the
approximation of asymmetric vortices in \eqref{ham_long} with the
corresponding phase-shift discrete solitons in
\eqref{e.NLdNLS} (cf. with Theorem 2.1 of \cite{BamPP10}).

The price one has to pay for the use of this strategy lies in the
restrictions in the regime of parameters for which the models
\eqref{ham_long} are well approximated by the corresponding averaged
normal forms \eqref{e.NLdNLS}. {However, Hamiltonian
  normal form theory provides the tools needed to modify the
  approximating models in different regimes of the parameters $E$ and
  $\eps$, as in \cite{PalP14,PelPP16}, thus leading to a new dNLS-type
  starting model for an indirect approach. Such a new nonlocal dNLS
  normal form would surely include additional linear and nonlinear
  corrections with respect to those present in \eqref{e.NLdNLS}.}
We conclude this section by remarking that there exist even more
degenerate models within the family~\eqref{ham_long}. One of these is
the Hamiltonian $\HH_{101}$ (see~\eqref{e.KG.deg}) which has $k_2=0$
and $k_3=1$ and it is used for the study of vortex-like configurations
in two-dimensional square lattices. This system admits, at the level
of the persistence condition, three vortex families, having the
symmetric vortex configuration in their triple intersection, giving
thus a complete degeneration. We stress that the corresponding dNLS
model is exactly the one studied in~\cite{PenSPKK16}, but within the
scheme implemented in the present paper, the higher degeneration of
$\HH_{101}$ does not allow us to transfer the nonexistence result
proved in~\cite{PenSPKK16} to the corresponding KG chain. We are
presently exploring a different normal form strategy which works
directly on the original KG model and interpret the problem in the
classical sense of breaking of a completely resonant low-dimensional
torus \cite{PenSD17}.

This paper is structured as follows. The numerical explorations are
reported in Section~\ref{s:NBA}, where we perform a study of the
persistence conditions of both the zigzag~\eqref{e.KG.zz} and the $\HH_{101}$ system as well as of
the full system~\eqref{ham_long} in the $(k_2, k_3)$-parameter values neighborhood which correspond to the zigzag and the $\HH_{101}$ systems.  The mathematical
strategy and the main nonexistence result for the KG
model~\eqref{e.KG.zz} is given in Section~\ref{s:cont}, while in
Section~\ref{s:break} we use the first order expansion of the Density
Current and the sufficient condition on the linearized bifurcation
equation in order to prove nonexistence of vortex solutions in the
zigzag-dNLS system \eqref{e.NLdNLS.zz}.  Finally,
Section~\ref{s:concl} includes some concluding remarks about possible
future directions on the topic.

\section{Study of the persistence conditions}
\label{s:NBA}

In the present Section we investigate the possibility of existence of
multibreather solutions by following a procedure similar to the one
introduced in \cite{KouKCR13} using the results of
\cite{KouI02,KouM05,KouK09}, and by numerical calculations.

%
%

\subsection{The zigzag system}
\label{s:21}
The {\it persistence conditions} for the zigzag system are the
equations that provide the candidate configurations among the ones in
the anticontinuous limit $\epsilon=0$, that could be continued, for
$\epsilon$ nonzero but small enough, to provide multibreather
solutions for this system.  These equations are derived through an
averaging procedure which is described in \cite{{KouKCR13}} and the
main points will be presented in what follows.  For $\epsilon=0$, we
consider four ``central'' oscillators moving (with the same frequency
but arbitrary initial phases) while the rest lie at their
equilibrium. The motion of each one of these central oscillators is
described by the cosine Fourier expansion
  \begin{equation}
    x(w,J)=\sum_{n=1}^\infty A_{2n-1}(J)\cos[(2n-1)w]\ ,
    \label{frm:fouexp}
  \end{equation}
where $(J,w)$ refer to the action-angle variables for the single
oscillator while the lack of the even terms $A_{2n}$ in the Fourier
expansion stems from the symmetry of the potential $V$.

Since we want to study structures with four central oscillators, only
three phase differences $\varphi_i$ between them are defined as in
\eqref{e.phi}. It has been proven in \cite{AhnMS02} that the
configurations $\pmb{\vphi}=(\vphi_1, \vphi_2, \vphi_3)$ that could be
continued for $\epsilon\neq0$ to provide multibreather solutions
correspond to critical points of the {\it effective Hamiltonian},
which at first order of approximation is given by
$\HH^{\text{eff}}=\HH_0(J_i)+\epsilon\langle \HH_1 \rangle(\vphi_i,
J_i)$. The average value of the coupling term of the Hamiltonian
\eqref{ham_long} $\langle \HH_1 \rangle$ is calculated along the
unperturbed orbit and reads
\begin{align*}
  \langle \HH_1 \rangle=
  -\frac{1}{2}\sum_{m=1}^\infty &A_m^2\Bigl( \cos(m\phi_1)+ \cos(m\phi_2)+\cos(m\phi_3)+\\
  &+k_2\bigl(\cos(m(\phi_1+\phi_2))+\cos(m(\phi_2+\phi_3))\bigr)\\
  &+ k_3\cos(m(\phi_1+\phi_2+\phi_3)) \Bigr)\ .
\end{align*}

The persistence conditions are obtained from the relation
$\frac{\pa\langle\HH_1\rangle}{\pa\vphi_i}=0$ and for the case of the
zigzag system~\eqref{e.KG.zz}, i.e., with $k_2=1$ and $k_3=0$, read
\begin{equation}
  {\cal P}_{110}(\pmb{\vphi})\equiv
  \begin{cases}
    M(\varphi_1)+M(\varphi_1+\varphi_2)=0                   \\
    M(\varphi_2)+M(\varphi_1+\varphi_2)+M(\varphi_3+\varphi_2)=0   \\
    M(\varphi_3)+M(\varphi_2+\varphi_3)=0
  \end{cases}
  \label{e.per4_KG.zz}
\end{equation}
with 
\begin{equation}
  \label{e.M_phi}
  M(\varphi)\equiv\sum_{m=1}^{\infty}(2m-1)A_{2m-1}^2\sin((2m-1)\varphi) \ ,
\end{equation}
with $A_i$ as in~\eqref{frm:fouexp}.

Note that, in the dNLS case, the conditions (\ref{e.per4_KG.zz}) hold but
with 
\begin{equation}
M(\varphi)\equiv \sin(\varphi)\ ,
\label{e.M_phi.DNLS}
\end{equation}
due to the rotational symmetry of the model, i.e., only the first
Fourier mode contributes.

Taking under consideration the symmetries of $M(\phi)$ 
\begin{align*}
  M(\pi+\varphi) &= - M(\varphi)\,,
  &M(-\varphi)&= -M(\varphi) = M(2\pi-\varphi)\,,\\
  M(\pi-\varphi) &= + M(\varphi)\,,
  &M(0)&=M(\pi)=0\ ,
\end{align*}
it is straightforward to check that the persistence conditions both
for the zigzag-KG case, i.e.,~\eqref{e.per4_KG.zz}
and~\eqref{e.M_phi}, as well as the zigzag-DNLS case,
i.e.,~\eqref{e.per4_KG.zz} and~\eqref{e.M_phi.DNLS}, admit two
families of solutions
\begin{equation}
  \label{e.fam.dnls.zz}
  F_1: \pmb{\vphi}=(\varphi,\pi,-\varphi)   \ ,
  \qquad
  F_2: \pmb{\vphi}=(\varphi,\pi,\pi+\varphi)\ ,  
\end{equation}
in addition to the other four standard isolated solutions
$F_{\text{iso}}=\big\{(0,0,0)$, $(0,0,\pi)$, $(\pi,0,0)$,
$(\pi,0,\pi)\bigr\}$.  In principle, all combinations of $0$'s and
$\pi$'s work trivially, since the persistence conditions simply
vanish.  We have to note here that the rest of the standard
multibreather solutions are part of the $F_1$ and $F_2$ families.

It is important to stress that conditions (\ref{e.per4_KG.zz}) are
necessary but not sufficient for the existence of multibreather
solutions. Indeed, in order to continue to real solutions
of~\eqref{e.KG.zz}, the corresponding Jacobian matrix
$D_\vphi(\cal{P})$ needs to be non-degenerate.  The matrix
$D_\vphi({\cal P}_{110})$ is given by
\begin{equation*}
  \tiny
  \begin{pmatrix}
    M'(\!\vphi_1\!) \!+\! M'(\!\vphi_1\!\!+\!\vphi_2\!) & M'(\!\vphi_1\!\!+\!\vphi_2\!) & 0\\
    M'(\!\vphi_1\!\!+\!\vphi_2\!) & M'({\!\vphi_2\!})\!+\! M'(\!\vphi_1\!\!+\!\vphi_2\!)\!+\! M'(\!\vphi_2\!\!+\!\vphi_3\!)   & M'(\!\vphi_2\!\!+\!\vphi_3\!)\\
    0 & M'(\!\vphi_2\!\!+\!\vphi_3\!) & M'(\!\vphi_3\!) \!+\! M'(\!\vphi_2\!\!+\!\vphi_3\!)\\
  \end{pmatrix},
\end{equation*}
where
$M'(\varphi)\equiv\sum_{m=1}^{\infty}(2m-1)^2A_{2m-1}^2\cos((2m-1)\varphi)$. By
using the symmetries of $M'(\varphi)$
\begin{equation*}
  \begin{aligned}
  &M'(2\pi -\varphi) = M'(\varphi) =M'(-\varphi)\ , \\
  &M'(\pi-\varphi) = M'(\pi+\varphi) = -M'(\varphi)\ , \\
  &M'\left(\frac{3\pi}{2}\right)=M'\left(\frac{\pi}{2}\right) = 0\ ,
  \end{aligned}
\end{equation*}
it is easy to check that for the isolated solutions $F_{\text{iso}}$
the matrix $D_\vphi({\cal P}_{110})$ is non-degenerate so these
solutions will be continued for $\epsilon\neq0$ to provide
multibreathers.

On the other hand, for the $F_1, F_2$ families, which correspond to
asymmetric vortices, $D_\vphi({\cal P}_{110})$ is degenerate
possessing one zero eigenvalue, reflecting the freedom of these
solutions with respect to variations in $\varphi$. So, we cannot know
if these solutions are also true multibreather solutions of the
system.

In particular, for the configurations where the two families cross
each other and correspond to the two symmetric vortices, i.e.,
$\pmb{\vphi}=\pm\Phi^{({\rm sv})}\equiv\pm(\pi/2, \pi, -\pi/2)$, the
matrix $D_\phi({\cal P}_{101})$ reads
\begin{displaymath}
  \left. D_\vphi({\cal P}_{101})\right|_{\varPhi^{({\rm sv})}} = \begin{pmatrix}
    0 & 0 & 0\\
    0 & M'({\pi}) & 0\\
    0 & 0 & 0\\
  \end{pmatrix}\ .
\end{displaymath}
This means that its degeneracy is even higher since the dimension of its kernel is exactly two,
i.e., given by the tangent directions to the two independent families
in the vortex solutions.


\subsection{The full system close to $k_2=1, k_3=0$}

Given the above analysis, it seems interesting to study {
  the role of the two families $F_1$ and $F_2$ of vortex solutions of
  the persistence conditions}
as the full system \eqref{ham_long}
tends to the zigzag model \eqref{e.KG.zz}. This occurs close to the
$(k_2,k_3)=(1,0)$ point of the $k$-parameter space.

The persistence conditions of the full system \eqref{ham_long} read

\begin{equation}
  {\cal P}(\pmb{\vphi})\equiv
  \begin{cases}
    M(\varphi_1)+k_2M(\varphi_1+\varphi_2)+k_3M(\varphi_1+\varphi_2+\vphi_3)=0                   \\
    M(\varphi_2)+k_2M(\varphi_1+\varphi_2)+k_2M(\varphi_3+\varphi_2)\\\hfill+k_3M(\varphi_1+\varphi_2+\vphi_3)=0   \\
    M(\varphi_3)+k_2M(\varphi_2+\varphi_3)+k_3M(\varphi_1+\varphi_2+\vphi_3)=0
  \end{cases}
\label{e.per4_KG.full}
\end{equation}
where $M$ is defined as in \eqref{e.M_phi}.

The proper illustration of the solutions
of~\eqref{e.per4_KG.full} would require a three-dimensional
plot for every phase-difference $\varphi_{i}$ as a function of both $k_2$ and
$k_3$. Since this surface is difficult to be properly illustrated, we prefer to
present some sections, first for fixed
$k_3$, varying $k_2$, and then by reversing the
roles between $k_2$ and $k_3$.


\subsubsection{The ${k_3< 0}$ case}
\begin{figure}[htbp]
  \centering
  \includegraphics[width=0.7\linewidth]{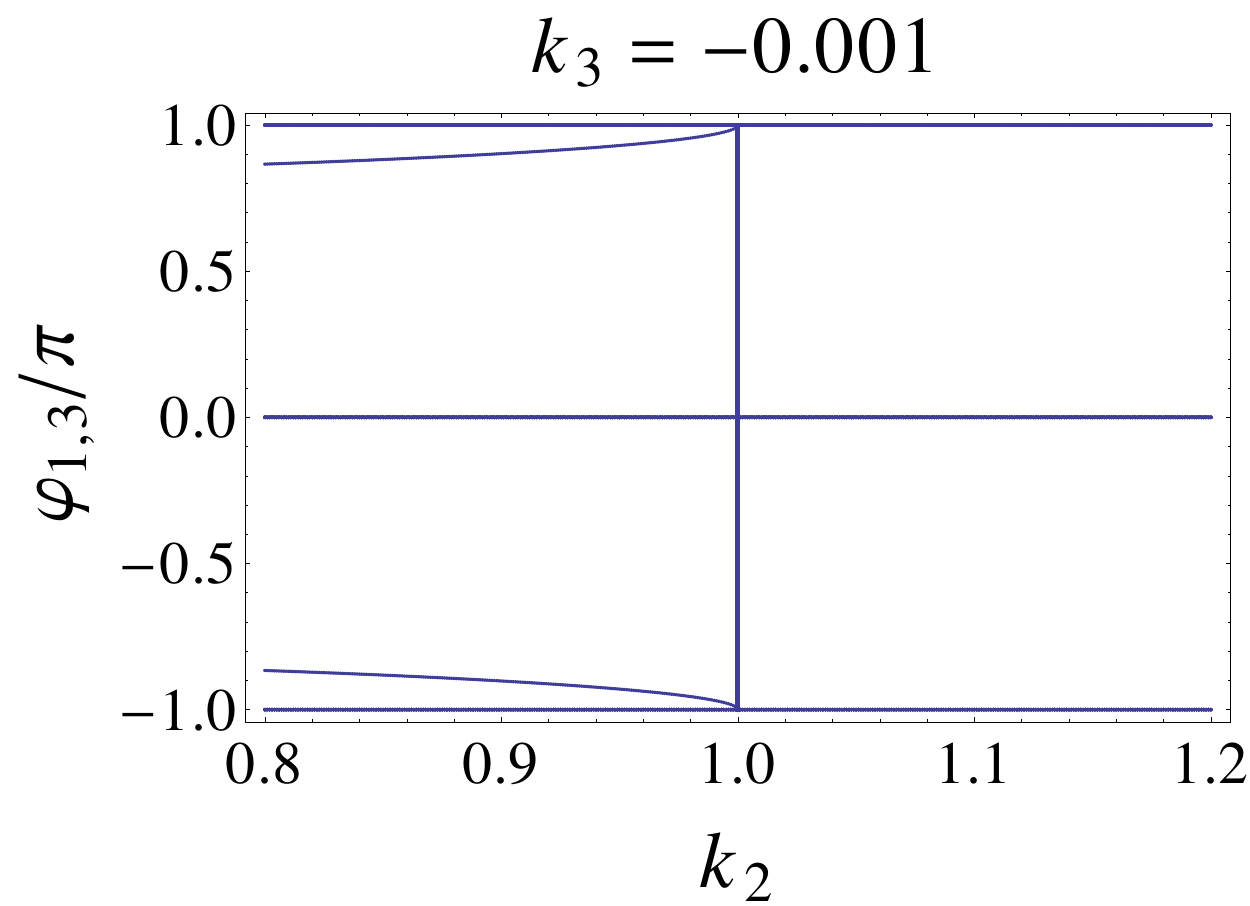}
  \includegraphics[width=0.7\linewidth]{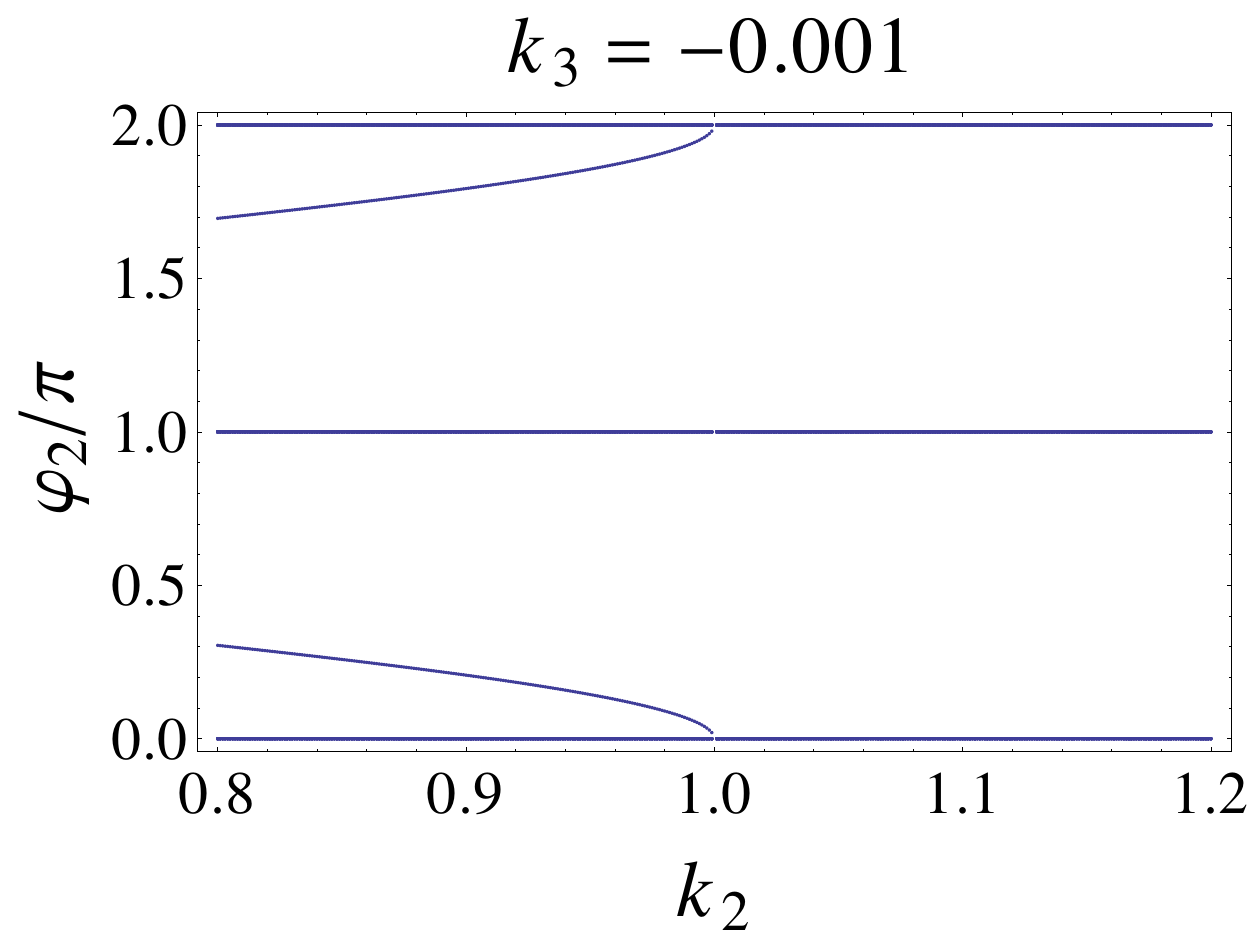}
  \caption{The complete bifurcation diagrams for $k_3=-0.001$. In this
    diagrams all the solution families are shown. There are no vortex-related families for $k_3<0$ except the $F_1$ one which exists for every $k_3$.}
  \label{fig:global_k3_m_0_001}
\end{figure}

A representative bifurcation diagram of this parameter region is shown
in Fig.~\ref{fig:global_k3_m_0_001} where the solutions of the
persistence conditions \eqref{e.per4_KG.full} are depicted for
$k_3=-0.001$ and $0.8\leqslant k_2\leqslant 1.2$. In this diagram all
the solution families are shown, where we can distinguish both
standard and phase-shift configurations. The family of solutions $F_1:
\pmb{\vphi}=(\varphi,\pi,-\varphi) $ , discussed in Section~\ref{s:21}
(see (\ref{e.fam.dnls.zz})), exists also for the persistence
conditions of the full system for every value of $k_3$ and $k_2=1$ as
it is easy to realize by substituting the values of $\pmb{\vphi}$
which correspond to $F_1$ into \eqref{e.per4_KG.full}. This family is
depicted as a vertical line at $k_2=1$ in the top panel of
Fig.~\ref{fig:global_k3_m_0_001}. In this diagram another phase-shift
solution family is shown which lies close to $\vphi_{1,3}=\pi$,
$\vphi_2=0$. {This family is not related to vortex
  solutions in the sense that it does not converge to any of the $F_1$
  or $F_2$ families as $(k_2, k_3)\to (1, 0)$: indeed, for $F_{1,2}$
  one has $\vphi_2=\pi$.} On the contrary it remains almost invariant
in the parameter region under consideration.
  
\subsubsection{The ${k_3\geq 0}$ case}
\begin{figure}[htbp]
  \centering
  \includegraphics[width=0.7\linewidth]{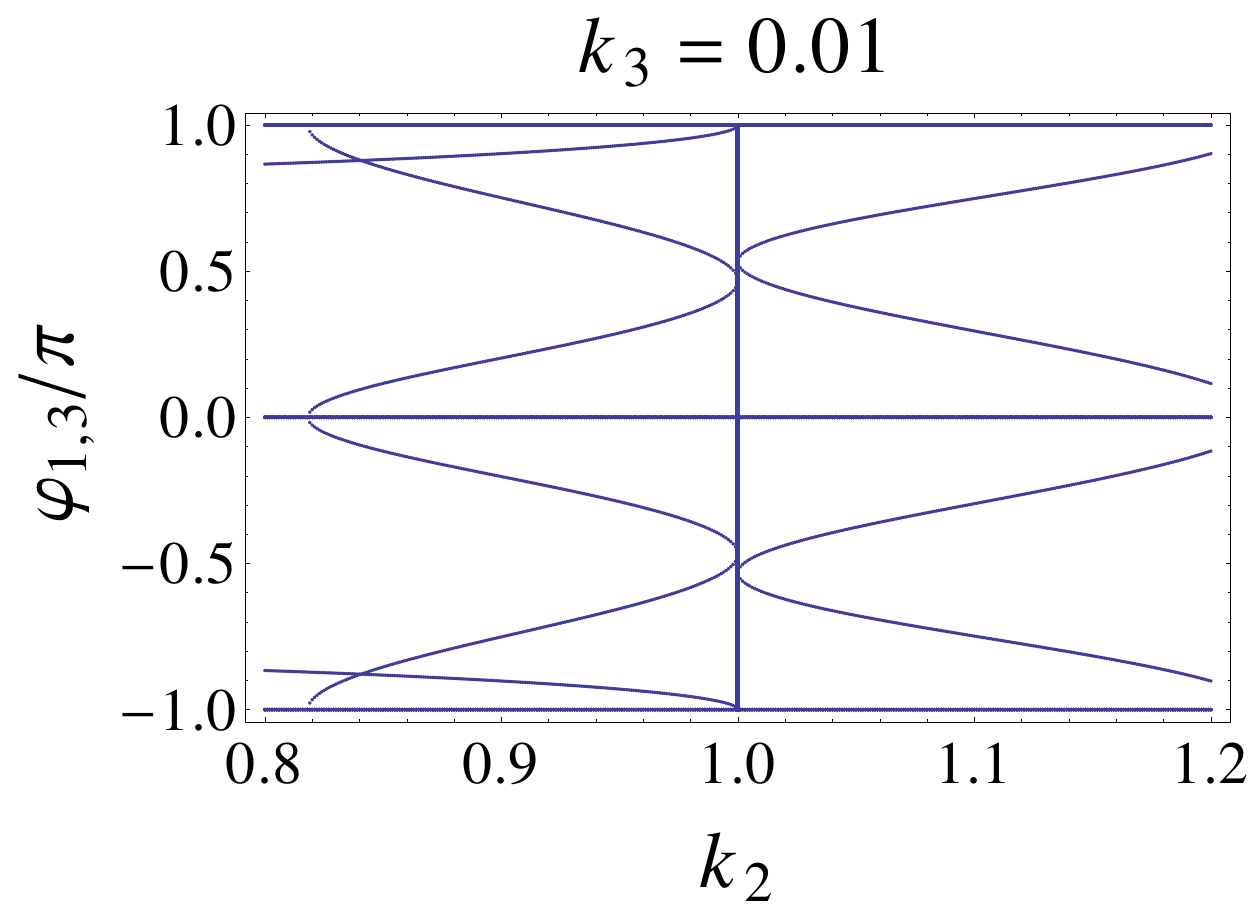}
  \includegraphics[width=0.7\linewidth]{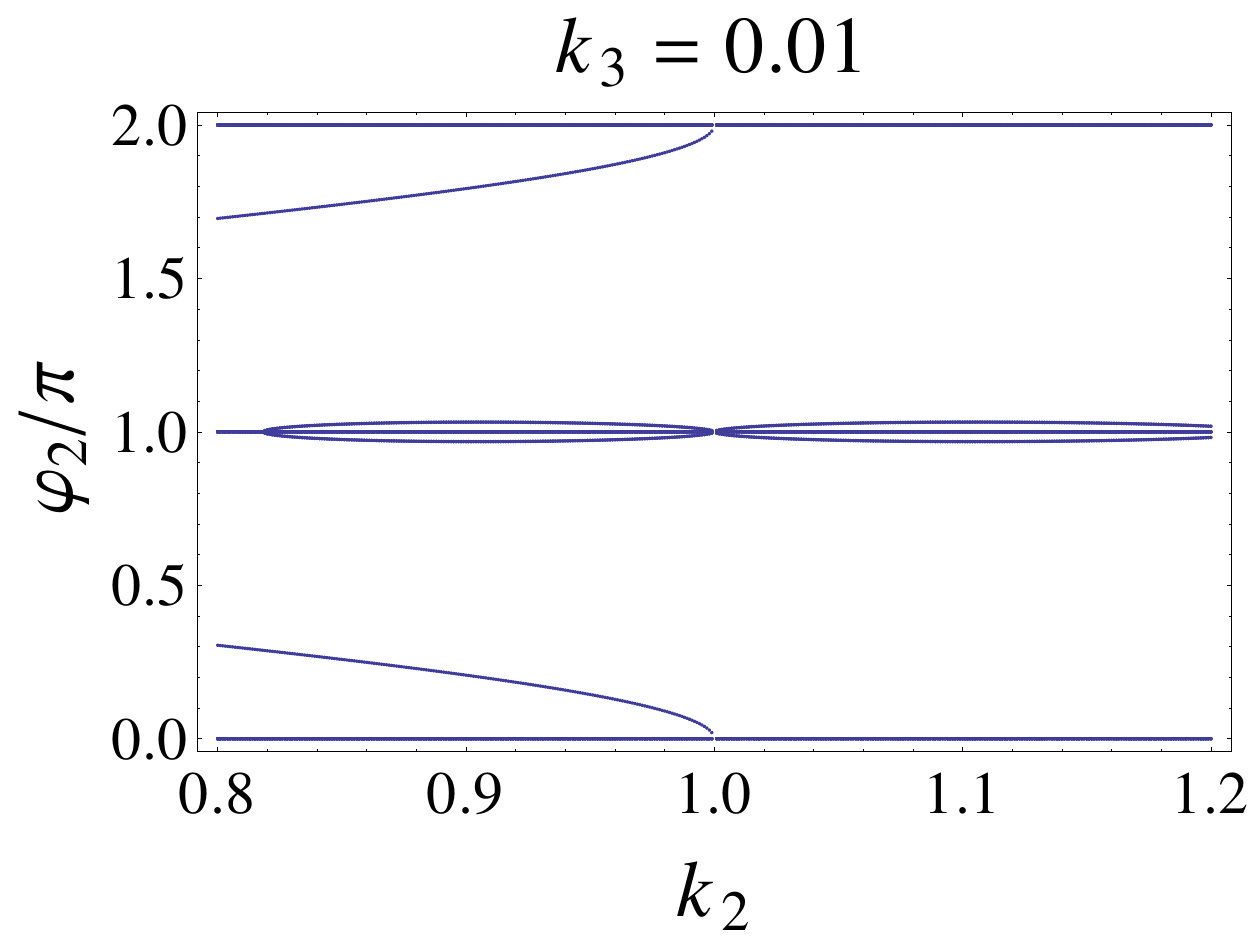}
  \caption{The complete bifurcation diagrams for $k_3=0.01$. In
    addition to the already existing families which for $k_3< 0$, in
    the $k_3\geq 0$ case two more phase-shift appear, which bifurcate
    from $\pmb{\vphi}=\varPhi^{({\rm sv})}$ (see also the figures
    below).}
  \label{fig:global_k3_m_0_01}
\end{figure}

\begin{figure*}[t]
  \centering
  \includegraphics[width=4.5cm]{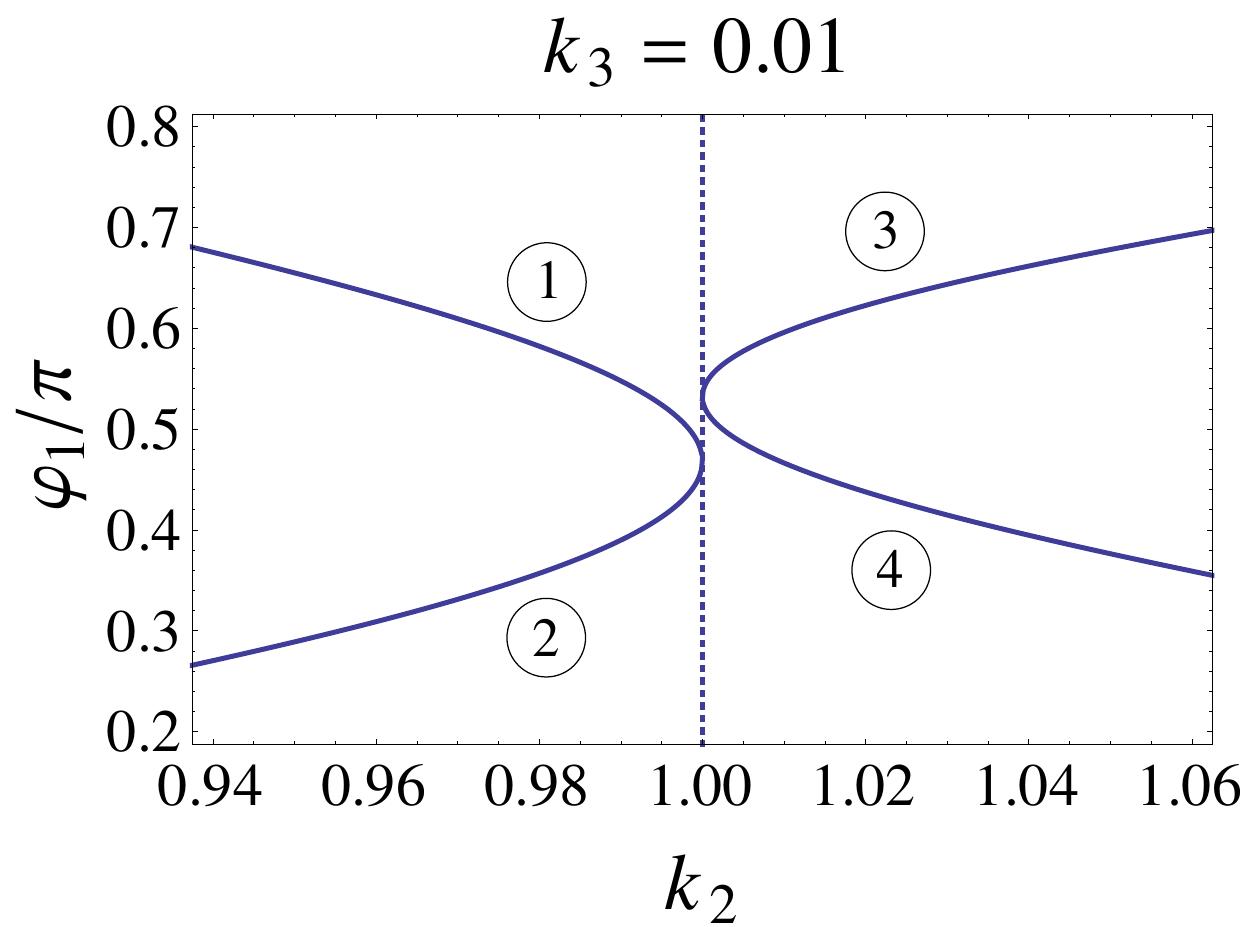}
  \hspace{0.1cm}
  \includegraphics[width=4.5cm]{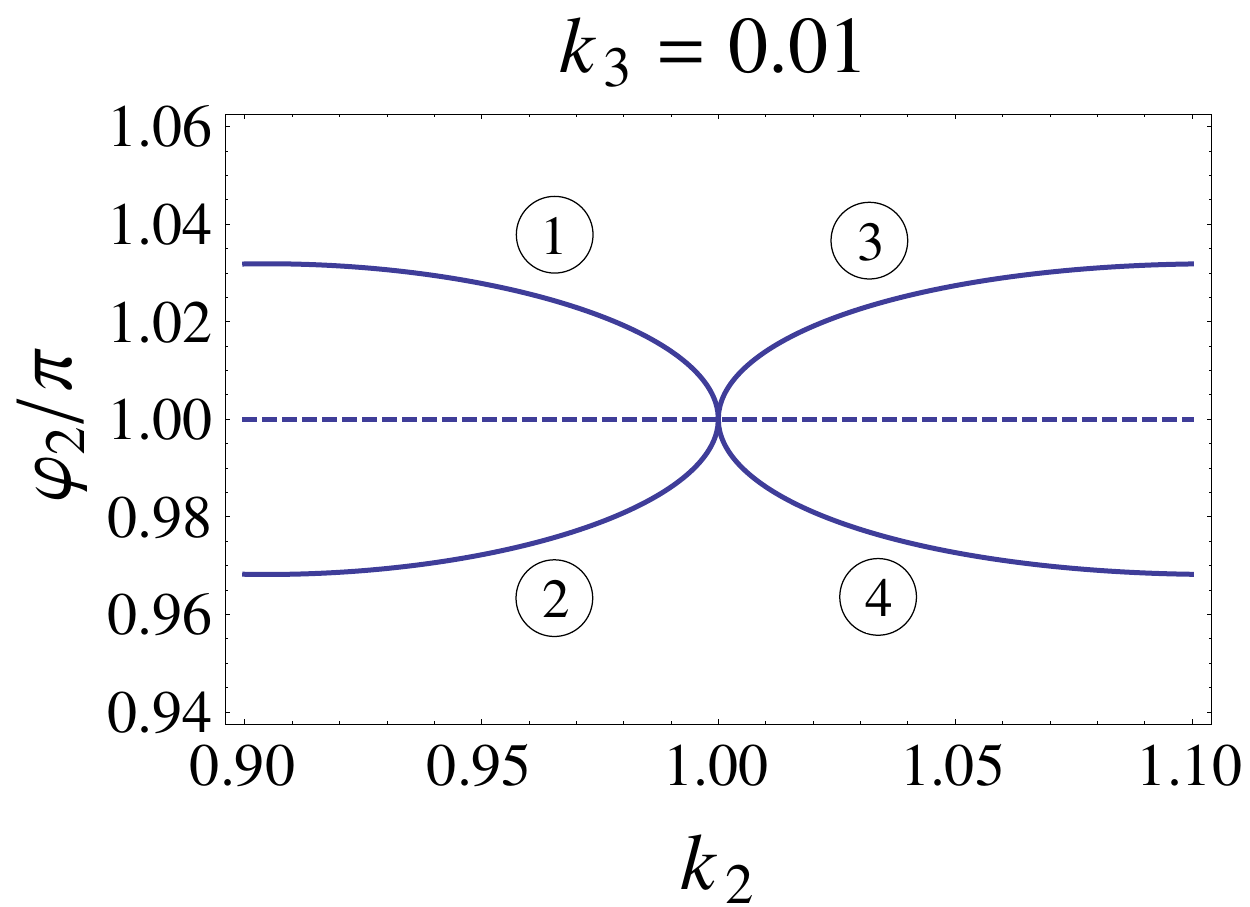}
  \hspace{0.1cm}
  \includegraphics[width=4.5cm]{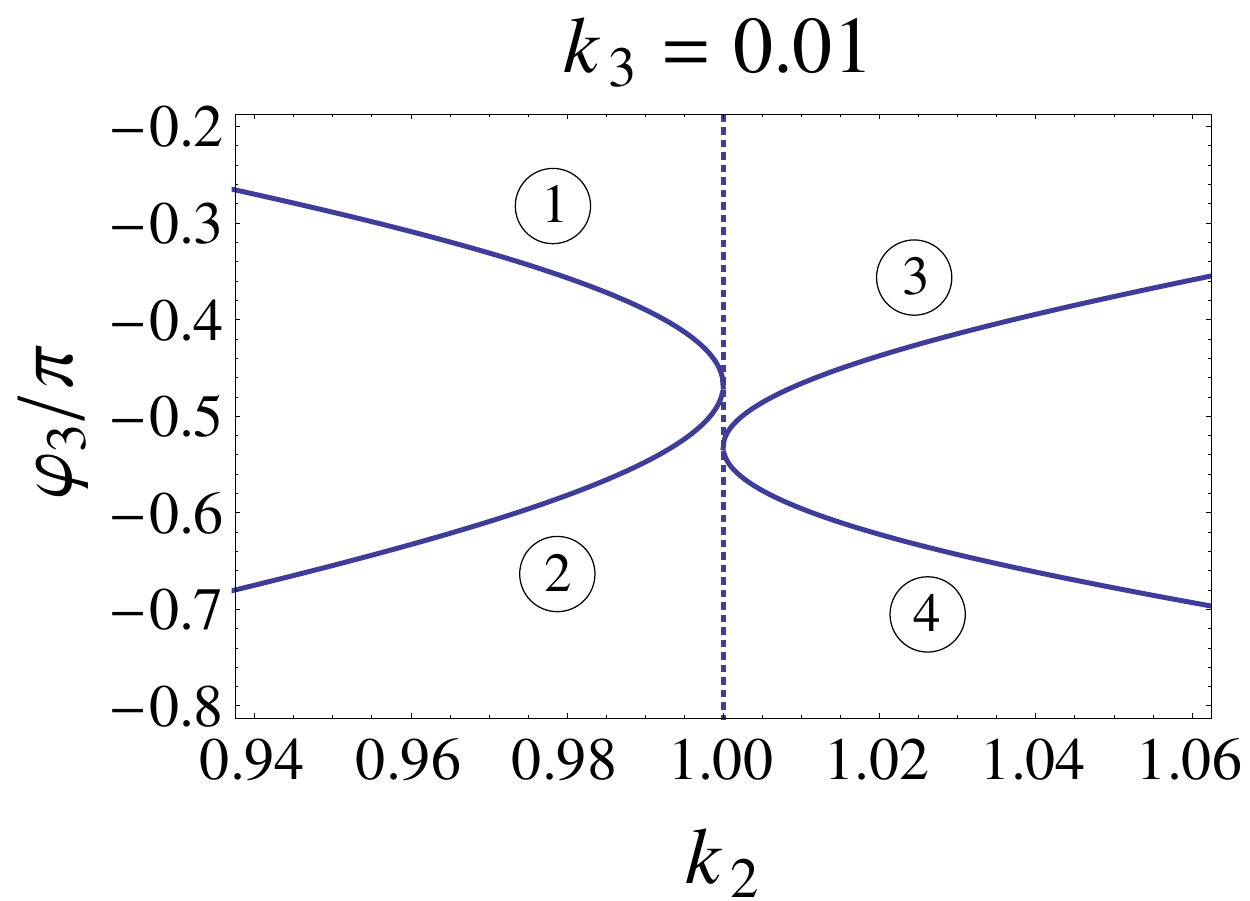}
  \\
  \includegraphics[width=4.5cm]{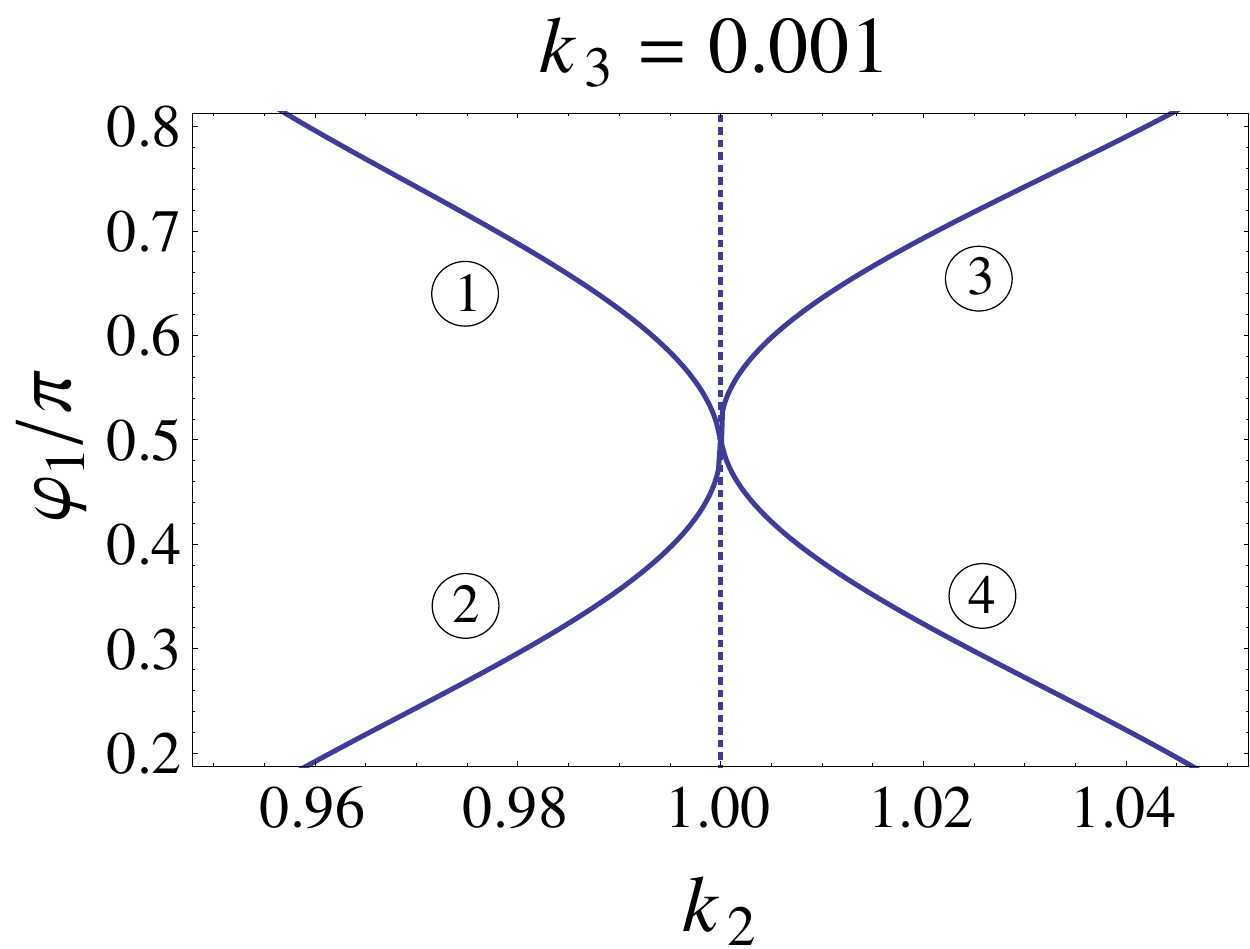}
  \hspace{0.1cm}
  \includegraphics[width=4.5cm]{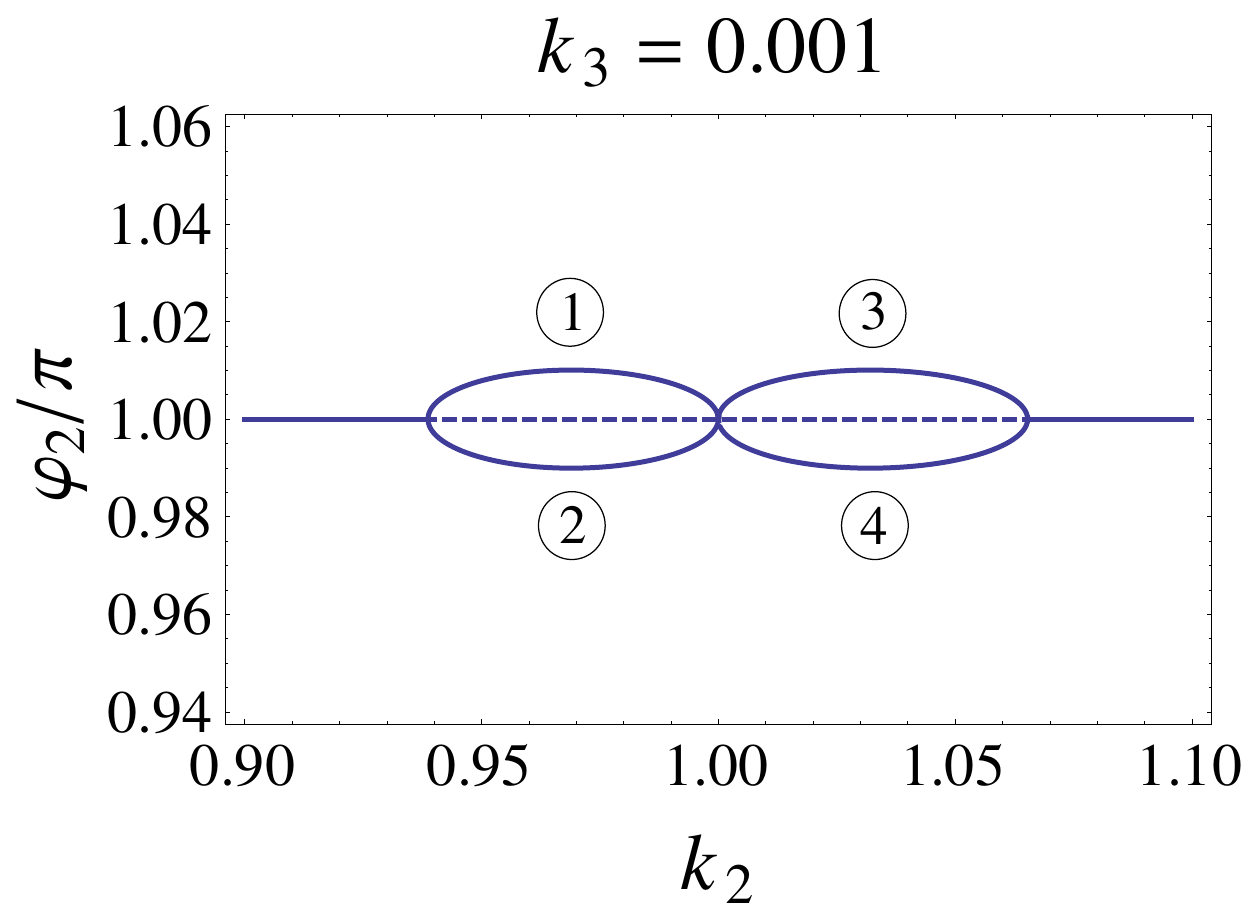}
  \hspace{0.1cm}
  \includegraphics[width=4.5cm]{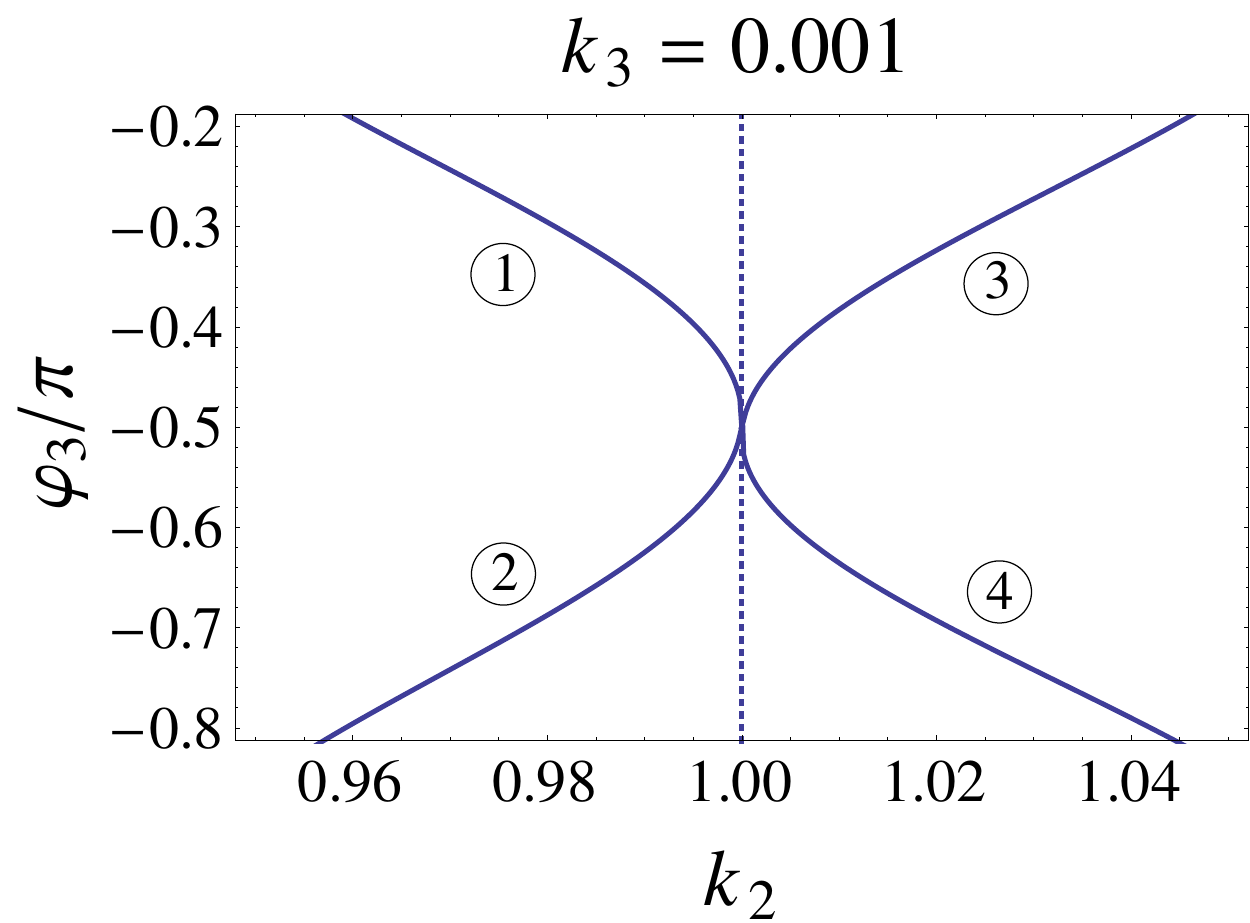}
  \\
  \includegraphics[width=4.5cm]{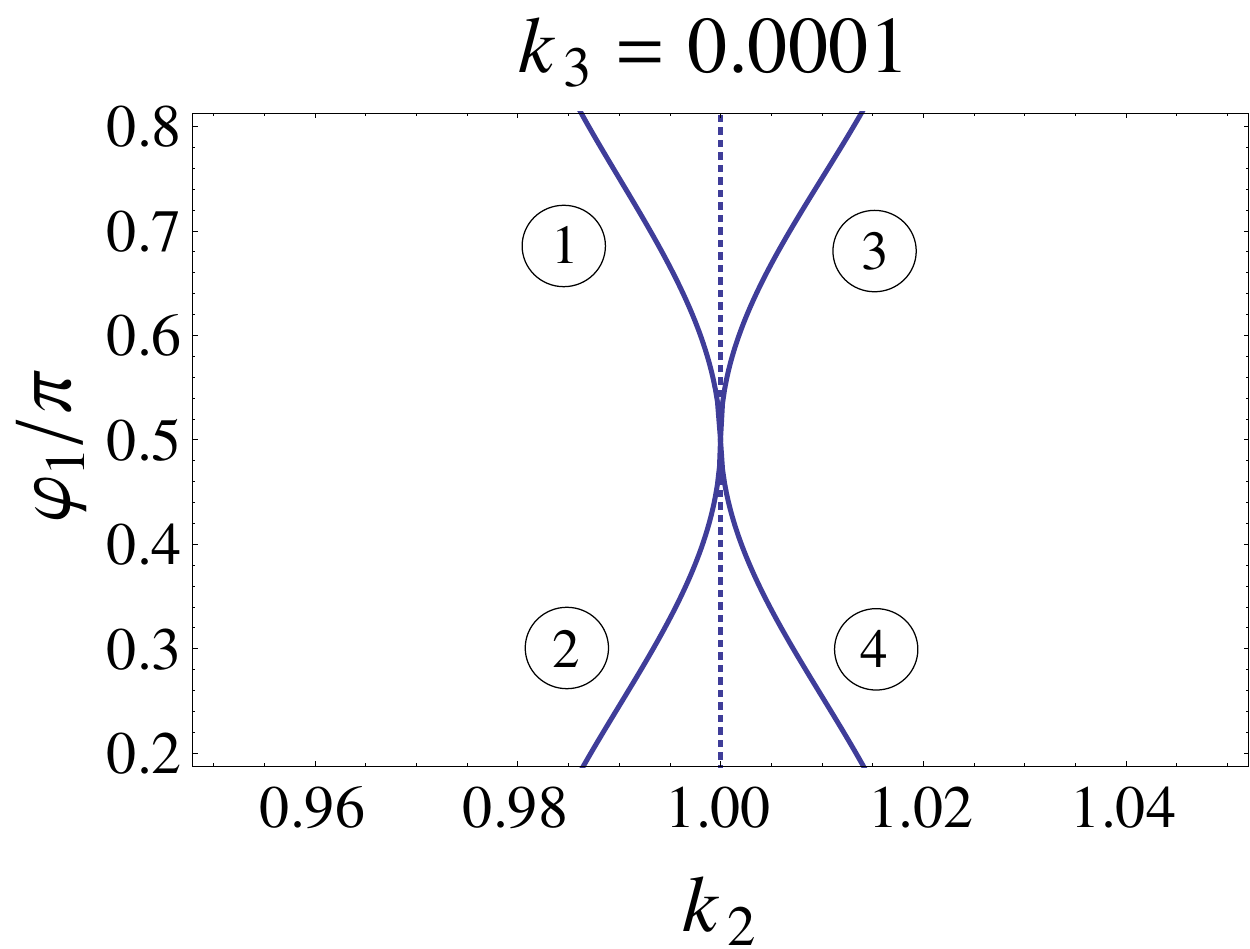}
  \hspace{0.1cm}
  \includegraphics[width=4.5cm]{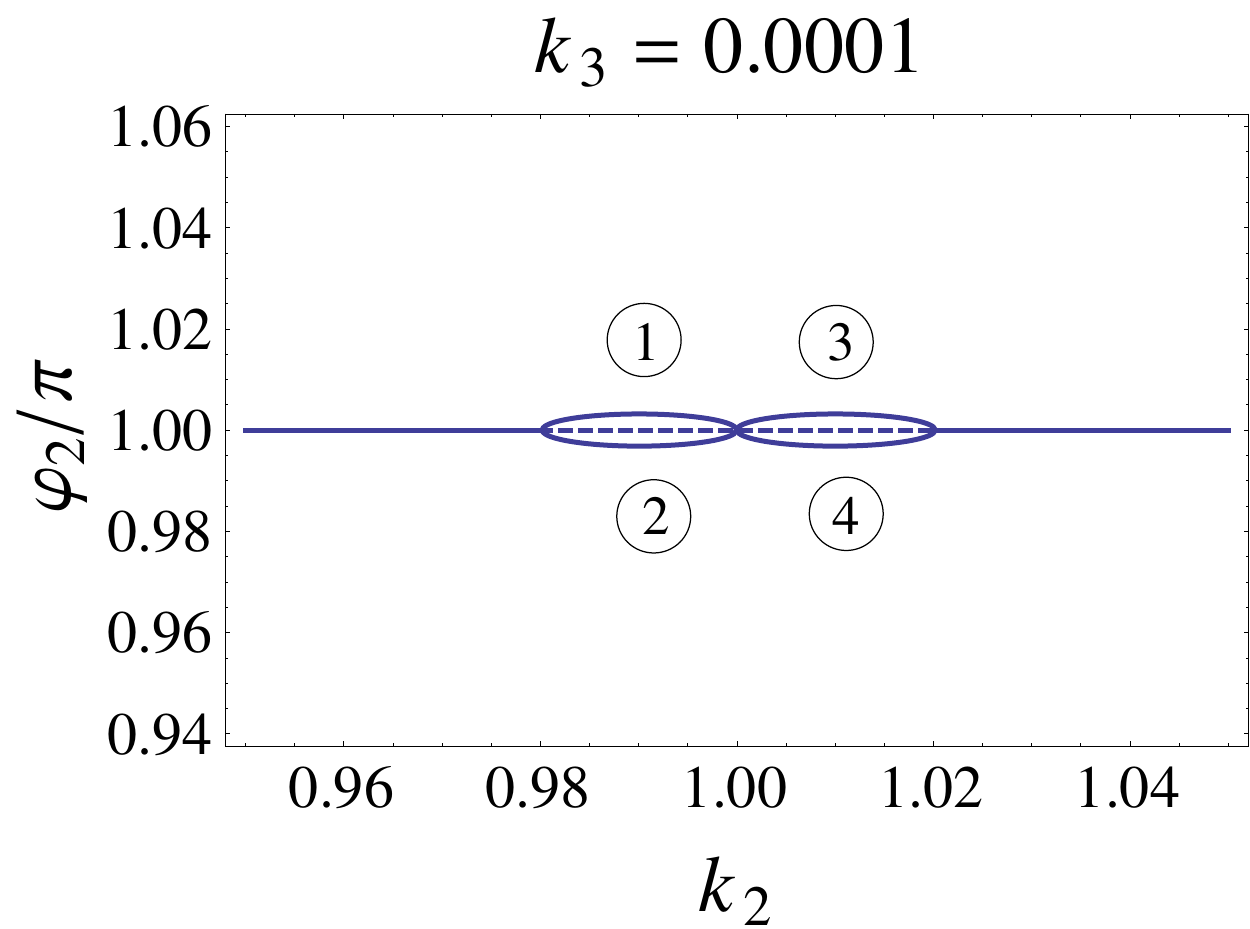}
  \hspace{0.1cm}
  \includegraphics[width=4.5cm]{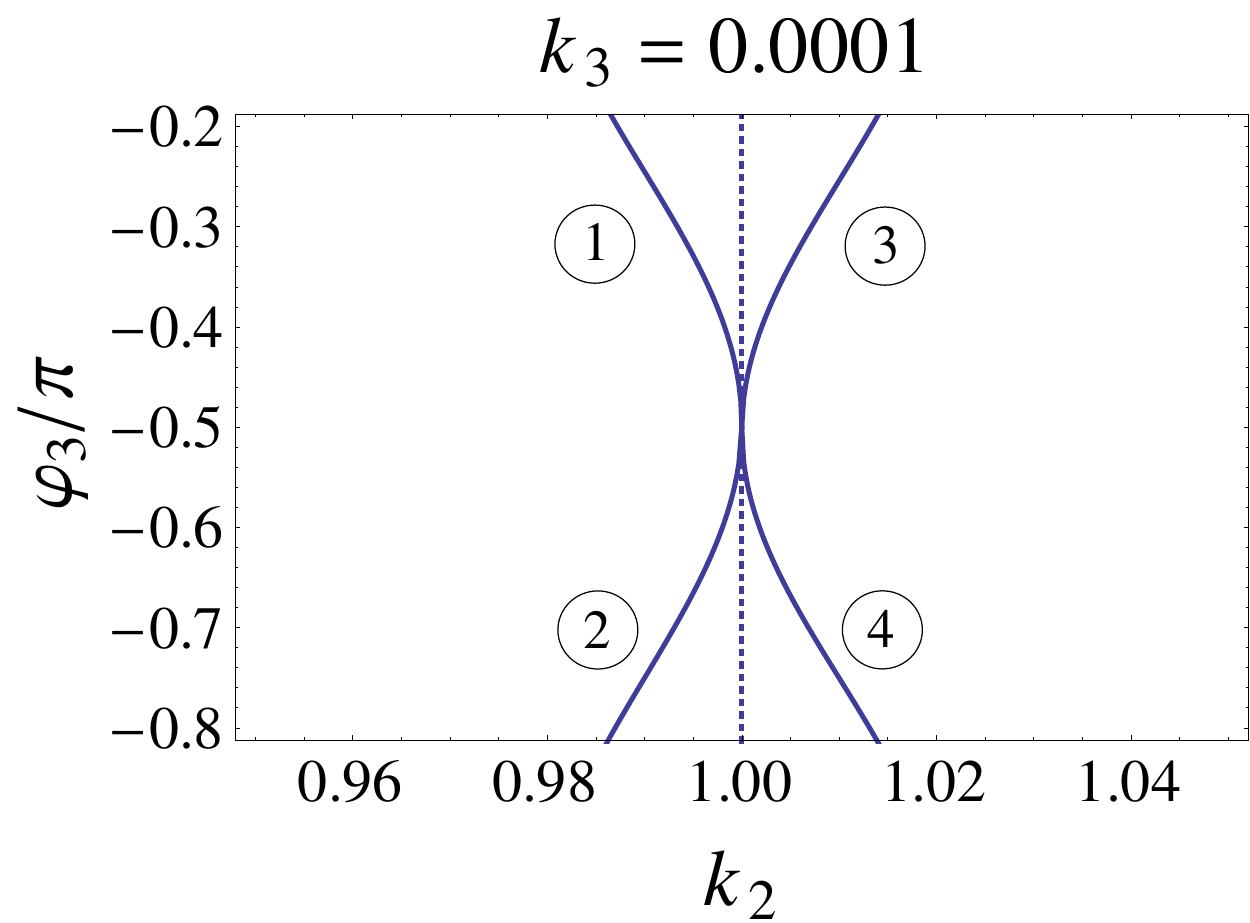}
  \\
  \includegraphics[width=4.5cm]{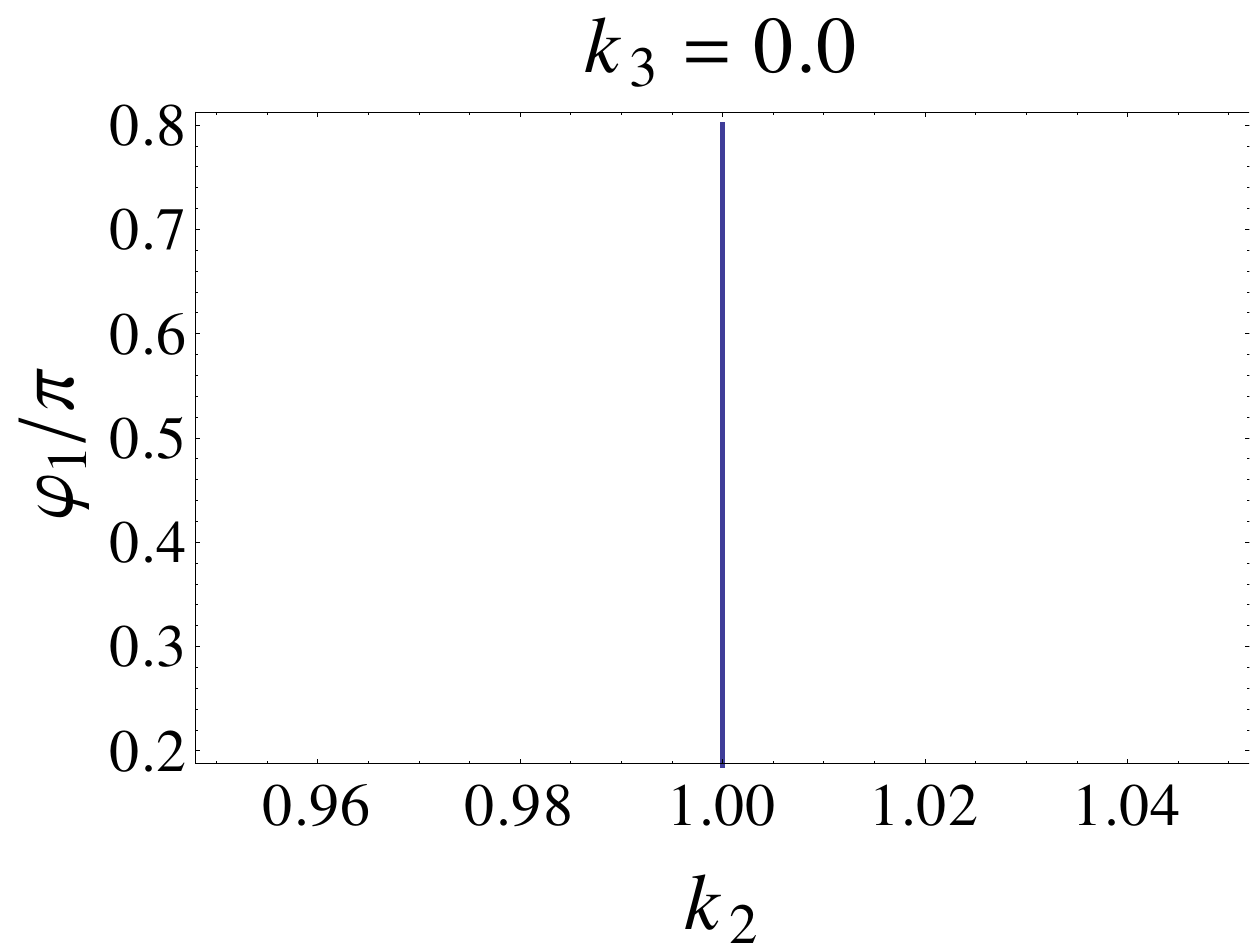}
  \hspace{0.1cm}
  \includegraphics[width=4.5cm]{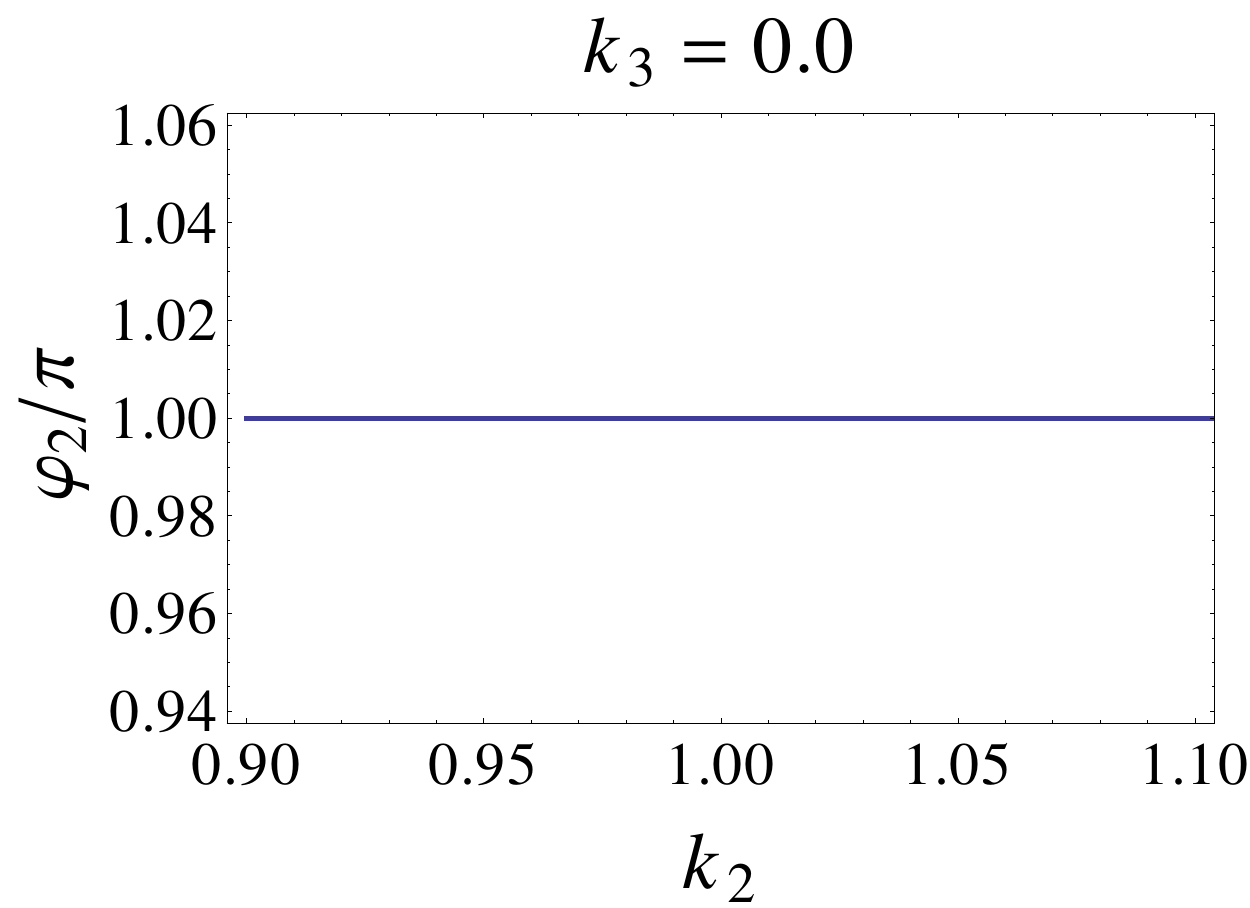}
  \hspace{0.1cm}
  \includegraphics[width=4.5cm]{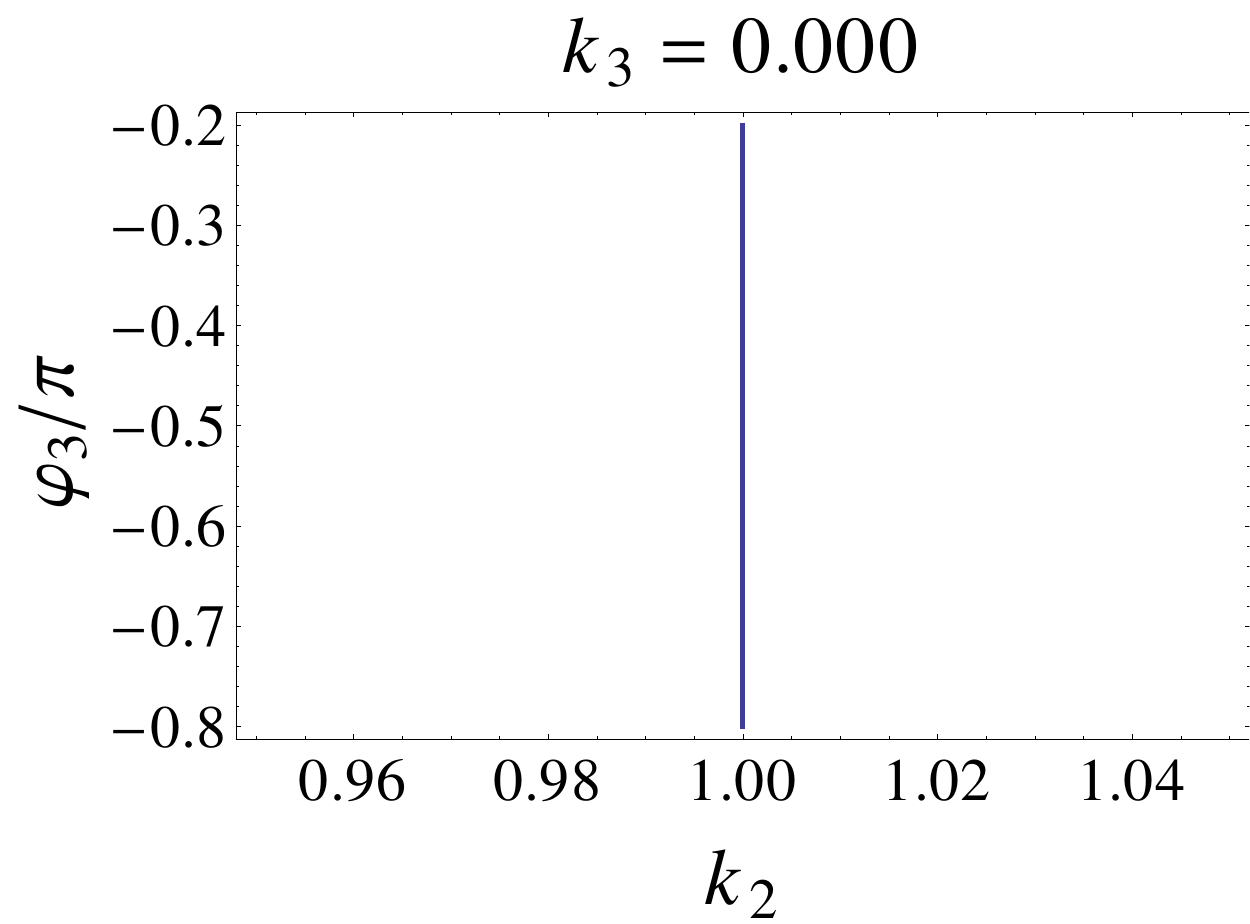}
  \caption{The bifurcation diagrams in the neighborhood $k_2=1$ and
    $\pmb{\vphi}=\varPhi^{({\rm sv})}$ are shown,
    for $k_3=0.01, 0.001, 0.0001, 0$, respectively. The family
    $F_1:{\pmb\varphi}=(\varphi,\pi,-\varphi)$ is
    degenerate (possesses one 0 eigenvalue) and it is represented by a
    dotted line at $k_2=1$.}
  \label{fig:k3_0_seq}
\end{figure*}

A representative example of this parameter region is depicted in
Fig.~\ref{fig:global_k3_m_0_01} for $k_3=0.01$.  We observe here that
two new phase-shift families appear, bifurcating from the symmetric
vortex configuration $\pmb{\vphi}=\Phi^{({\rm sv})}=(\pi/2, \pi,
-\pi/2)$, as will become more transparent through our parametric
variations below. Indeed, in order to acquire a better understanding
of the bifurcating families, in Fig.~\ref{fig:k3_0_seq} we perform a
magnification of the area around the bifurcation point, for the values
of $k_3$ close to zero. Each family is determined by its values of
$\vphi_i$'s and it is detailed in Table~\ref{t:1} (e.g., family 1
consists of the $\vphi_1=\circled{1}$, $\vphi_2=\circled{2}$,
$\vphi_3=\circled{2}$ in Fig.~\ref{fig:k3_0_seq}, etc.).
\begin{table}[h!]
\begin{center}
  \begin{tabular}{cc}
  \toprule  
    \multirow{2}{*}
             {\parbox[c]{1.5cm}{\centering $\sharp$ of Family}} & Branch description\\
    \cmidrule(r){2-2}
     & $\varphi_1 \qquad  \varphi_2 \qquad  \varphi_3$   \\
    \midrule
    1 & $\circled{1} \qquad \circled{2} \qquad \circled{2}$  \\
    2 & $\circled{2} \qquad \circled{1} \qquad \circled{1}$  \\
    3 & $\circled{3} \qquad \circled{3} \qquad \circled{4}$  \\
    4 & $\circled{4} \qquad \circled{4} \qquad \circled{3}$ \\
    \bottomrule
  \end{tabular}
\end{center}
\caption{The solution families depicted in Fig.~\ref{fig:k3_0_seq}.}
\label{t:1}
\end{table}

 In this sequence of figures we can see that the bifurcation points of
 the phase-shift families under consideration approach $\Phi^{({\rm
     sv})}$ and the families themselves tend to coincide with the
 $F_2$ family (\ref{e.fam.dnls.zz}) as $k_3\rightarrow0$. For $k_3=0$ the families coincide
 with $F_2$ which visually coincides also with $F_1$. The $F_1$ and
 $F_2$ families really cross each other at
 $\Phi^{({\rm sv})}$.  
 
\begin{figure*}[t]
  \centering
  \includegraphics[width=4.5cm]{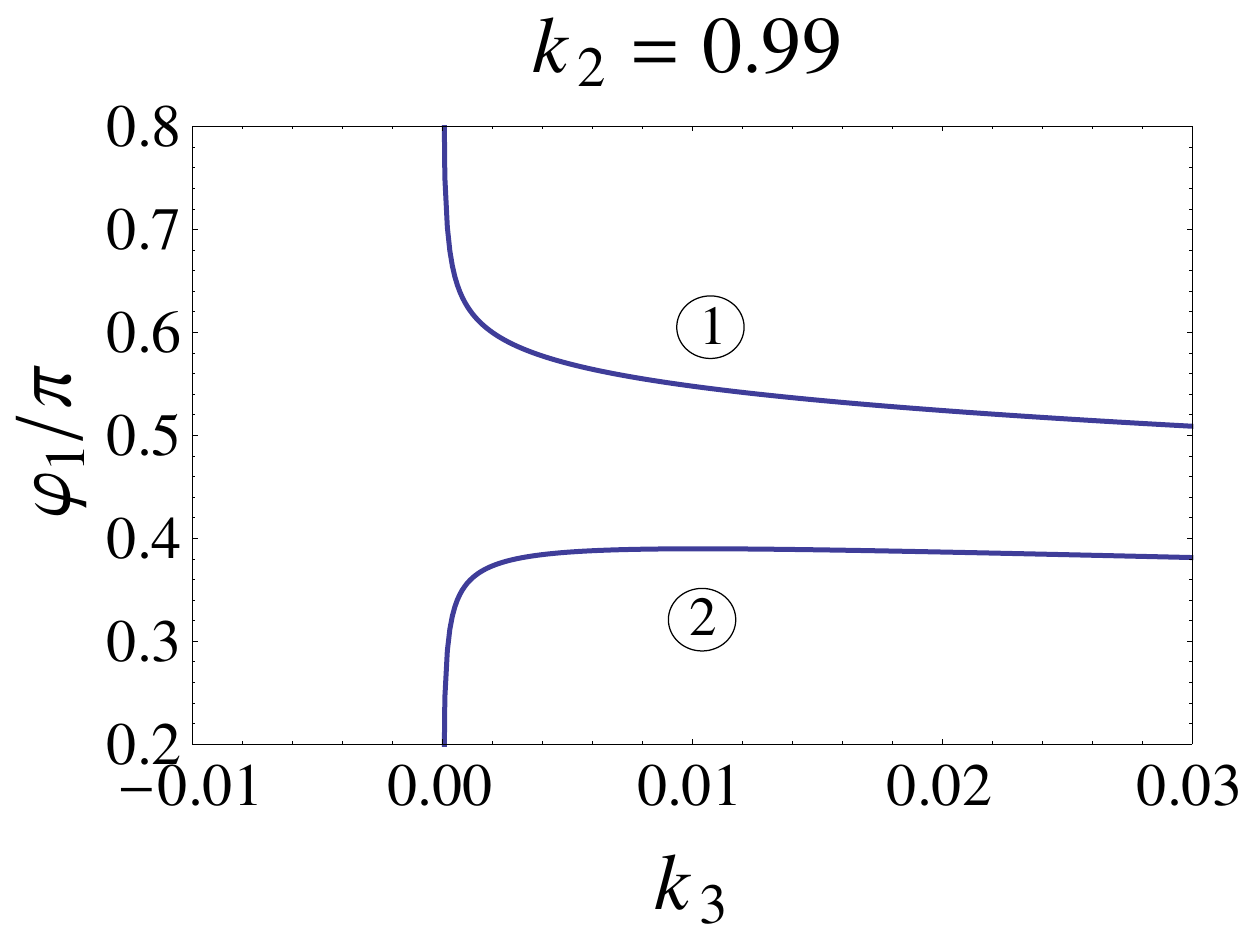}
  \hspace{0.1cm}
  \includegraphics[width=4.5cm]{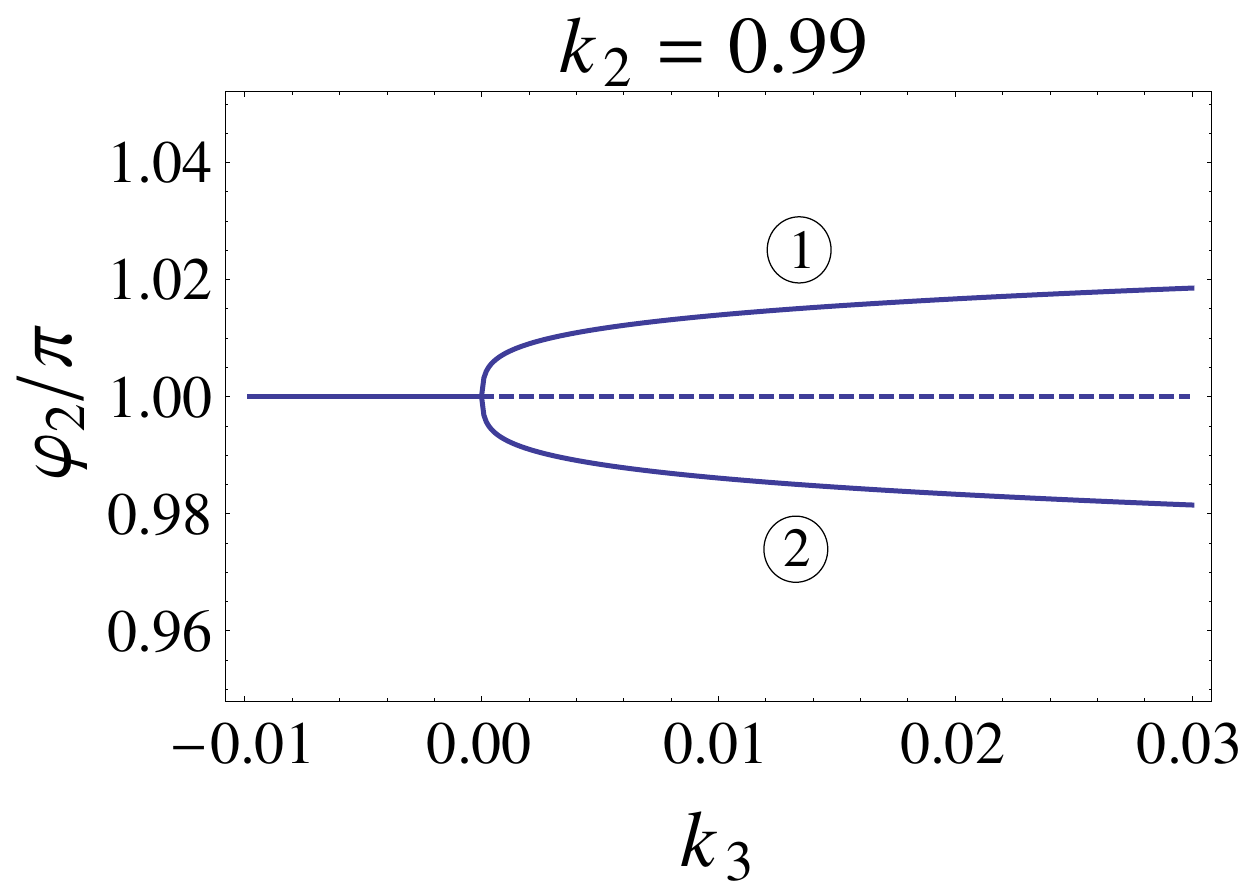}
  \hspace{0.1cm}
  \includegraphics[width=4.5cm]{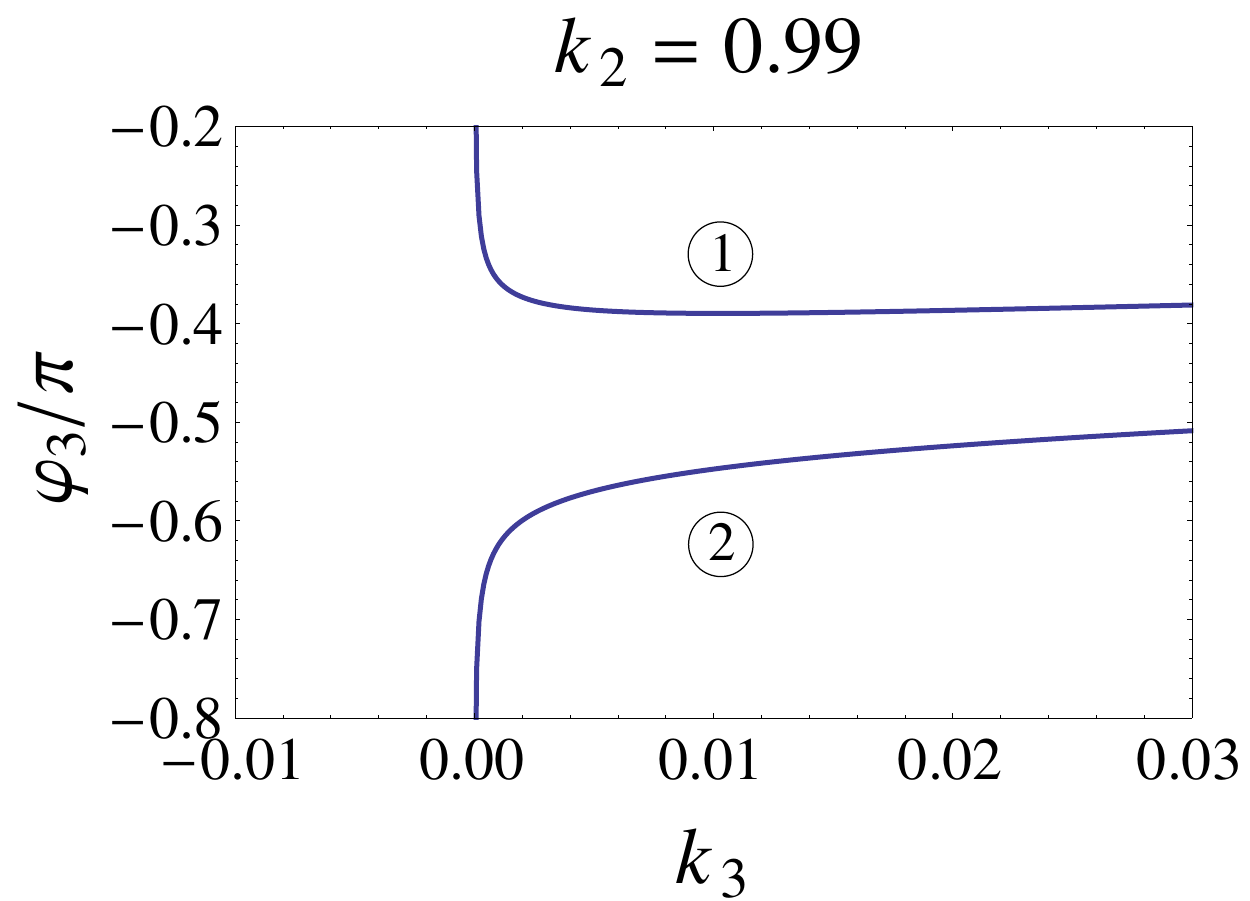}
\\
\includegraphics[width=4.5cm]{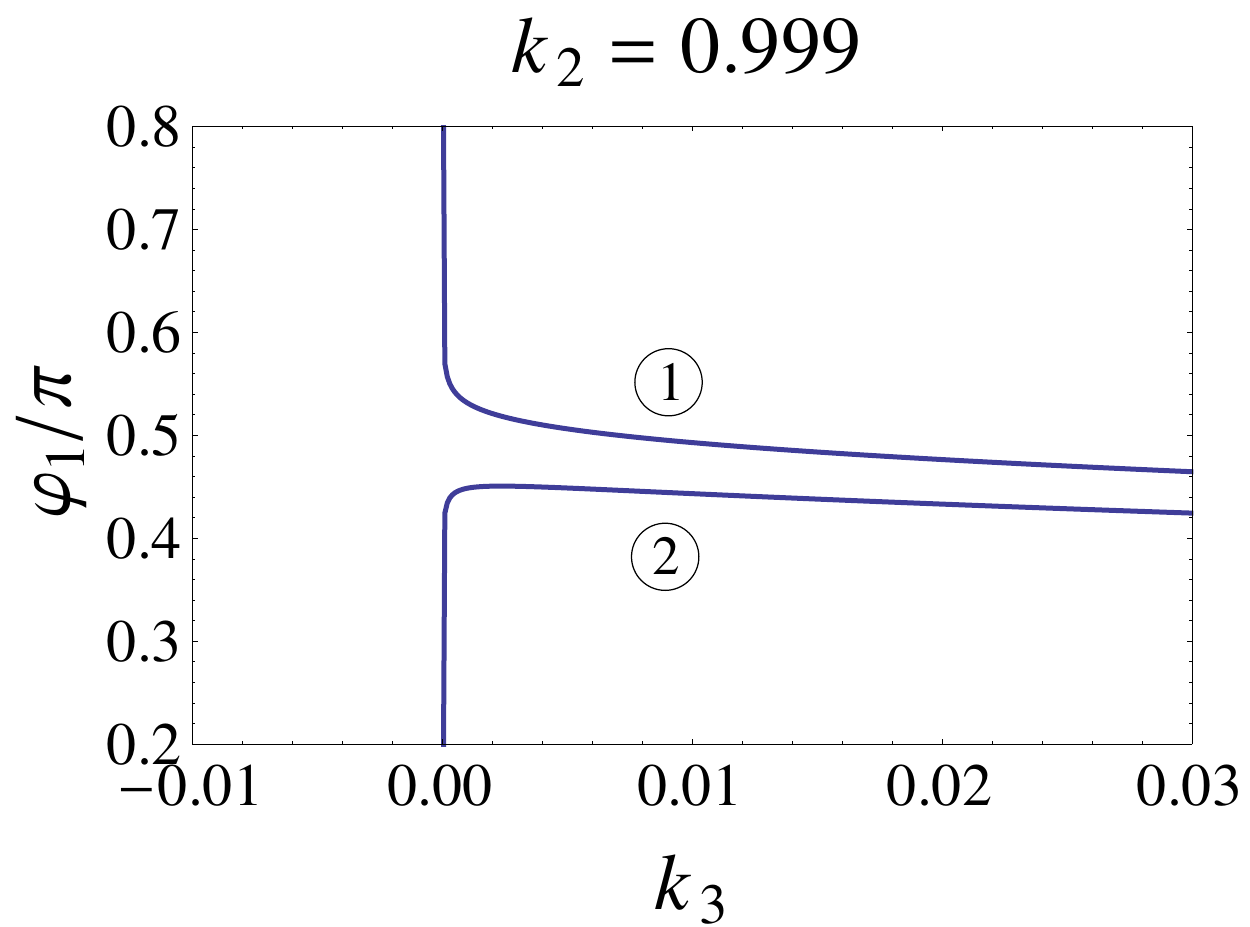}
  \hspace{0.1cm}
  \includegraphics[width=4.5cm]{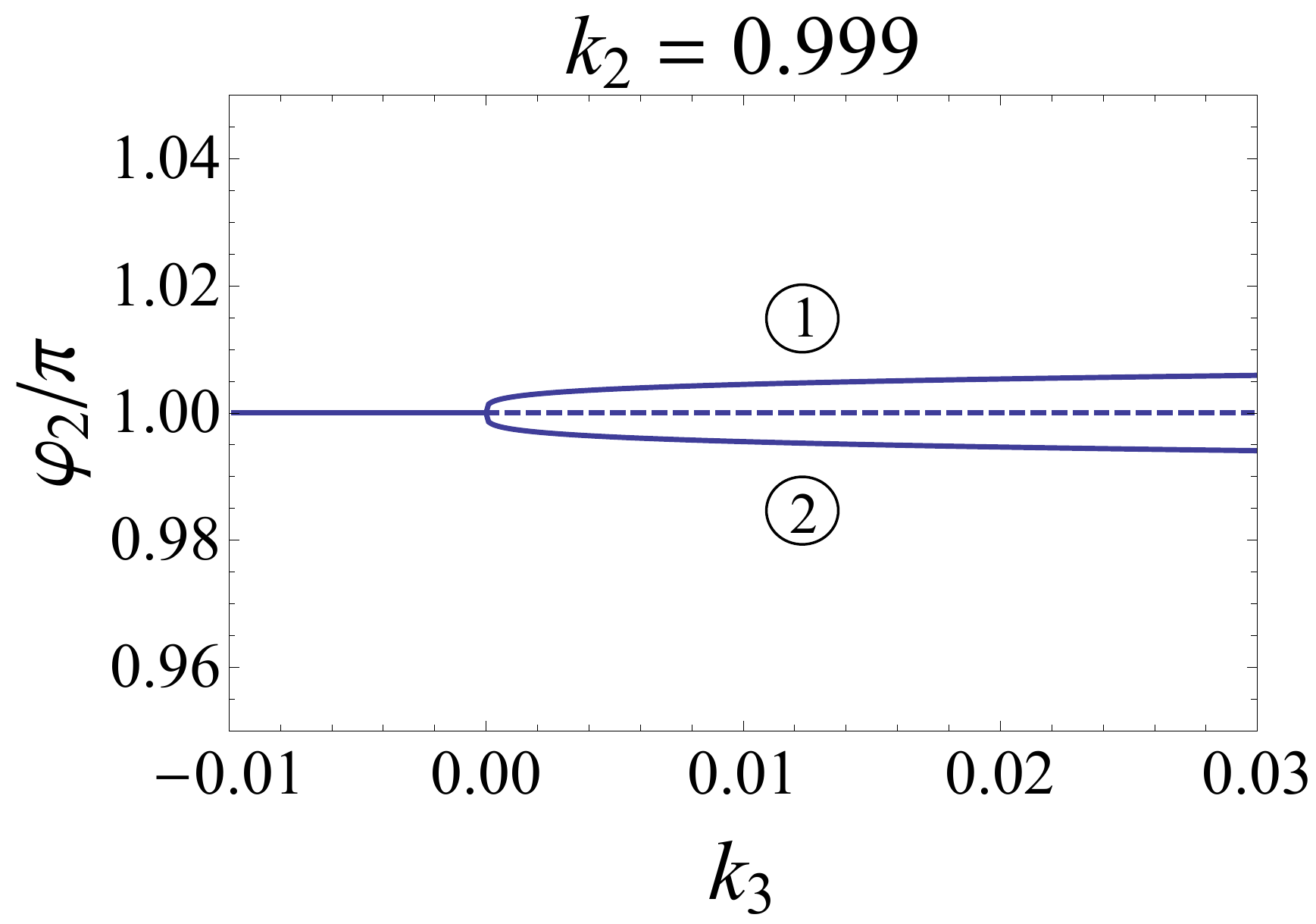}
  \hspace{0.1cm}
  \includegraphics[width=4.5cm]{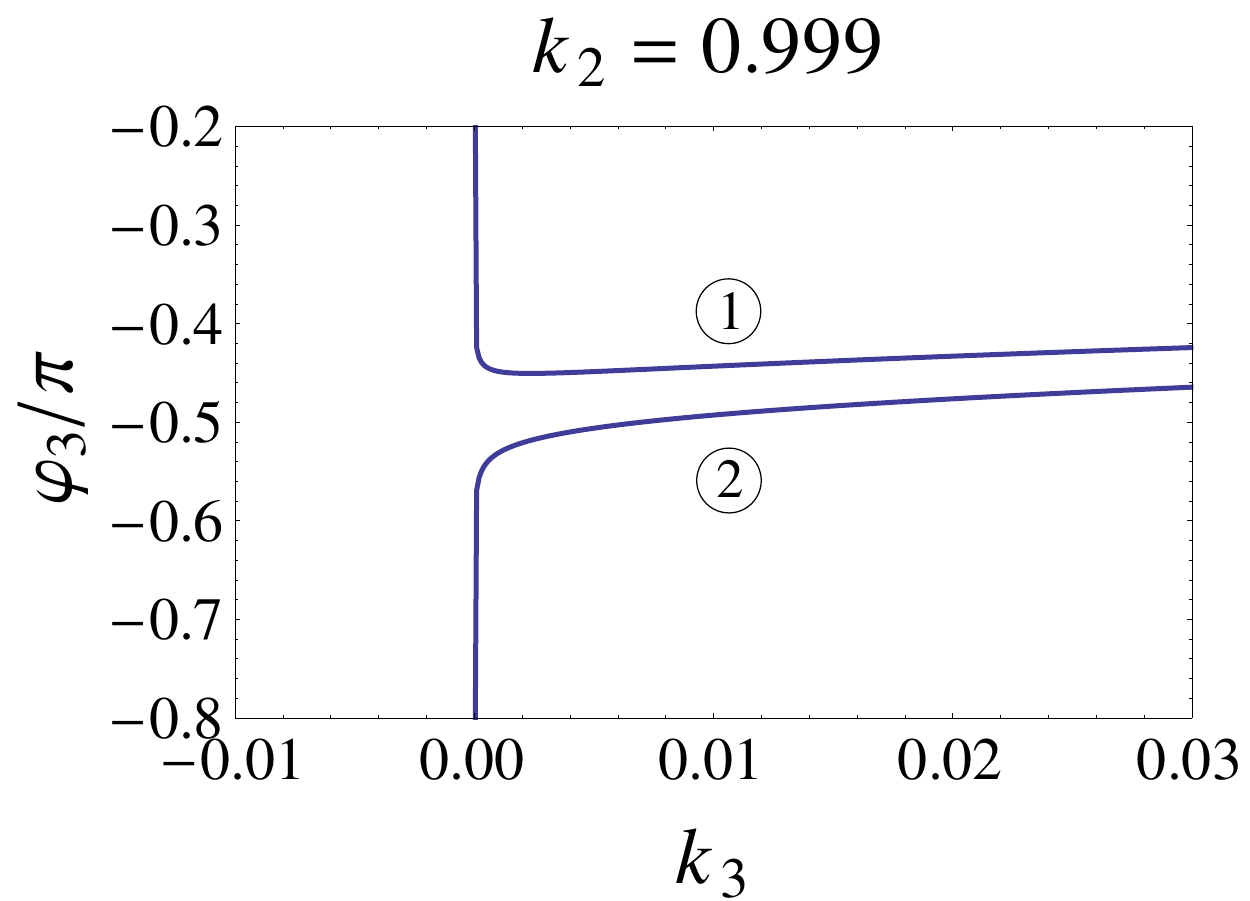}
\\
\includegraphics[width=4.5cm]{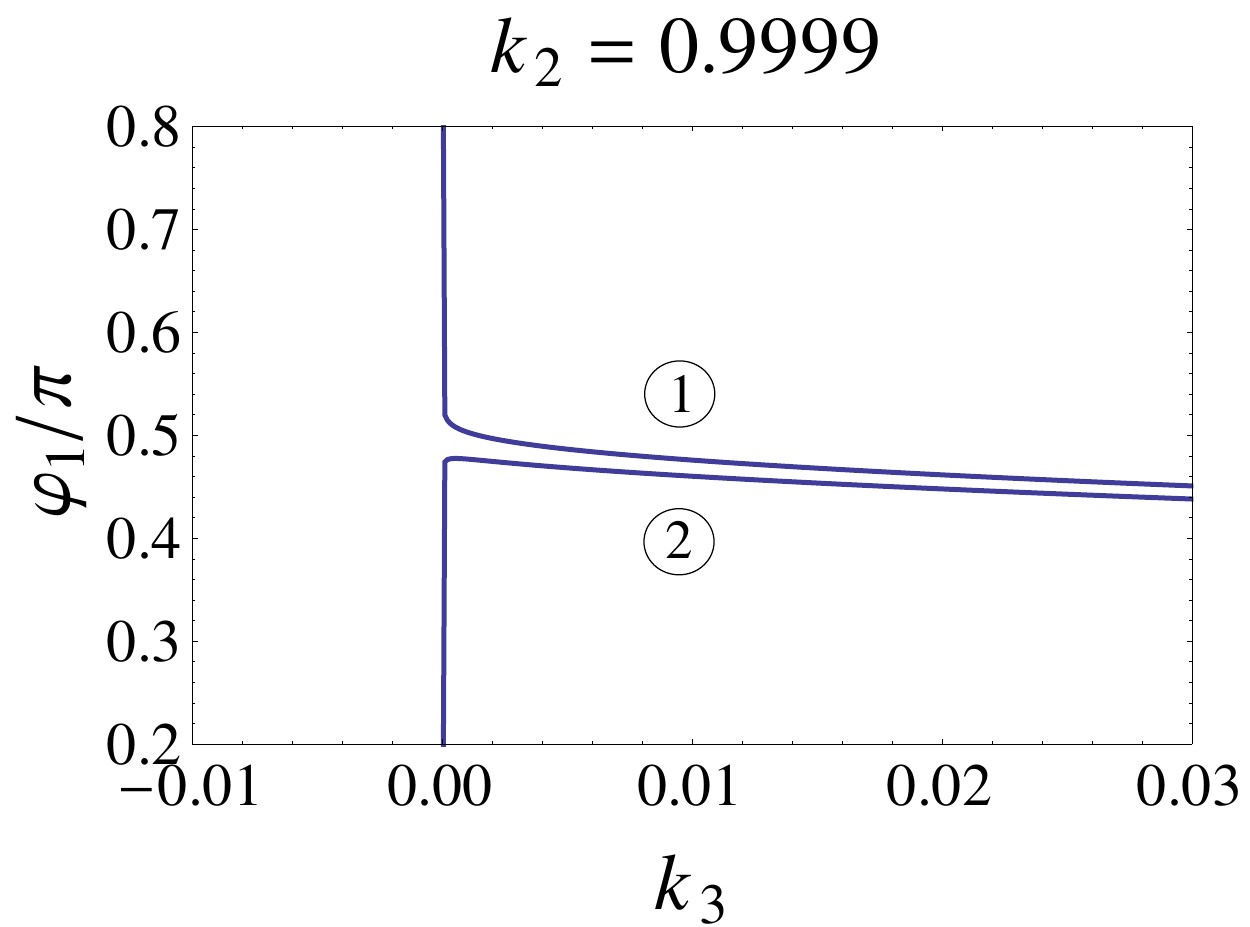}
  \hspace{0.1cm}
  \includegraphics[width=4.5cm]{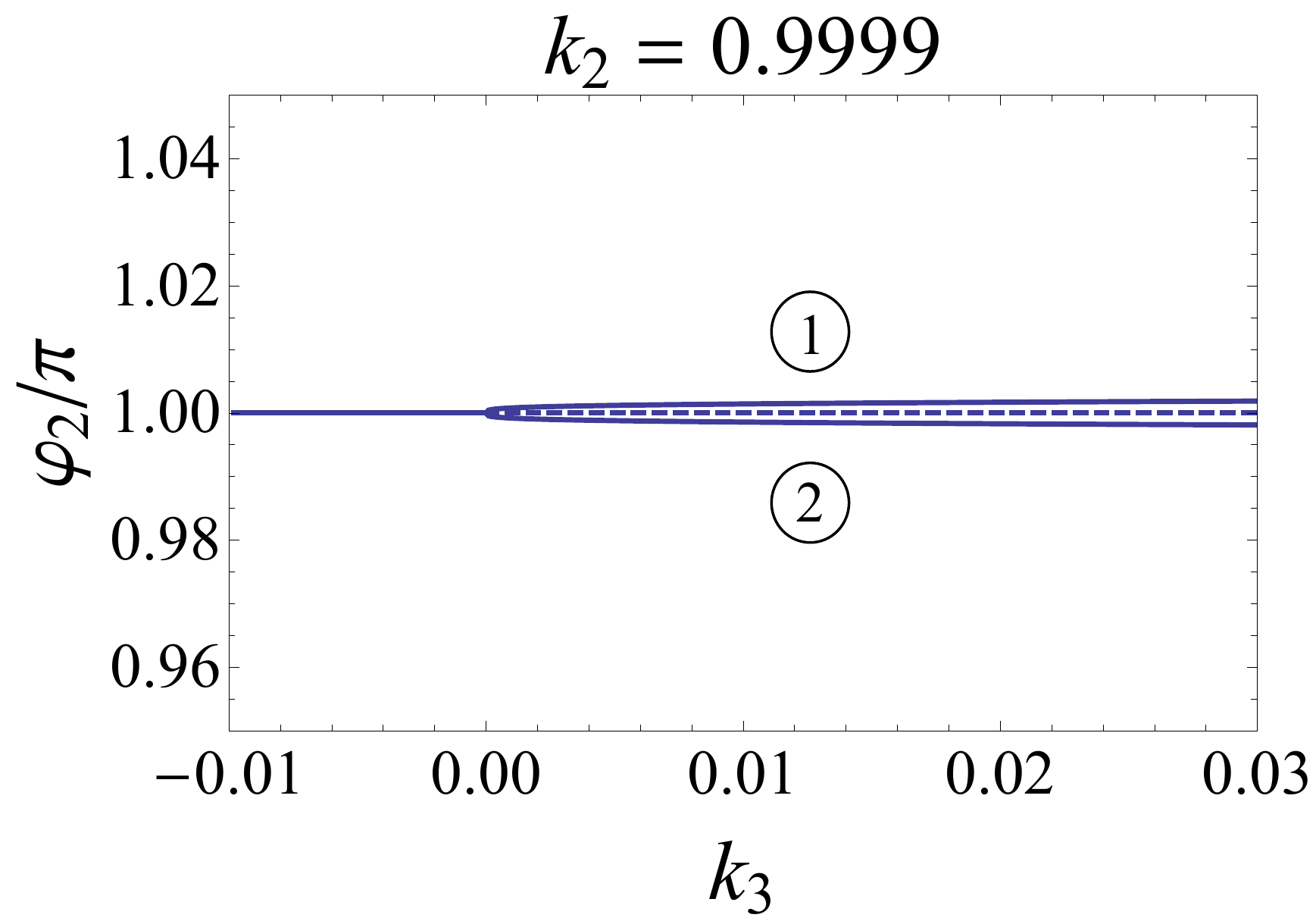}
  \hspace{0.1cm}
  \includegraphics[width=4.5cm]{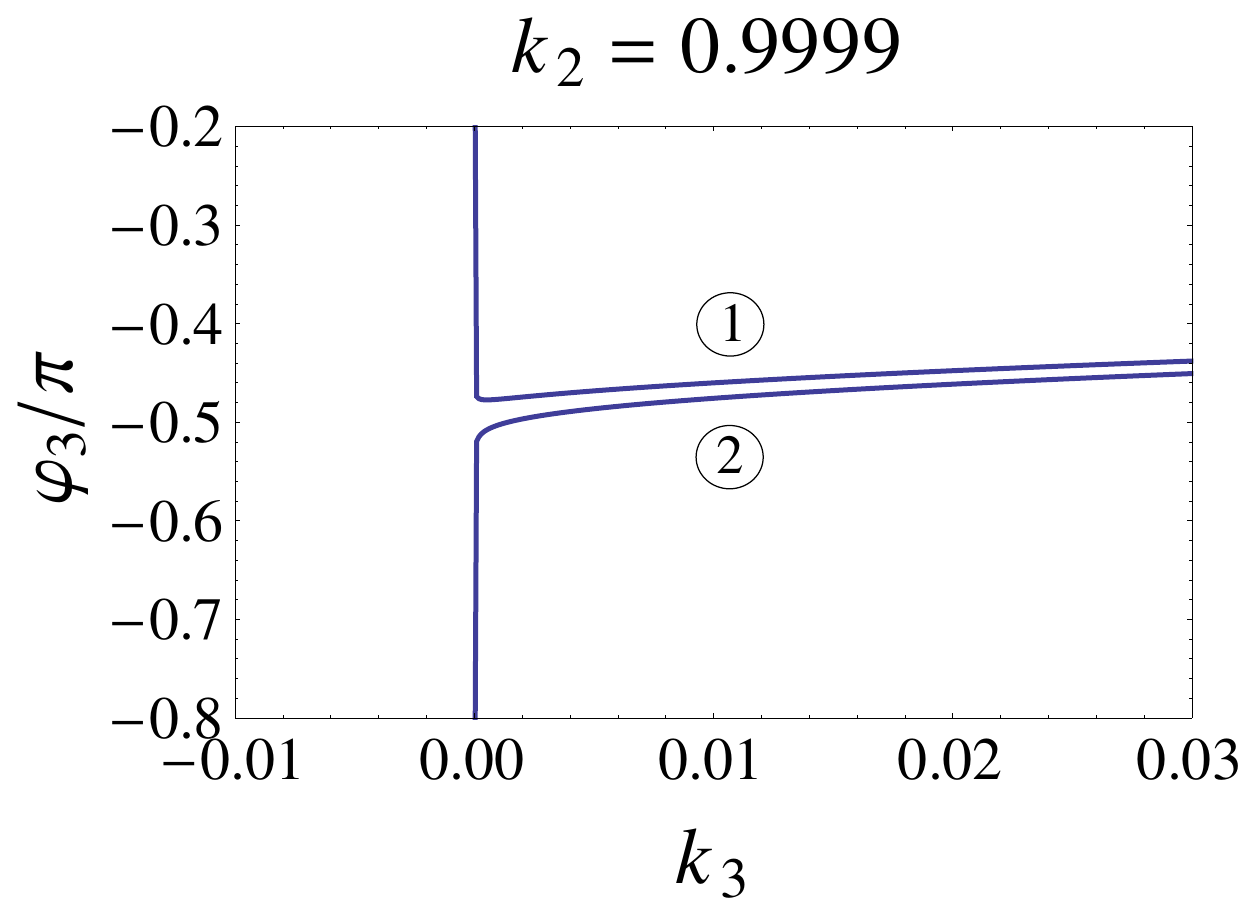}
  \caption{The bifurcation diagrams are shown, for three different
    values of $k_2$ close to 1, respectively $k_2=0.99, 0.999\ \text{and}\ 0.9999$.}
  \label{fig:k2_1_seq}
\end{figure*}

 This is
 also suggested in Fig.~\ref{fig:k2_1_seq}, where the role of $k_2$ and $k_3$ has been reversed. Here, $k_2$ has been
 chosen close to, but less than, $1$ and $k_3$ left free to vary
 around $0$. The two families which are depicted in Fig.~\ref{fig:k2_1_seq} are the ones shown in Table~\ref{t:2}. We can observe in a more clear way the difference between
 the $k_3\leq 0$ case and the $k_3>0$ case, in terms of phase-shift
 solutions. When $k_3>0$ there are branches connecting (apparently) to
 $0$ and $\pi$: although the situation very close to $k_3=0$ is not
 perfectly shown, it is anyway evident that the branches in the upper and lower parts of the frames
get closer and closer as $k_2\to 1$,
 like converging to a curve which emerges from
 $\Phi^{({\rm sv})}$. At
 exactly $k_2=1$, one should observe a full band for the phase
 differences $\vphi_{1,3}$. The picture is completely symmetrical to the one of Fig.~\ref{fig:k2_1_seq} in the $k_2>1$ case.
\begin{table}[h!]
\begin{center}
  \begin{tabular}{cc}
  \toprule  
    \multirow{2}{*}
             {\parbox[c]{1.5cm}{\centering $\sharp$ of Family}} & Branch description\\
    \cmidrule(r){2-2}
     & $\varphi_1 \qquad  \varphi_2 \qquad  \varphi_3$   \\
    \midrule
    1 & $\circled{1} \qquad \circled{2} \qquad \circled{1}$  \\
    2 & $\circled{2} \qquad \circled{1} \qquad \circled{2}$  \\
    \bottomrule
  \end{tabular}
\end{center}
\caption{The solution families depicted in Fig.~\ref{fig:k2_1_seq}.}
\label{t:2}
\end{table}

The overall picture emerging from the above numerical exploration is the
following. Whenever $k_2 \neq 1$, solutions appear to be
isolated, thus non-degenerate and suitable to be continued. Nonetheless, as
$k_2\to 1$, their non-degeneration gets weaker and weaker, so that the domain of
continuation in the coupling parameter $\eps$ is expected to vanish, according
to the standard estimate given by the implicit function theorem. The degenerate
scenario which appears at $k_2=1$, due to the existence of a one-parameter
family of solutions $F_1$ for generic values of $k_3$, becomes richer at
$k_3=0$, since a second family $F_2$ arises which intersects the already
existing $F_1$ at $\Phi^{({\rm sv})}$. The possibility to continue such degenerate solutions requires a
more accurate mathematical analysis, that we develop in the forthcoming
Sections~\ref{s:cont} and~\ref{s:break}.


\subsection{The $\HH_{101}$ system}
\label{s:101}

Before entering the more mathematical part of the paper, in the
present section we will study the persistence conditions of the $\HH_{101}$ model, i.e.,
the model \eqref{ham_long} with $k_2=0$ and $k_3=1$ which is described
by the Hamiltonian
\begin{equation}
  \label{e.KG.deg}
  \begin{aligned}
  \HH_{101} &= \sum_{j\in\ZZ} \left(\frac12 y^2_j + V(x_j)\right)\\
  &+\frac\eps2\sum_{j\in\ZZ} \quadr{(x_{j+1}-x_j)^2 +
    (x_{j+3}-x_j)^2}\ .
  \end{aligned}
\end{equation}
Such a system represents a first order approximation of a square NN lattice and
a four-site multibreather solution of
\eqref{e.KG.deg} can be thought of
as representing a one-dimensional analogue of a four-site vortex for the
two-dimensional square KG lattice and as it will be shown
it constitutes a more degenerate case then the one of the $\HH_{110}$ model.

In order to begin our investigation, we consider the persistence conditions for
this system, which, following again \cite{KouKCR13}, are given by
\begin{equation}
  {\cal P}_{101}(\pmb{\vphi})\equiv
  \begin{cases}
    M(\varphi_1)+M(\varphi_1+\varphi_2+\varphi_3)=0 \\ 
    M(\varphi_2)+M(\varphi_1+\varphi_2+\varphi_3)=0 \\ 
    M(\varphi_3)+M(\varphi_1+\varphi_2+\varphi_3)=0
  \end{cases}
  \label{e.per4_KG.deg}
\end{equation}
where $M(\vphi)$ is given by \eqref{e.M_phi}. The corresponding NL-dNLS system
\begin{equation}
  \label{e.NLdNLS.deg}
  \begin{aligned}
  {H_{\rm 101}} &=
  \sum_j|\psi_j|^2 + \frac38\sum_j|\psi_j|^4\\
  &+\frac{\eps}2\sum_j\quadr{|\psi_{j+1}-\psi_j|^2+|\psi_{j+3}-\psi_j|^2} \ ,
  \end{aligned}
\end{equation}
has been the subject of a detailed investigation performed in
\cite{PenSPKK16}. Note that \eqref{e.NLdNLS.deg} possesses the same
persistence conditions \eqref{e.per4_KG.deg} but with $M(\vphi)\equiv
\sin(\vphi)$, see~\eqref{e.M_phi.DNLS}. In \cite{PenSPKK16}, it is
showed that Eqs.~\eqref{e.per4_KG.deg} and~\eqref{e.M_phi.DNLS} admit
three families of {\it asymmetric vortex} solutions
\begin{equation}
  \label{e.fam.dnls.deg}
  \begin{aligned}
  F_1&:\pmb{\varphi}=(\varphi,\varphi,\pi-\varphi) \ ,\\
  F_2&:\pmb{\varphi}=(\varphi,\pi-\varphi,\varphi) \ ,\\
  F_3&:\pmb{\varphi}=(\varphi,\pi-\varphi,\pi-\varphi) \ ,
  \end{aligned}
\end{equation}
in addition to the two isolated standard solutions
$F_{\text{iso}}:\varphi=\bigl\{(0,0,0)$,
$(\pi,\pi,\pi)\bigr\}$. Again, the rest of the standard configurations
of this case are part of the $F_1$, $F_2$, $F_3$ families. By using
the symmetries of $M(\vphi)$, it is easy to prove that also the
persistence equations~\eqref{e.per4_KG.deg} and~\eqref{e.M_phi} admit
the same families of solutions.  These families are degenerate since
the corresponding Jacobian $D_\vphi({\cal P}_{101})$ possesses a zero
eigenvalue, while the {\it symmetric vortex} solutions
\begin{displaymath}
{\Phi}^{(\rm sv)}_{101}\equiv\pmb{\varphi}=\pm\tond{\frac\pi2,\frac\pi2,\frac\pi2}
\end{displaymath}
are fully degenerate, since $D_\vphi({\cal P}_{101})$ equals the null
matrix. The latter can be seen both by a direct computation, or by observing
that in these solutions we have three independent Kernel directions,
one for each family passing through the solution.

\subsection{The full system close to $k_2=0, k_3=1$}
In order to understand how the three above  families merge, we numerically study
the persistence of the full problem \eqref{ham_long} in the region of the
parameter point $(k_2, k_3)=(0,1)$
around the ${\Phi}^{({\rm sv})}_{101}$ configuration. Since we consider low amplitude solutions
the results for the two models (Klein-Gordon and dNLS) are equivalent
both qualitatively as well as quantitatively, since the differences in
the solutions are negligible\footnote{This fact holds of course also for the $\HH_{110}$ model studied before.}. 

In order to showcase the relevant results, we consider first some specific
values of $k_3$ (close to $1$) and we perform
a scan for solutions in an interval of $k_2$ (close to
  $0$) and then we reverse the roles of $k_2$ and $k_3$.


\subsubsection{The $k_3<1$ case}
\begin{figure*}
  \centering
  \includegraphics[width=0.32\textwidth]{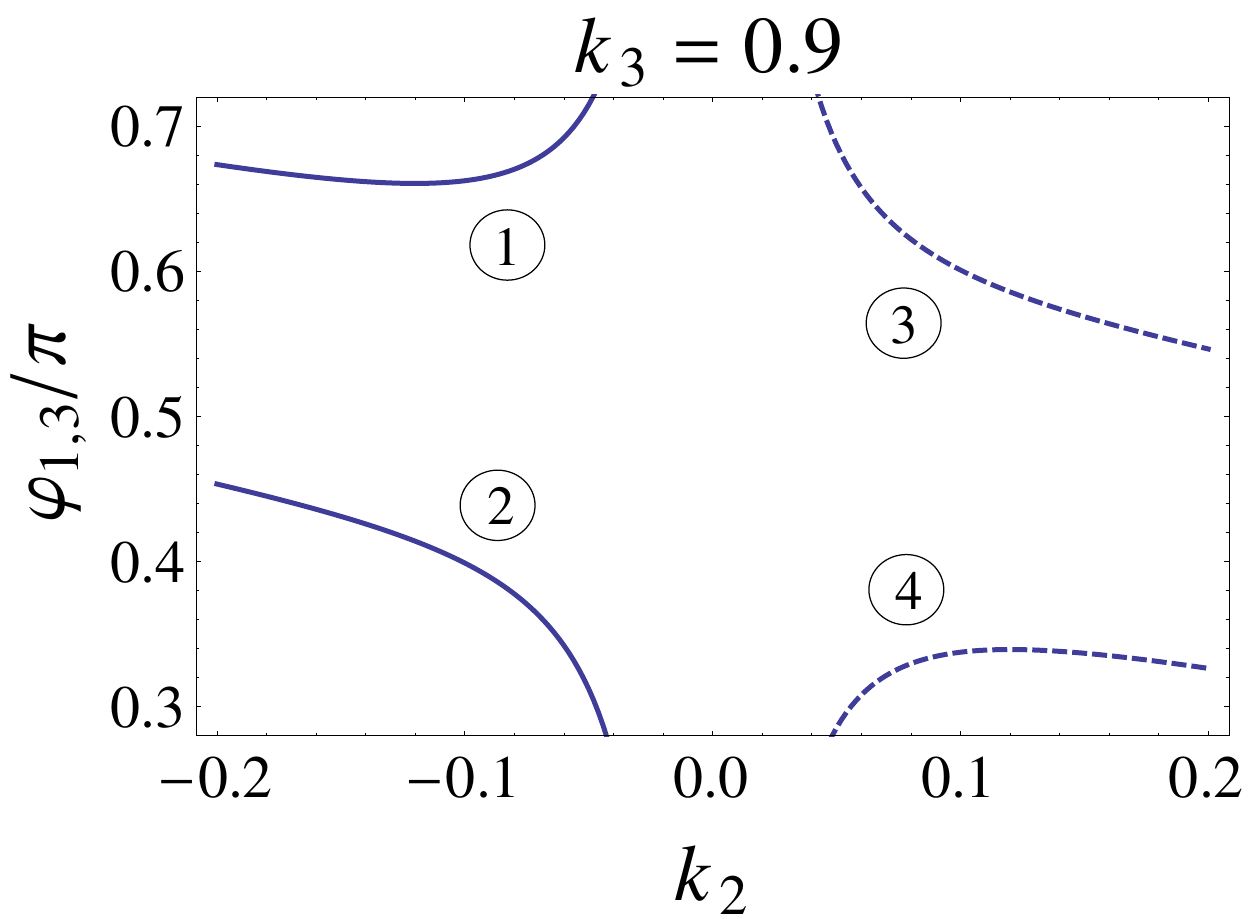}\; 
  \includegraphics[width=0.32\textwidth]{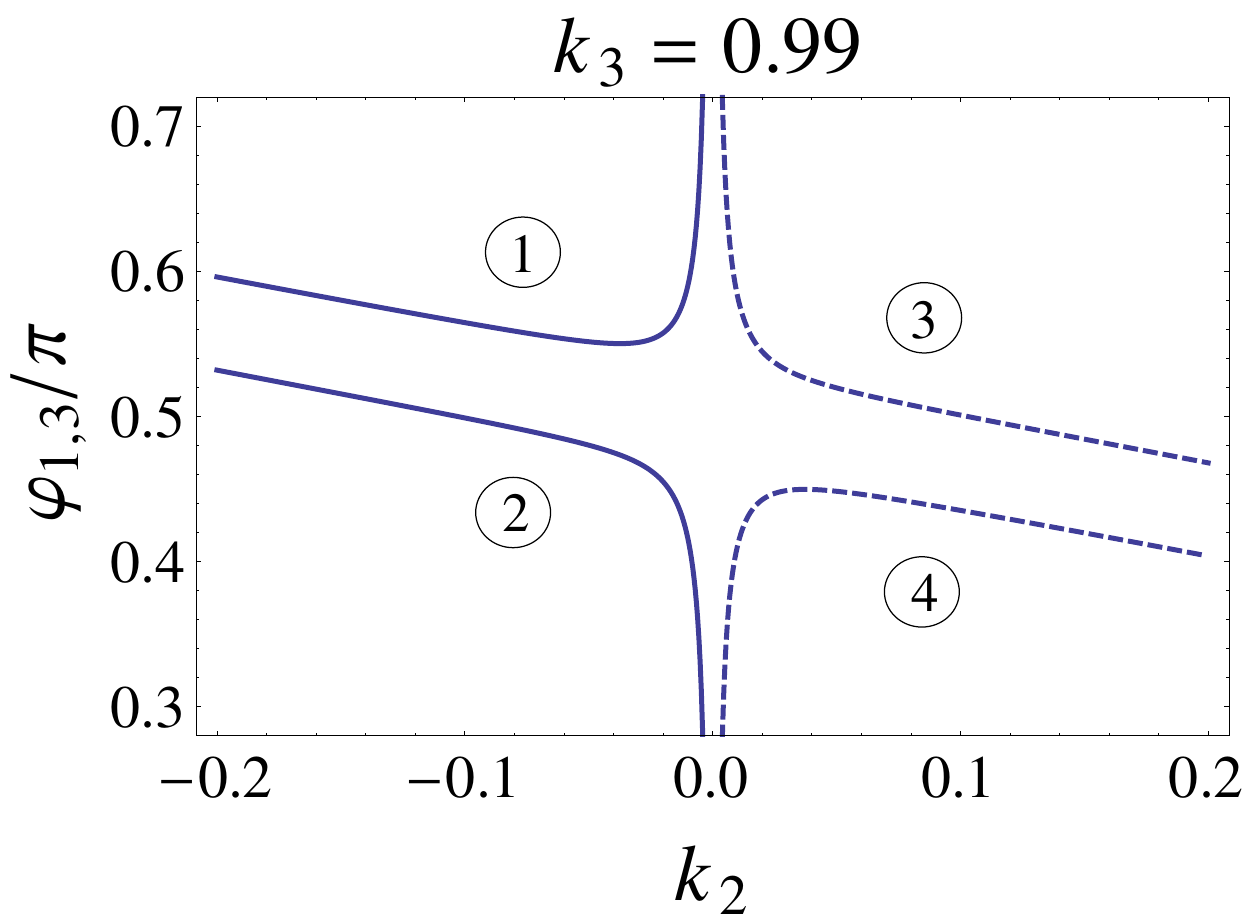}\; 
  \includegraphics[width=0.32\textwidth]{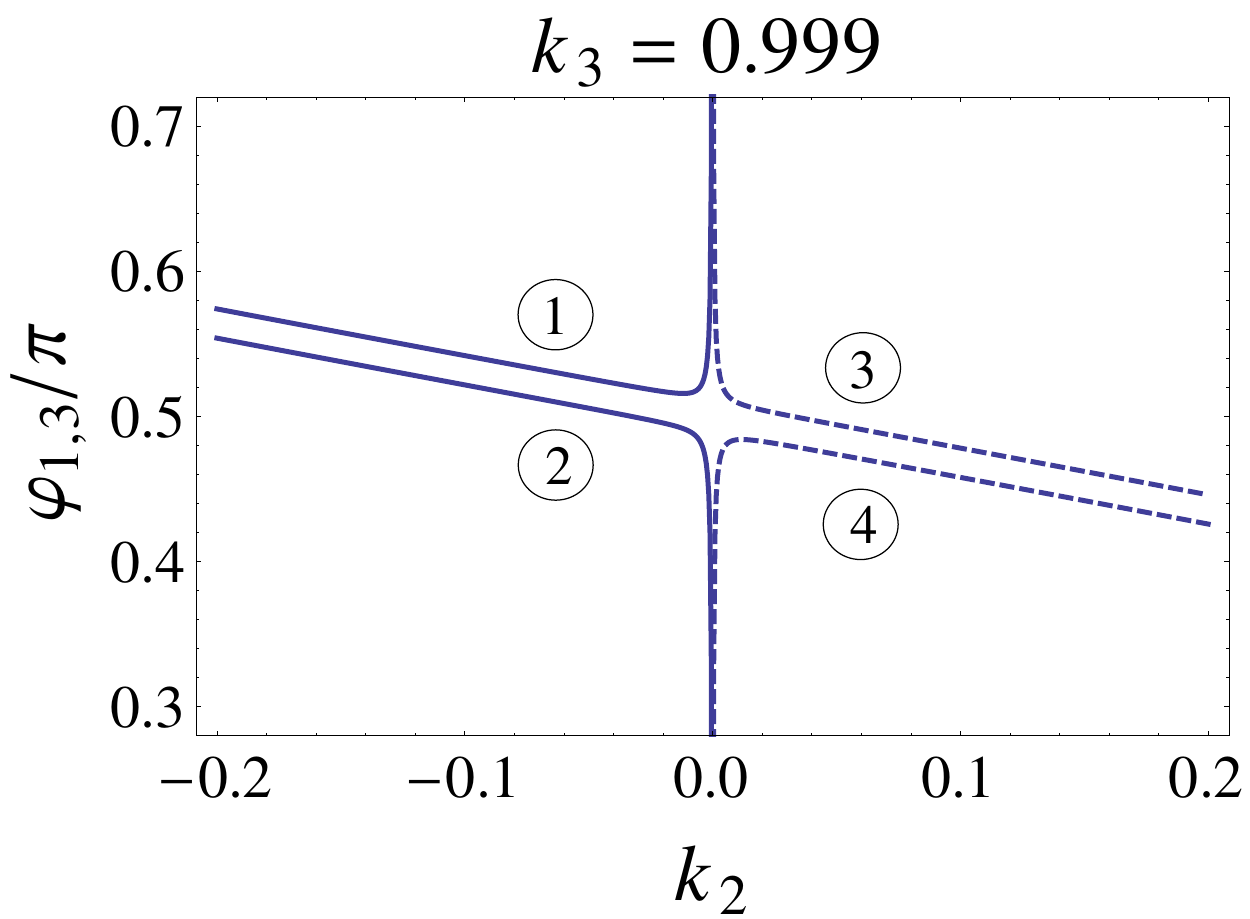}
  \vskip.2cm
  \includegraphics[width=0.32\textwidth]{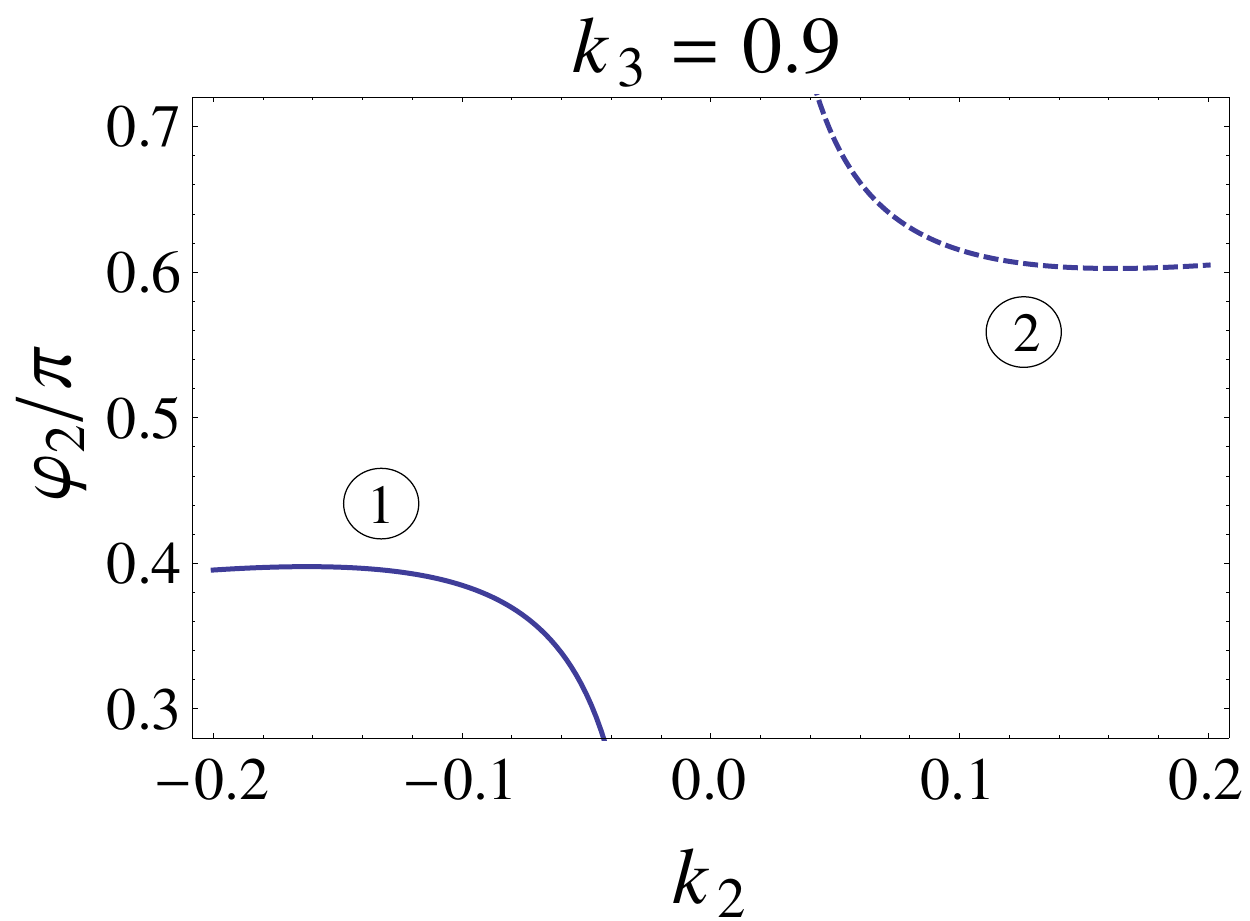}\; 
  \includegraphics[width=0.32\textwidth]{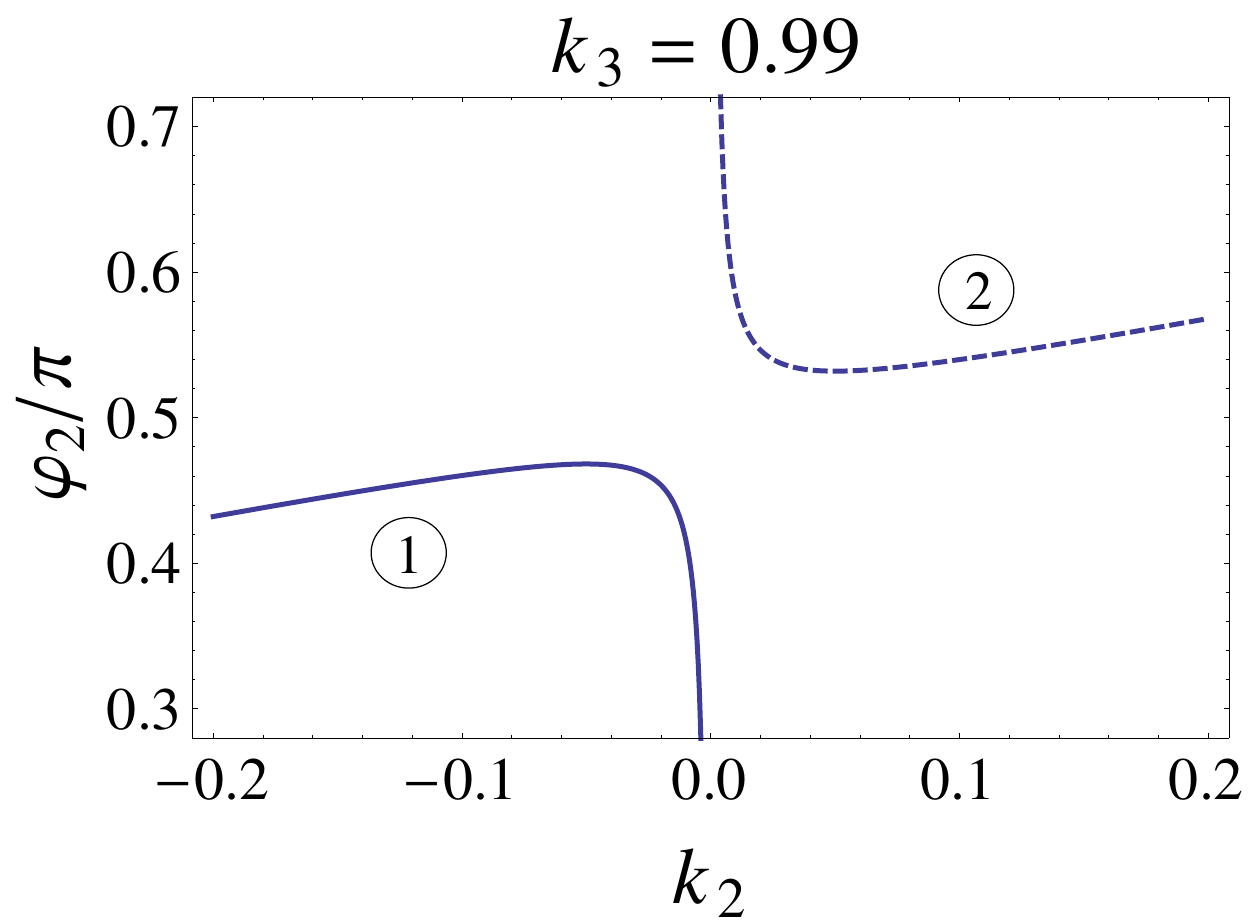}\; 
  \includegraphics[width=0.32\textwidth]{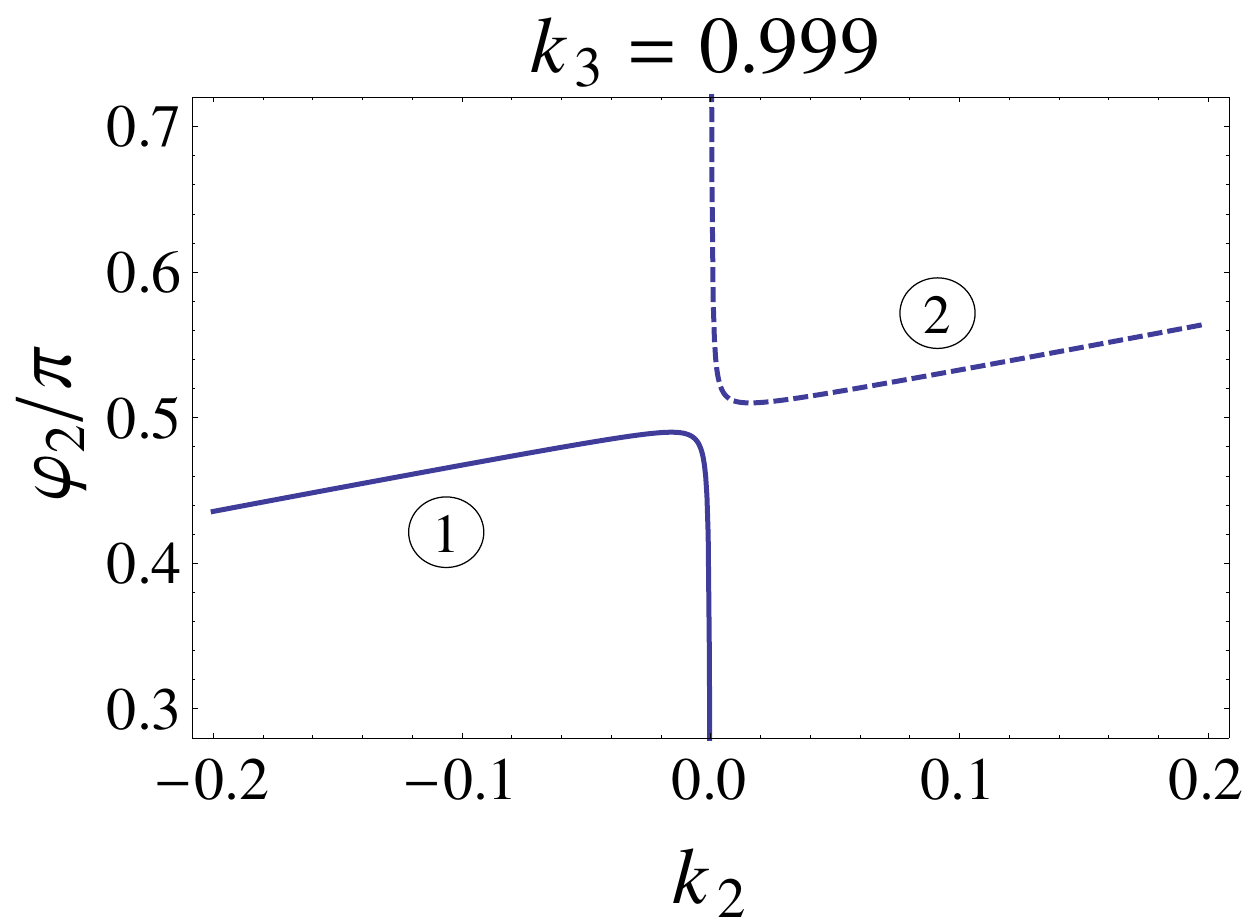}
  \caption{The bifurcation diagrams in the neighborhood $k_2=0$,
    $\varphi_i=\pi/2$ are shown, for $k_3=0.9$, $0.99$ and $0.999$. We can observe how the various solution families converge to the $k_2=1$ asymptote.}
  \label{fig:k3_0_900}
  \label{fig:k3_0_990}
  \label{fig:k3_0_999}

  \vspace{50pt}
  \centering
  \includegraphics[width=0.32\textwidth]{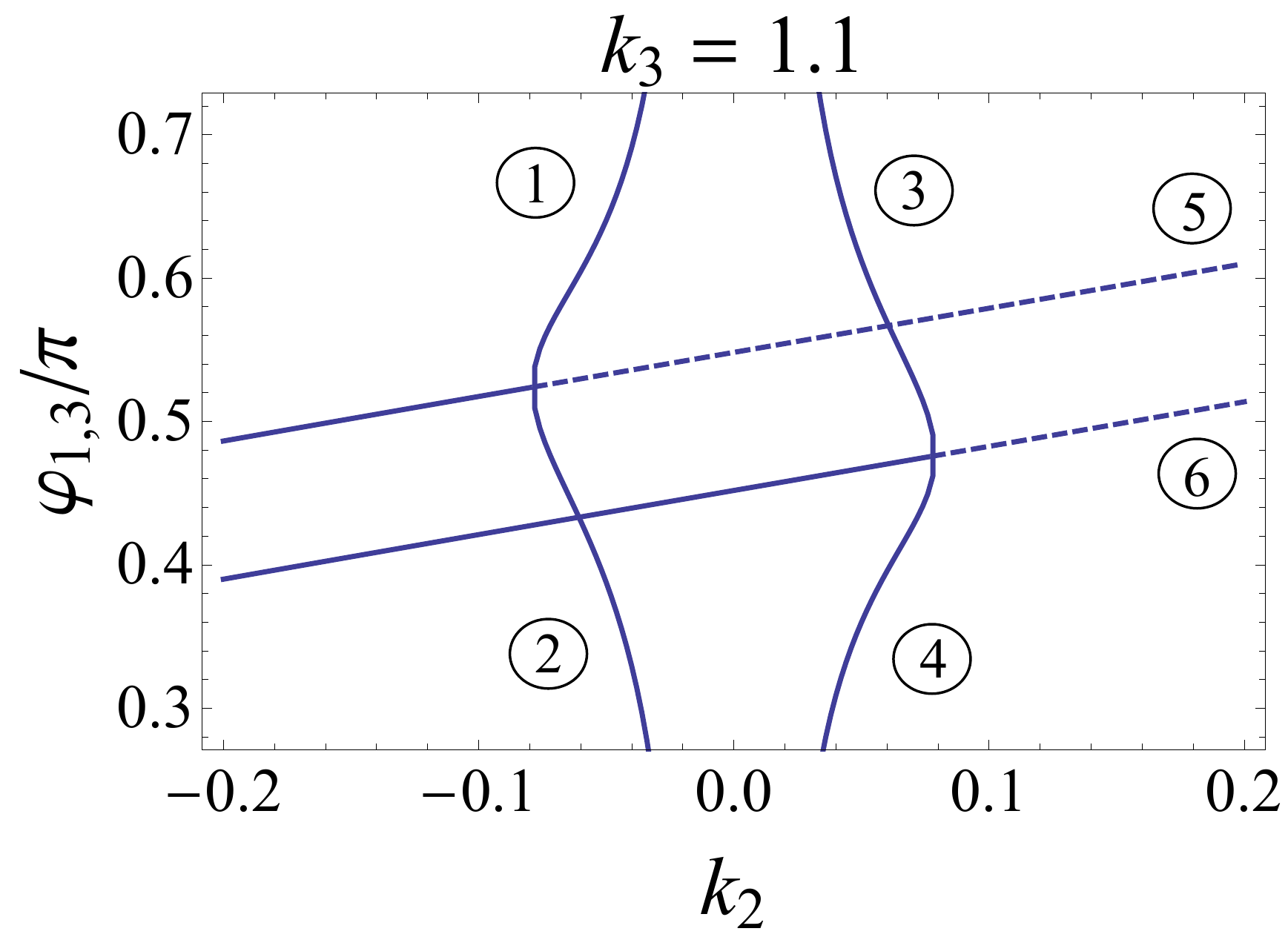}\;
  \includegraphics[width=0.32\textwidth]{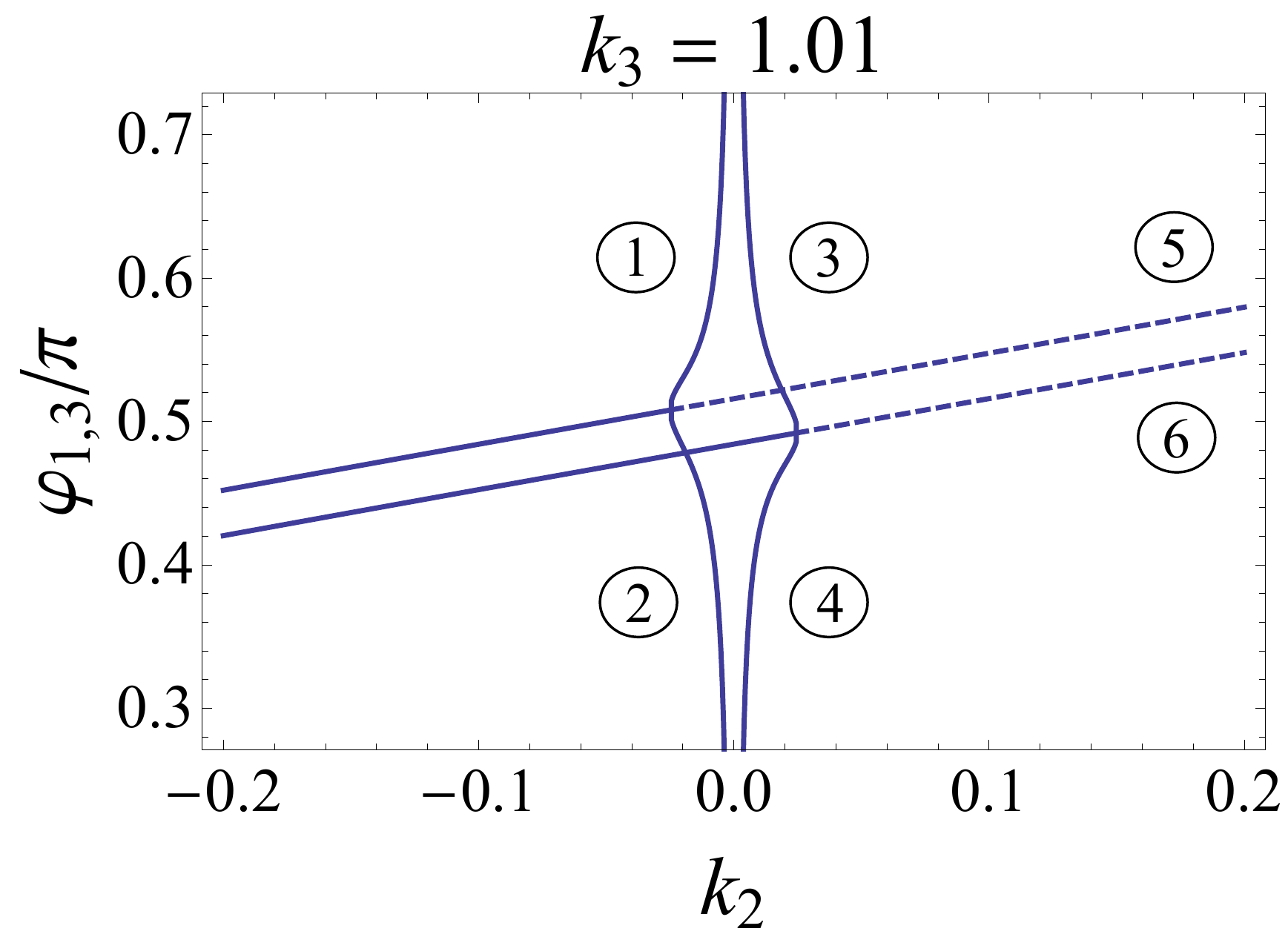}\;
  \includegraphics[width=0.32\textwidth]{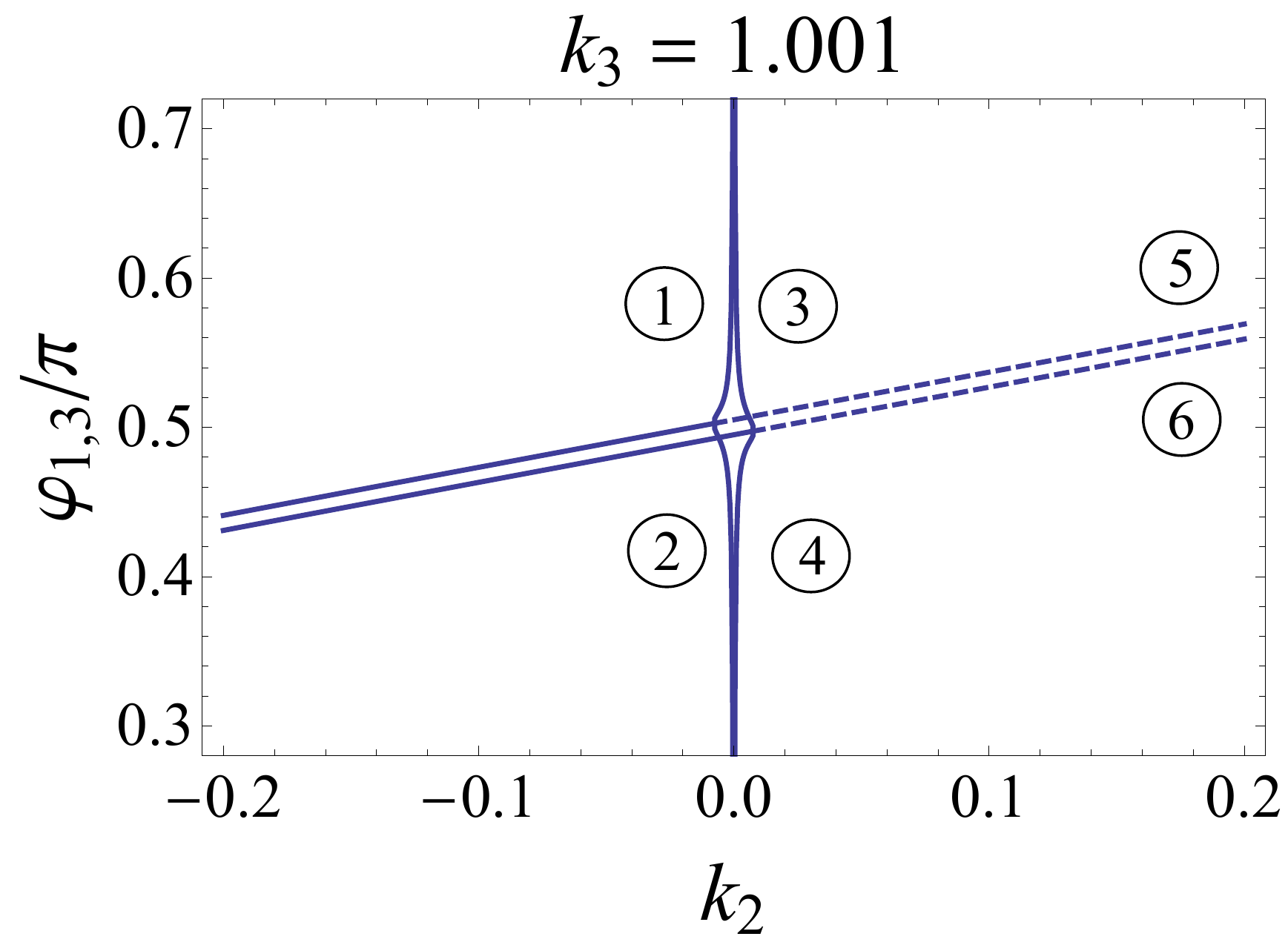}
  \vskip0.2cm
  \includegraphics[width=0.32\textwidth]{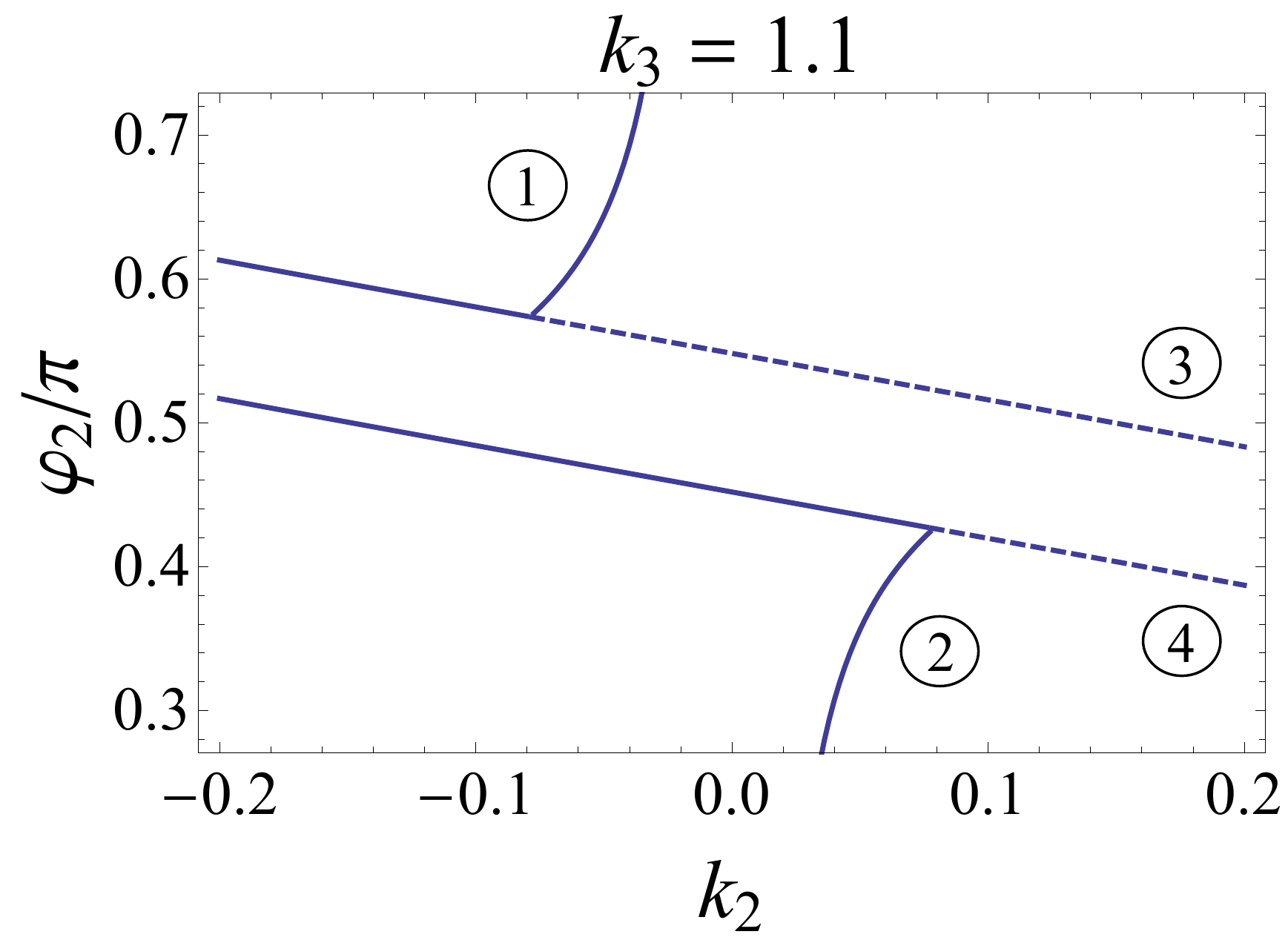}\;
  \includegraphics[width=0.32\textwidth]{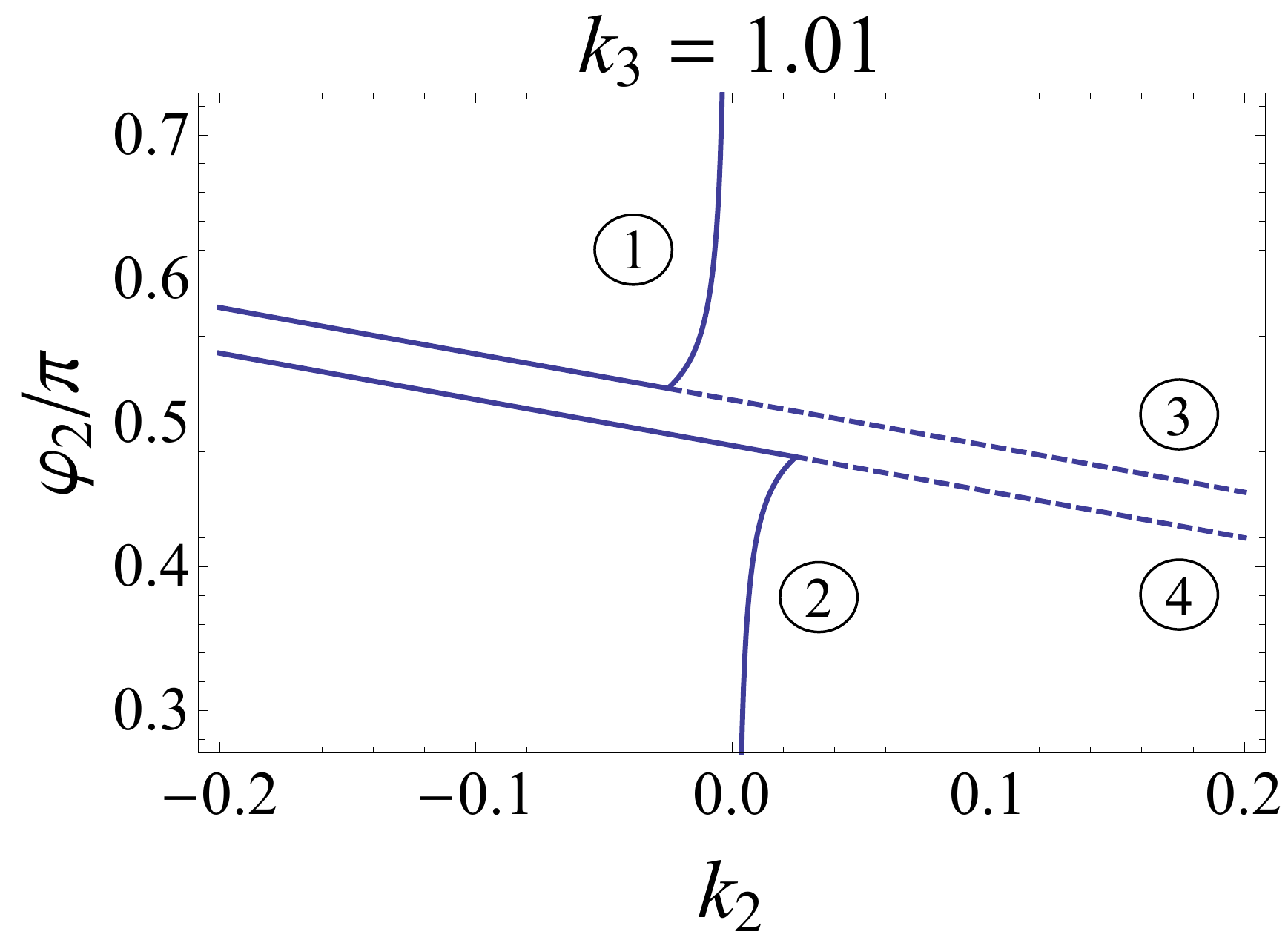}\;
  \includegraphics[width=0.32\textwidth]{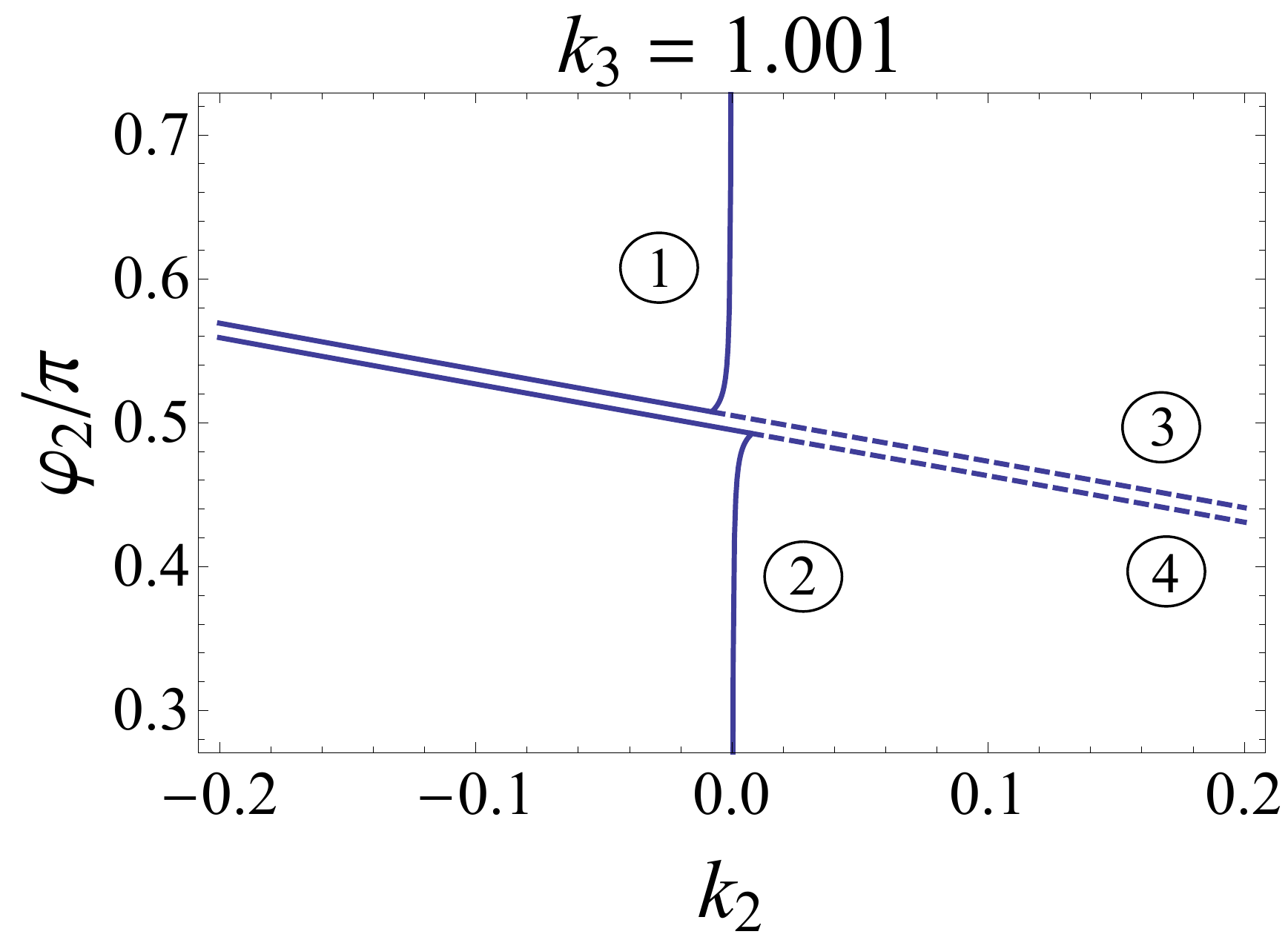}
  \caption{The bifurcation diagrams in the neighborhood $k_2=0$,
    $\varphi_i=\pi/2$ are shown, for $k_3=1.1$, $k_3=1.01$ and
    $1.001$. As $k_3\to1$ The families converge to the vortex families as $k_3\to1$.}
  \label{fig:k3_1_100}
  \label{fig:k3_1_010}
  \label{fig:k3_1_001}
\end{figure*}

In order to examine this parameter region we consider the values
$k_3=0.9, 0.99\ \text{and}\ 0.999$. The corresponding solution
families are shown in Fig.~\ref{fig:k3_0_900}. In the top row the
values of the angles $\vphi_1$ and $\vphi_3$ are shown while the
bottom row depicts the values of $\vphi_2$.  Although, the $\varphi_1$
and $\varphi_3$ angles are depicted in the same diagram, this does not
mean that $\varphi_1=\varphi_3$ for every value of
$k_2$. The four families which are shown in
  Fig.~\ref{fig:k3_0_900} are labeled with encircled numbers and are
  summarized in Table \ref{t:3} below (e.g., family 1 is defined as
$\vphi_1=\circled{1}$ of the upper row of the figure,
$\vphi_2=\circled{1}$ of the lower row and $\vphi_3=\circled{2}$ of
the upper row panels). We see in these diagrams how these families
converge to the $k_2=0$ asymptote. In particular, families 1 and 4
converge to $F_3$, while families 2 and 3 converge to $F_1$ (\ref{e.fam.dnls.deg}). The
different line symbols denote different linear stability of the
families. In particular a solid line corresponds to a family with one
unstable eigenvalue while the dashed line corresponds to two unstable
eigenvalues. As the families converge one of their stability eigenvalue converges to zero and it changes sign when $k_2$ crosses zero. Since the stability discussion lies outside the scope of
the present manuscript we will not refer further to these facts.

\begin{table}[h!]
\centering
  \begin{tabular}{cc}
  \toprule  
    \multirow{2}{*}
             {\parbox[c]{1.5cm}{\centering $\sharp$ of Family}} & Branch description\\
    \cmidrule(r){2-2}
     & $\varphi_1 \qquad  \varphi_2 \qquad  \varphi_3$   \\
    \midrule
    1 & $\circled{1} \qquad \circled{1} \qquad \circled{2}$  \\
    2 & $\circled{2} \qquad \circled{1} \qquad \circled{1}$  \\
    3 & $\circled{3} \qquad \circled{2} \qquad \circled{4}$  \\
    4 & $\circled{4} \qquad \circled{2} \qquad \circled{3}$  \\
    \bottomrule
  \end{tabular}
  \caption{The solution families depicted in Fig.~\ref{fig:k3_0_900}.}
  \label{t:3}
\end{table}

\subsubsection{The $k_3>1$ case}
The bifurcation-diagram for this case is depicted in
Fig.~\ref{fig:k3_1_100}.  We can clearly observe that families 1 and 4
of Table~\ref{t:4} below converge into $F_3$ as $k_3\to1$ while families 2
and 3 converge to $F_1$.  The main difference of this diagrams, with
respect to the ones of the $k_3<1$ case, is that in this case there
exist also the two new phase-shift solution families 5 and 6, where
the families 1-4 bifurcate from through pitchfork bifurcations. These
families have also the characteristic that they are the only ones that
exist for $k_2=0$ and for all $k_3>0$.
\begin{table}[h!]
\begin{center}
  \begin{tabular}{cc}
  \toprule  
    \multirow{2}{*}
             {\parbox[c]{1.5cm}{\centering $\sharp$ of Family}} & Branch description\\
    \cmidrule(r){2-2}
     & $\varphi_1 \qquad  \varphi_2 \qquad  \varphi_3$   \\
    \midrule
    1 & $\circled{1} \qquad \circled{1} \qquad \circled{2}$  \\
    2 & $\circled{2} \qquad \circled{1} \qquad \circled{1}$  \\
    3 & $\circled{3} \qquad \circled{2} \qquad \circled{4}$  \\
    4 & $\circled{4} \qquad \circled{2} \qquad \circled{3}$  \\
    5 & $\circled{5} \qquad \circled{3} \qquad \circled{5}$  \\
    6 & $\circled{6} \qquad \circled{4} \qquad \circled{6}$  \\
    \bottomrule
  \end{tabular}
\end{center}
\caption{The solution families depicted in Fig.~\ref{fig:k3_1_100}.}
\label{t:4}
\end{table}


\begin{figure*}
  \centering
  \includegraphics[width=6.5cm]{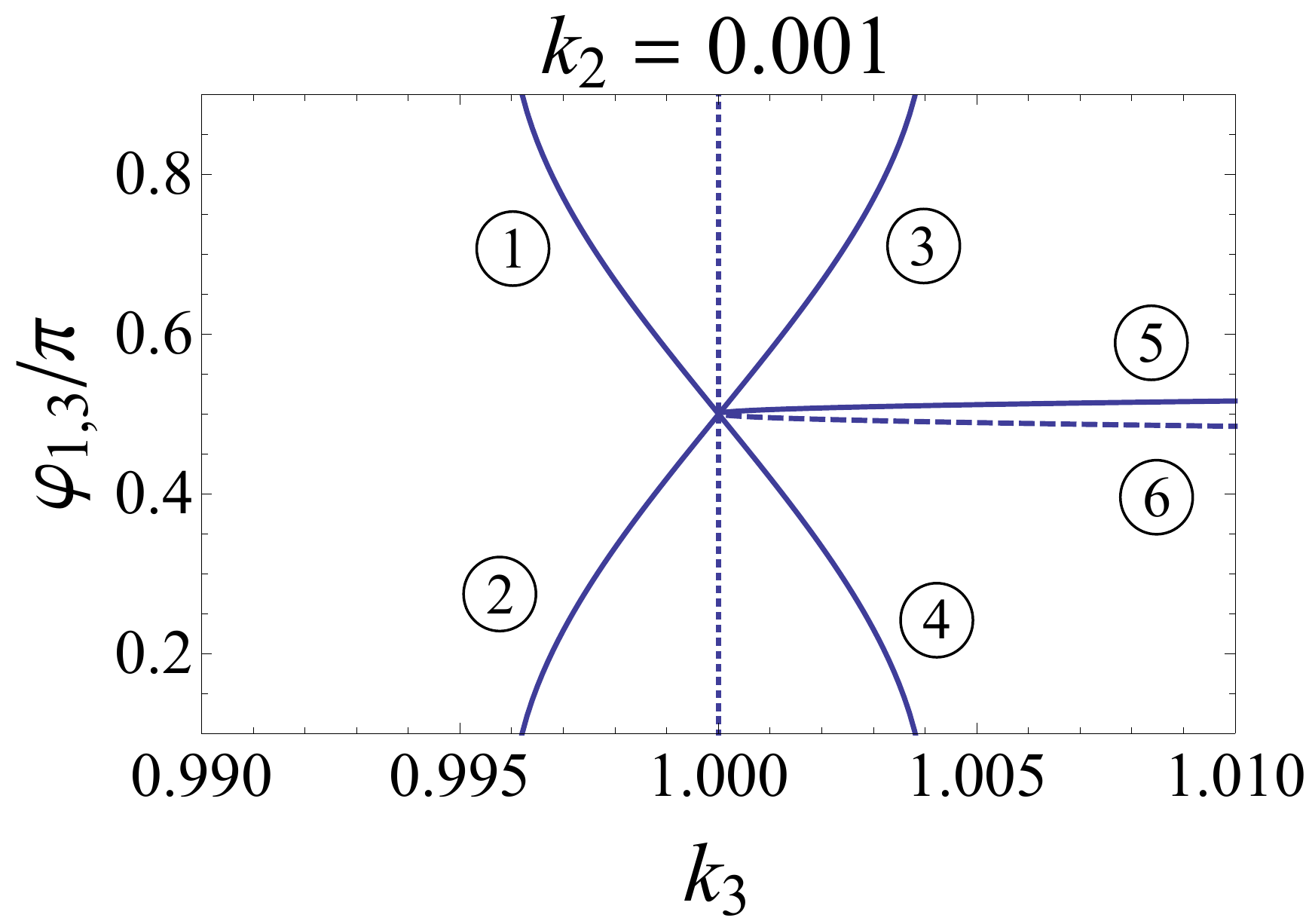}
  \hspace{0.5cm}
  \includegraphics[width=6.5cm]{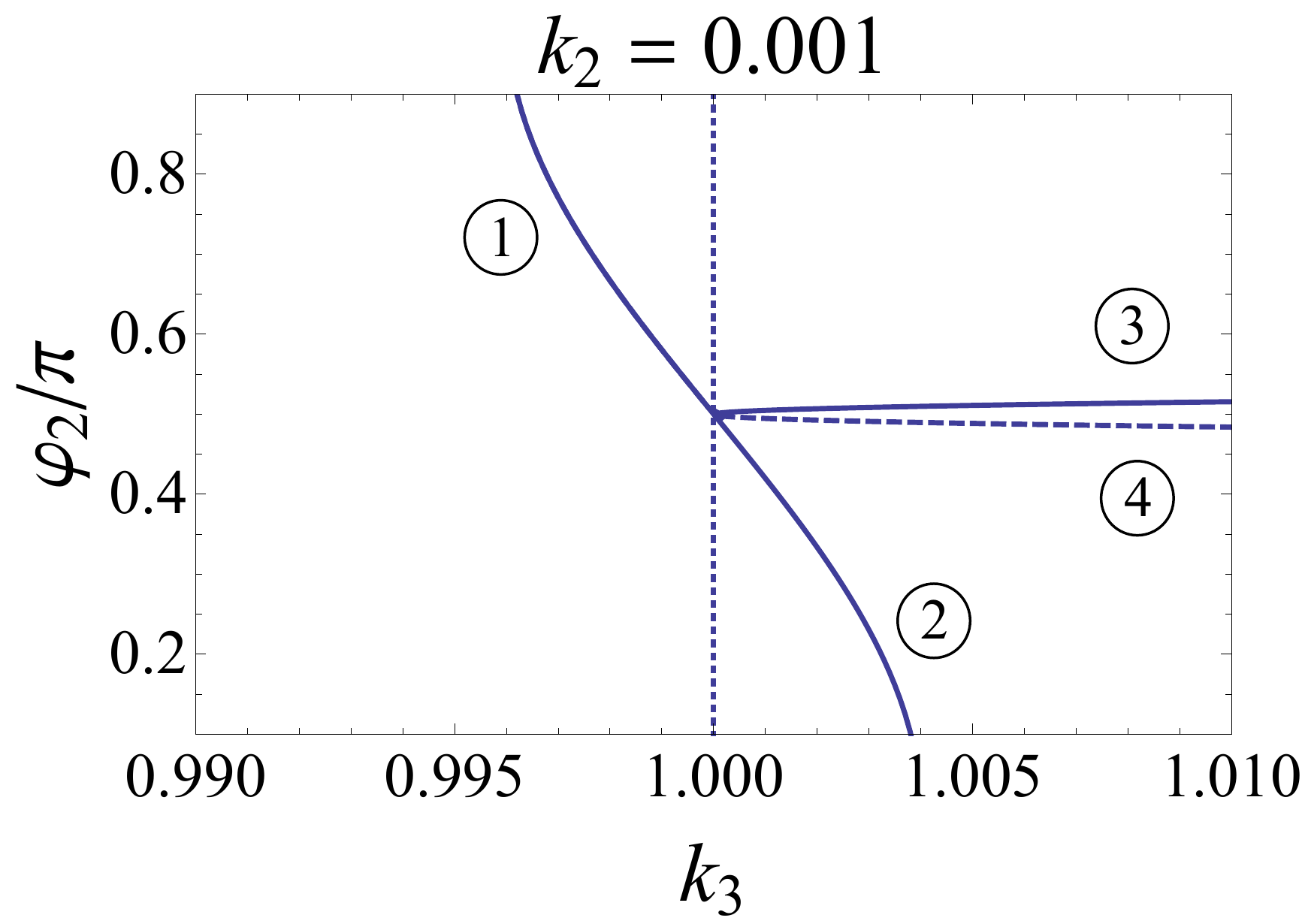}
  \\
  \includegraphics[width=6.5cm]{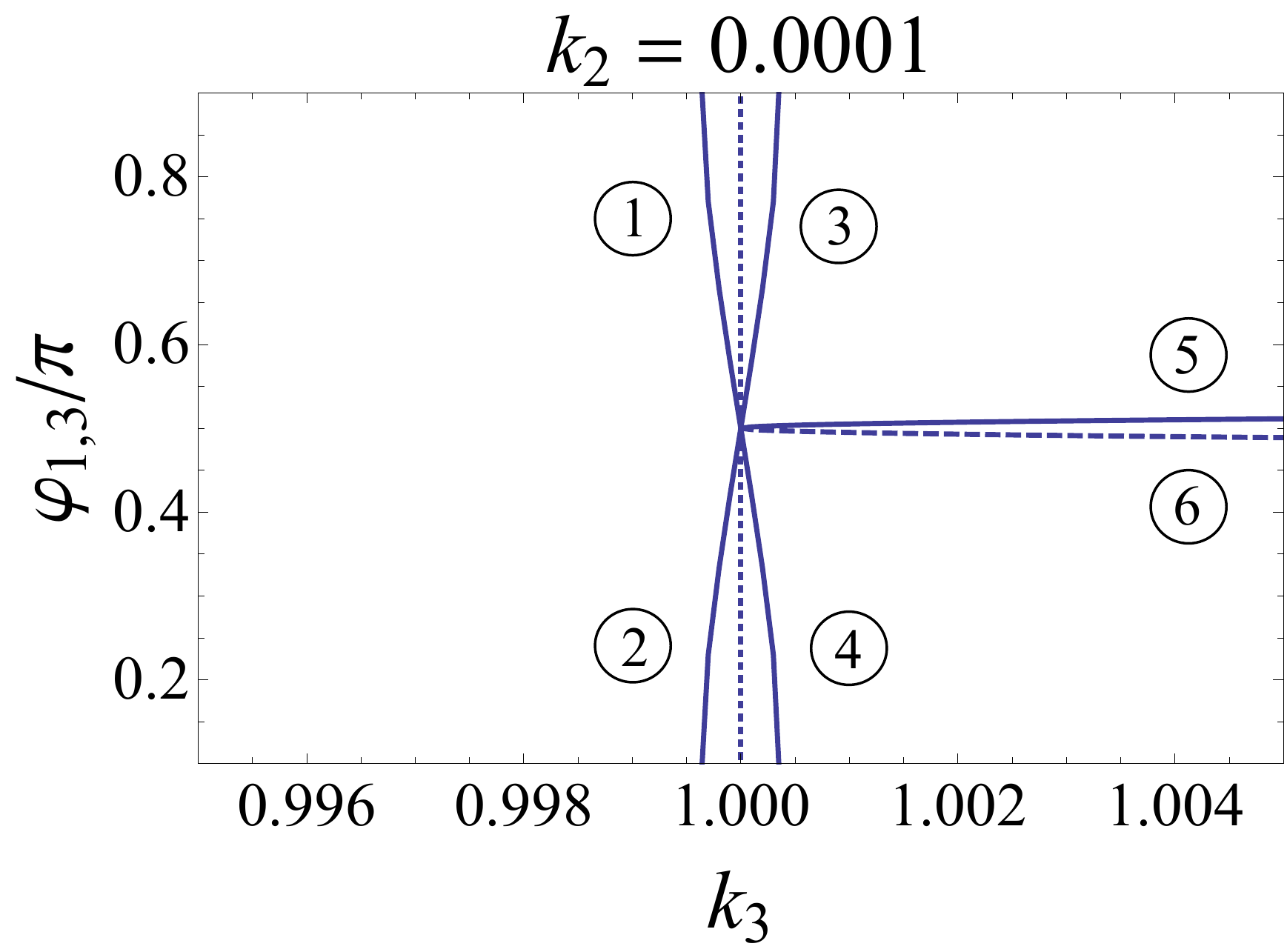}
  \hspace{0.5cm}
  \includegraphics[width=6.5cm]{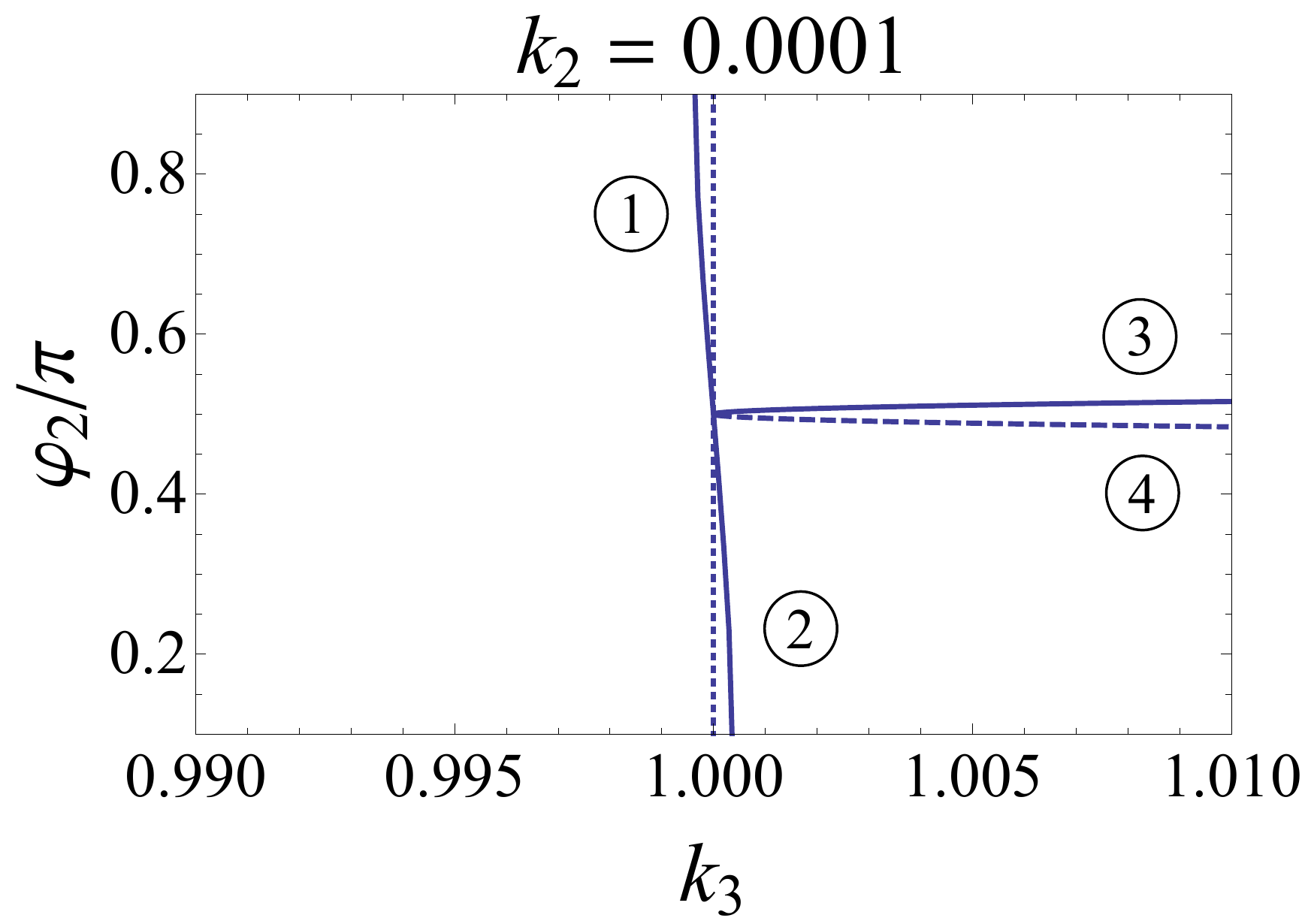}
\\
  \includegraphics[width=6.5cm]{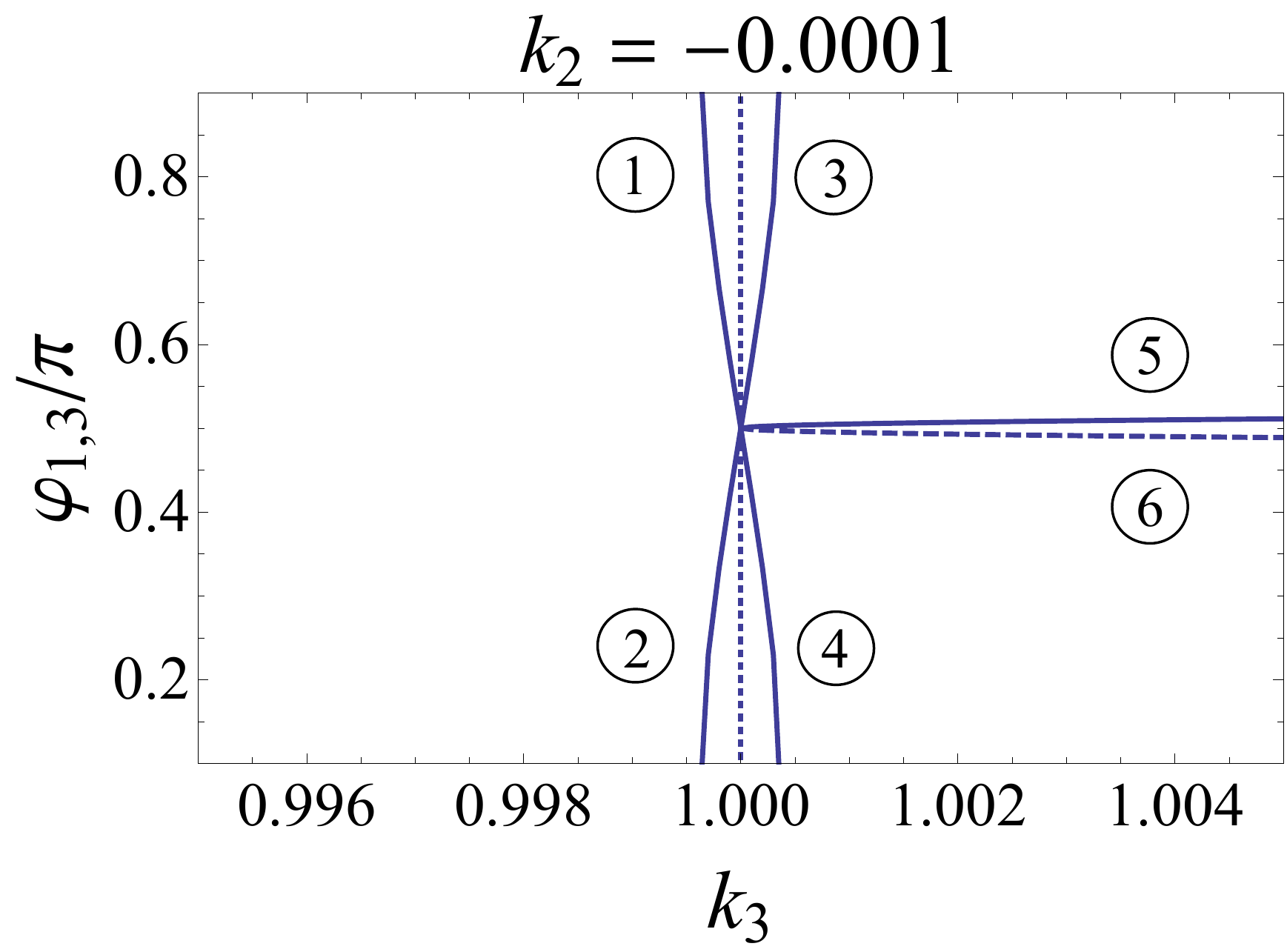}
  \hspace{0.5cm}
  \includegraphics[width=6.5cm]{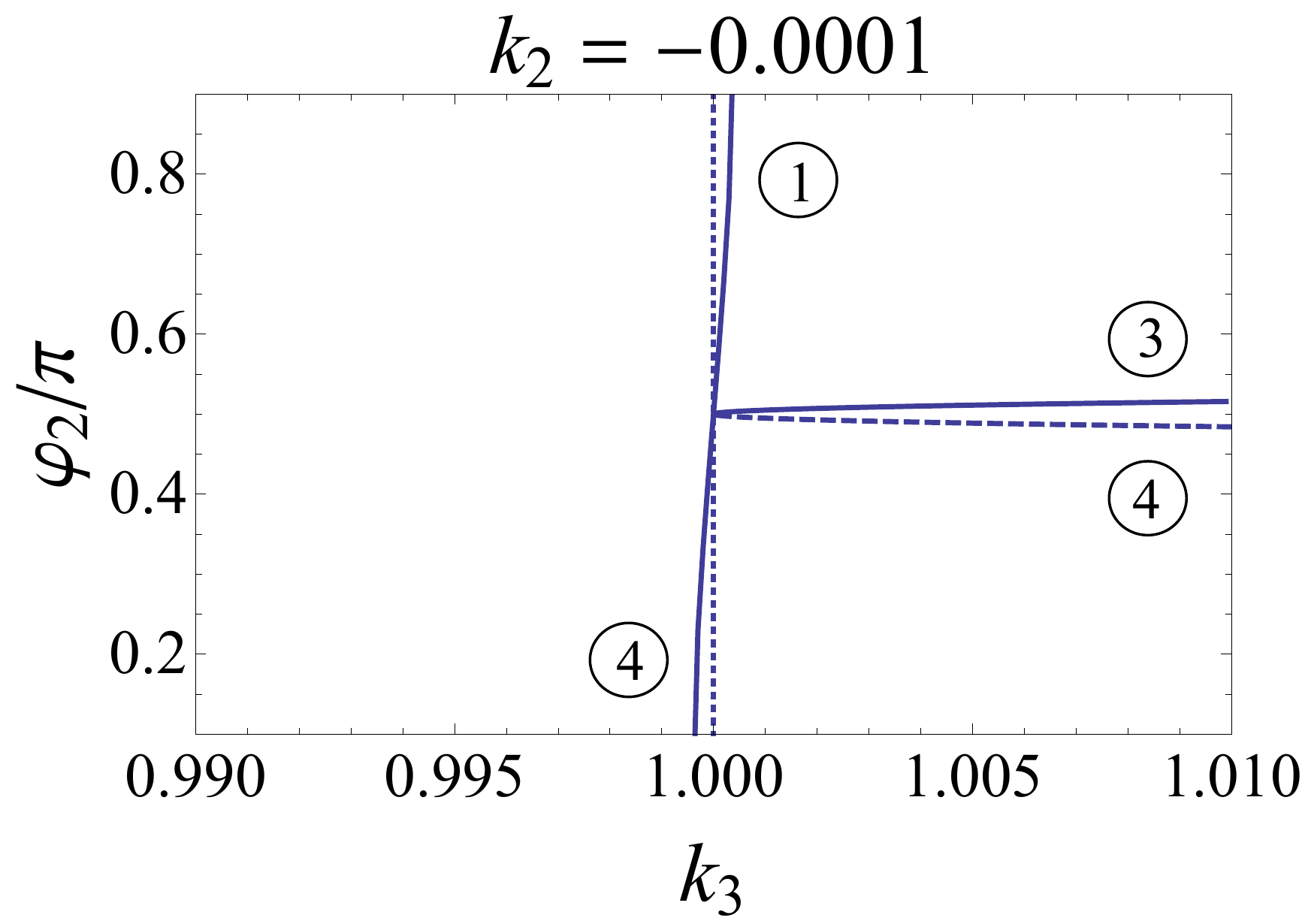}
\\
  \includegraphics[width=6.5cm]{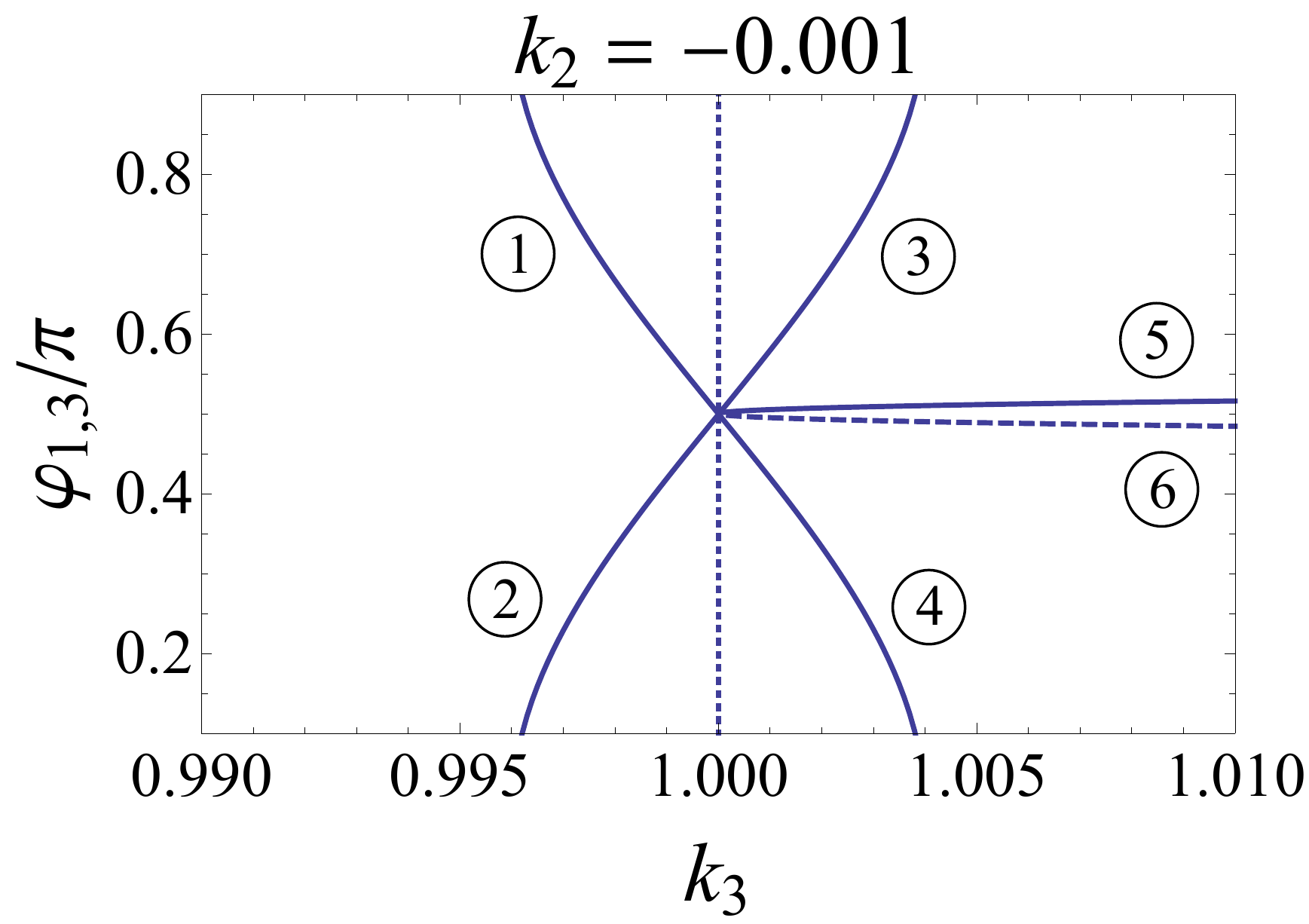}
  \hspace{0.5cm}
  \includegraphics[width=6.5cm]{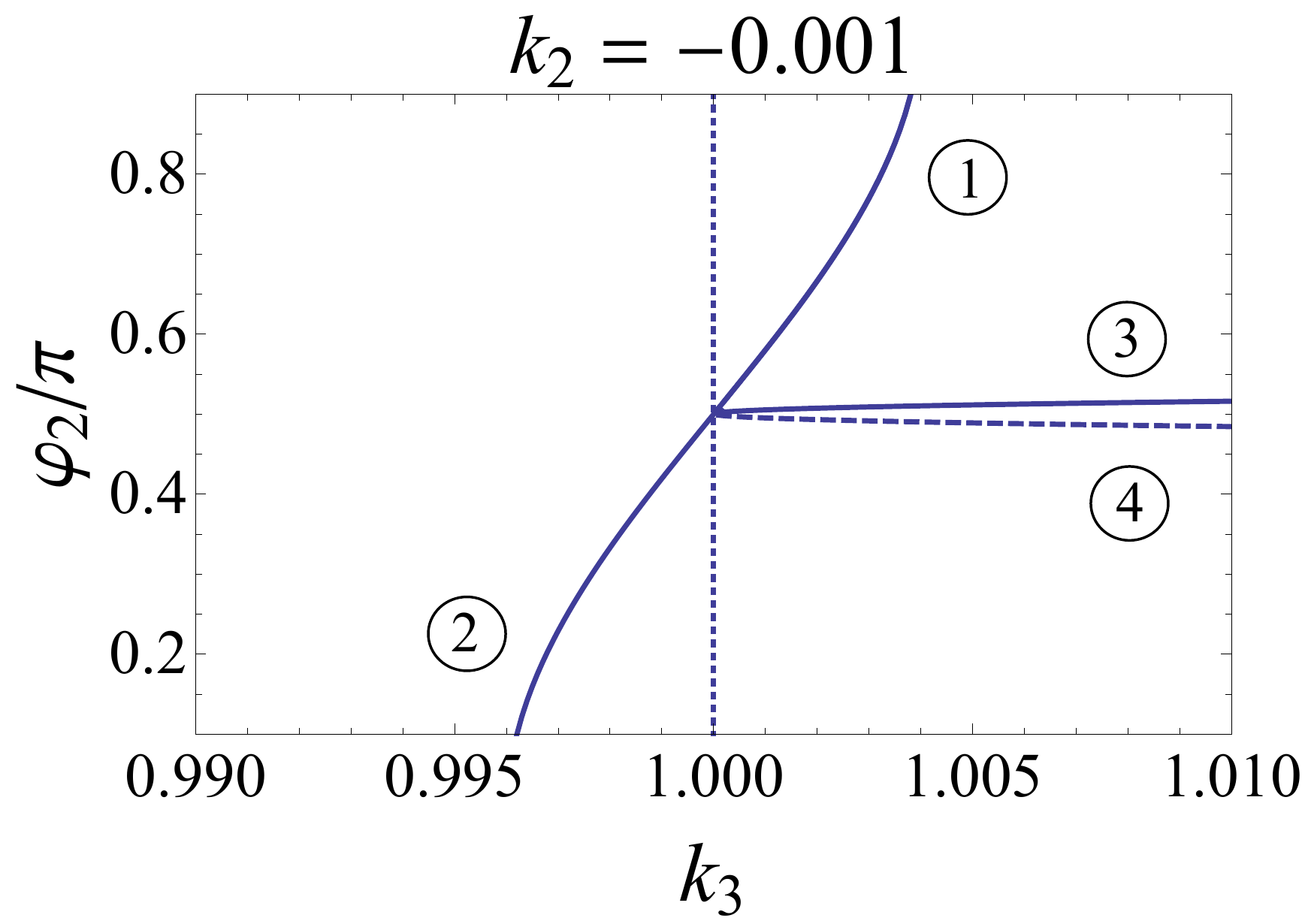}
  \caption{The bifurcation diagrams in the neighborhood $k_3=0$ and
    $\pmb{\vphi}={\Phi^{({\rm sv})}}_{101}$ are shown,
    for $k_2=-0.001, -0.0001, 0.0001, 0.001$, respectively. The family
    $F_2:(\varphi,\pi-\varphi, \vphi)$ is
    degenerate and it is represented by a
    dotted line at $k_3=1$.}
  \label{fig:near_k2_0_k3_1_flipped}
\end{figure*}

\subsubsection{The $k_3=1$ case}
For $k_3=1$, the Jacobian is highly degenerate and hence we show no
frame for this value of $k_3$. Nevertheless, it is straightforward to
see that the three families $F_1, F_2$ and $F_3$ coincide at $k_3=1$. 

In order to demonstrate this fact better, as well as to better show the role of
the families 5 and 6 of the $k_3>1$ case, we reverse that role of $k_2$ and
$k_3$ in the diagrams.

\subsubsection{The $k_2\neq0$ case}

We consider now specific values of $k_2$ close to $k_2=0$
(i.e.,~$k_2=-0.001, -0.0001, 0.0001, 0.001$) and an interval of values
around $k_3=1$. We numerically seek for solutions of the persistence
conditions \eqref{e.per4_KG.full} and the results are shown in
Fig.~\ref{fig:near_k2_0_k3_1_flipped}. First of all we can see the
family $F_2:\pmb{\vphi}=(\vphi, \pi-\vphi, \vphi)$ which exists for
$k_3=1$ and every value of $k_2$. Since this family is degenerate it
is depicted as a dotted line. The rest of the families depicted there
are shown in Table~\ref{t:5} below. We can see that families 1 and 4 tend to
$F_1$ while families 2 and 3 tend to $F_3$ as $k_2\to0$. Geometrically
this means that both tend to the $k_3=1$ asymptote. On the other hand,
there exist families 5 and 6 which correspond to the families 5 and 6
of Fig.~\ref{fig:k3_1_100}. We see that they exist only for
$k_3\geqslant1$ being a product of a saddle-node bifurcation occurring
at $k_3=1$. Although these are $k_2, k_3$-parameter solution families
for \eqref{e.per4_KG.full}, they constitute an isolated solution of
Eqs.~\eqref{e.per4_KG.deg}.
\begin{table}[h!]
\begin{center}
  \begin{tabular}{cc}
  \toprule  
    \multirow{2}{*}
             {\parbox[c]{1.5cm}{\centering $\sharp$ of Family}} & Branch description\\
    \cmidrule(r){2-2}
     & $\varphi_1 \qquad  \varphi_2 \qquad  \varphi_3$   \\
    \midrule
    1 & $\circled{1} \qquad \circled{1} \qquad \circled{2}$  \\
    2 & $\circled{2} \qquad \circled{1} \qquad \circled{1}$  \\
    3 & $\circled{3} \qquad \circled{2} \qquad \circled{4}$  \\
    4 & $\circled{4} \qquad \circled{2} \qquad \circled{3}$  \\
    5 & $\circled{5} \qquad \circled{3} \qquad \circled{5}$  \\
    6 & $\circled{6} \qquad \circled{4} \qquad \circled{6}$  \\
    \bottomrule
  \end{tabular}
\end{center}
\caption{The solution families depicted in Fig.~\ref{fig:near_k2_0_k3_1_flipped}.}
\label{t:5}
\end{table}

\begin{figure}
  \centering
  \includegraphics[width=0.4\textwidth]{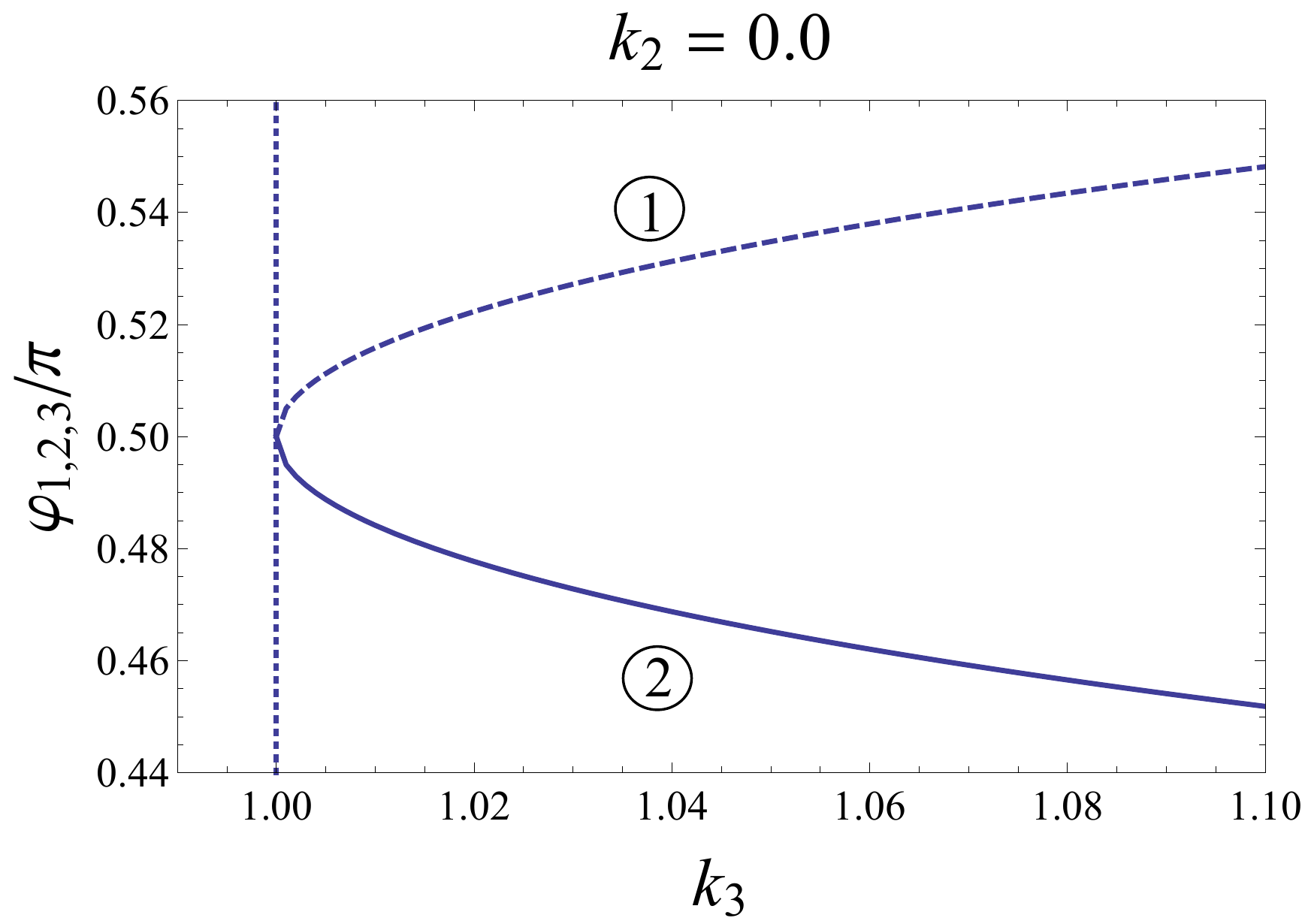}
  \caption{The bifurcation diagram for $k_2=0$ in the
    neighborhood $k_3=1$, $\varphi_i=\pi/2$ is shown.}
  \label{fig:k2_0}
\end{figure}

\subsubsection{The $k_2=0$ case}
To complement the picture set forth in the previous subsections, we separately
consider the special case $k_2=0$. For this value of $k_2$ the only families
that exist for $k_3\neq0$ are the families 5 and 6 of
Fig.~\ref{fig:near_k2_0_k3_1_flipped} as it can be shown better in
Fig.\ref{fig:k3_1_100}. The resulting bifurcation diagram is shown in
Fig.~\ref{fig:k2_0} since for this particular case it holds that
$\varphi_1=\varphi_2=\varphi_3$.

The fact that for this choice of $k_2$ and for $k_3=1$ we get the
symmetric vortex solution ${\Phi^{({\rm sv})}}_{101}$ both as a member
of the vertical families and as a member of the ``parabolic'' family,
numerically poses the question of the existence of the symmetric
vortex solution in the real system. This question is also triggered by
the fact that the two-dimensional analogue of our system in the dNLS
limit it has been proven to suport vortex solutions
\cite{PelKF05b}. 

We summarize the results of the previous numerical investigation, by saying that
the persistence conditions provide three one-parameter
families of candidate MBs, instead of the two families for
the $\HH_{110}$ case. Each family carries two standard in-phase/out-of-phase
solutions (whose existence is guaranteed via other approaches,\cite{PelS12}) and
the three intersect in two highly symmetric objects, having ${\pmb
  \vphi}=\Phi^{({\rm sv})}_{101}$ and emulating
two-dimensional vortices. The same kind of scenario and
consequent degeneracy is shared by its non-local discrete NLS (NL-dNLS)
approximation
\begin{equation}
  \label{e.NLdNLS.deg.101}
  \begin{aligned}
  {H_{\rm 101}} &=
  \sum_j|\psi_j|^2 + \frac38\sum_j|\psi_j|^4 \\ &+
  \frac{\eps}2\sum_j\quadr{|\psi_{j+1}-\psi_j|^2+|\psi_{j+3}-\psi_j|^2}\ ,
  \end{aligned}
\end{equation}
examined systematically in \cite{PenSPKK16}. It is thus natural to
attempt transfering the nonexistence results there obtained to the
corresponding $\HH_{110}$ model, by means of an accurate mathematical
analysis.  However, the techniques developed in the present paper in
the subsequent sections are tailored for less degenerate models.  More
comments on this are reported in Section~\ref{sbs:H101}.

\section{The zigzag KG model}\label{s:cont}
Motivated by the numerical results detailed in Section~\ref{s:NBA}, we
present here a complete description of the problem of continuing, from
the anti-continuous limit $\eps=0$, the phase shift solutions obtained
from the persistence condition \eqref{e.per4_KG.zz}{, thus
  focusing on the zigzag model \eqref{e.KG.zz}. We first develop a
  Lyapunov-Schmidt decomposition (see \cite{AmbP95}), which enables to
  link time-periodic solutions of the general class of KG models
  \eqref{ham_long} to time-periodic solutions of the corresponding
  resonant normal forms \eqref{e.NLdNLS}; there the continuation
  problem can be formulated, and solved, in a simpler way due to the
  rotational symmetry. To relate \eqref{ham_long} with
  \eqref{e.NLdNLS} we follow essentially the scheme developed in
  \cite{BamPP10}.} The proof will be divided in several steps, the
last one illustrated in a separated Section. Here we will also provide
the more detailed version of Theorem~\ref{t.nonexistence}; the
corresponding statements (Theorems~\ref{T.2} and~\ref{T.1}) are
presented after the first step of the proof in order to make reference
to the objects introduced at that stage. {In particular,
  Theorem~\ref{T.2} is formulated for the general class of
  Hamiltonians \eqref{ham_long}, while Theorem~\ref{T.1} applies in
  the restricted context of the zigzag KG model \eqref{e.KG.zz} and
  its normal form \eqref{e.NLdNLS.zz}.}

Let us consider the KG Hamiltonian \eqref{ham_long} and its equations
of motion
\begin{equation}
  \label{e.Kg.eq}
  \ddot x_j = -x_j - x_j^3 + \eps\tond{Lx}_j\ ,
\end{equation}
with
\begin{equation*}
  \begin{aligned}
    L &:= \Delta_1+k_2\Delta_2 + k_3 \Delta_3 \ ,
    \\
    (\Delta_m x)_j &:= x_{j-m} - 2x_j + x_{j+m} \ .
  \end{aligned}
\end{equation*}
We look for a periodic orbit with frequency $\gamma$; hence by
introducing the time scaling $u_j(\tau):=x_j(t)$, where $\tau:=\gamma
t$, we get
\begin{equation}
  \label{e.scale.Kg.eq}
  \gamma^2 u'' + u + u^3 - \eps Lu = 0\ .
\end{equation}
We define
\begin{equation}
  \label{e.def.op}
  L_0 := \gamma^2\partial_\tau^2 + \Id\ ,\qquad L_{\eps}:= L_0 - \eps
  L\ ,\qquad N(u):= u^3\ .
\end{equation}
The equation for a generic periodic orbit becomes
\begin{equation}
  \label{e.po}
  \FF(\eps,u):= L_{\eps}u + N(u) = 0\ ,
\end{equation}
with
\begin{displaymath}
  \FF:\RR\times X_2:=H^2\tond{[0,2\pi],\ell^2}\to
  X_0:=L^2\tond{[0,2\pi],\ell^2}\ ,
\end{displaymath}
where $X_{0,2}$ are endowed with the usual norms (see Sections 3.2 and
3.3 of \cite{BamPP10}).

As it is stated in the Introduction, we consider, in the unperturbed case
$\eps=0$, a periodic orbit $\bar u(\tau)$ which lies on the
four-dimensional completely resonant torus~\eqref{e.u0} with amplitude
$\rho$ (see~\eqref{e.nonlin.osc}). We wish to continue this periodic
orbit for $\eps\not=0$, thus we look for a function
$u(\rho,\eps,\tau)$ such that $u(\rho,0,\tau) = \bar u(\tau)$ and
\eqref{e.po} is solved for $\eps$ small enough
\begin{equation}
  \label{e.problem}
  \FF(\eps,u(\rho,\eps,\tau))= 0\ ,\qquad |\eps|<\eps^*(\rho)\ ,
\end{equation}
with $\gamma$ kept fixed.

The problem has been partially solved in \cite{PelS12} by restricting
to time-reversible solutions $u(-\tau)=u(\tau)$; i.e., by considering
only standard phase-differences $\varphi_j=\{0,\pi\}$ for all $j\in
S$.  Indeed, with this strategy the problem reduces to non-degenerate
critical points where the implicit function theorem can be applied,
like in the averaging approach of \cite{AhnMS02,KouK09,KouI02}. In the
case of other phase-differences, like the vortex (or phase-shift
multibreather) solutions we consider here, it is not possible to make
such a restriction, which ensures invertibility of the linearized
operator $\FF_u(0,\bar u)$ on the subspace of even periodic
solutions. In other words, in our case, the approximate solution $\bar
u$ is a degenerate critical point; thus a small perturbation may in
principle destroy the solution. In order to see that the linearized
operator $\FF_u(0,\bar u)$ has a non-trivial Kernel, let us recall
some facts presented in the first part of
\cite{PelS12}\footnote{ Geometrically, the main idea is
  that a small displacement on the four-dimensional torus from a given
  unperturbed periodic solution, leads to a new unperturbed periodic
  solution with the same frequency.}.

First, notice that
\begin{displaymath}
  \FF_u(0,\bar u)[\zeta] = \begin{cases}
    L_0 \zeta_j                           & j\not\in S\ ,\\
    L_0 \zeta_j + 3\bar u^2(\tau)_j \zeta_j   & j\in S\ .
  \end{cases}
\end{displaymath}
The non-resonant condition $j\gamma\not=\pm 1$ allows to invert
$L_0$ on the space of $2\pi$-periodic functions. On the other hand,
differentiating the nonlinear oscillation equation w.r.t. both $\tau$
and the energy $E$, one sees that
\begin{displaymath}
  {\rm Ker}\tond{\gamma^2\partial^2_\tau + 3x^2(\tau + \varphi_j)} =
  \inter{x'(\tau+\varphi_j), \tau\derp{\gamma}{E}x'(\tau+\varphi_j)};
\end{displaymath}
as a consequence, the non-degeneracy condition of the
frequency $\derp{\gamma}{E}\not=0$ guarantees that only the time derivatives
$x'(\tau+\varphi_j)$ are $2\pi$-periodic solutions. Thus the differential
$\FF_u(0,\bar u)$ has a four-dimensional Kernel
\begin{displaymath}
  {\rm Ker}\tond{\FF_u(0,\bar u)} = \inter{f_j(\tau)}\ ,\qquad j\in S\ ,
\end{displaymath}
generated by the velocities of the nonlinear oscillations\footnote{We
  will use the notation $[\ldots|\cdot,\cdot,\cdot,\cdot|\ldots]$ to
  denote values along the chain: in particular, the two vertical bars
  enclose the sites belonging to $S$.}
\begin{displaymath}
  f_j = \quadr{0\big|x'(\tau+\varphi_j)\big|0}\ .
\end{displaymath}
For the above reason an implicit function theorem cannot be applied,
unless (as in~\cite{PelS12}) we
restrict\footnote{Indeed, let us set $u(\tau):=x'(\tau+\pi)$. Using
  $2\pi$ periodicity of $x(\tau)$ and its even-parity we immediately
  get $u(-\tau) = x'(-(\tau-\pi)) = -x'(\tau-\pi) = -x'(\tau+\pi) =
  -u(\tau)$. Hence the velocities $x'(\tau+\varphi_j)$ are not even
  functions, i.e., the Kernel is transversal to the subspace of even
  solutions.} to $\varphi_j\in\graff{0,\pi}$, and a Lyapunov-Schmidt
decomposition represents a natural approach to the problem.



\subsection{The first Lyapunov-Schmidt decomposition}
We consider \eqref{e.scale.Kg.eq} and we introduce the time-Fourier
expansion for the solution of the uncoupled anharmonic oscillator $x(\tau)$ in
\eqref{e.nonlin.osc}
\begin{equation}
  \label{e.x.Fou}
  x(\tau) = \sum_{k\geq 1} a_k\cos(k\tau)\ ;
\end{equation}
then, from \eqref{e.u0}, we get
\begin{displaymath}
  \bar u_j(\tau) = \begin{cases}
    0                                        &j\not\in S\\
    \sum_{k\geq 1} a_k\cos(k\tau + k\varphi_j)  &j\in S\\
  \end{cases}\ ,
\end{displaymath}
thus we can write $ \bar u_j = \sum_{k\geq 1} a_k \Big(
\cos(k\varphi_j)\cos(k\tau) -\sin(k\varphi_j)\sin(k\tau) \Big)$, for
any $j\in S$.  Let us now introduce the Fourier base
\begin{equation}
  \label{e.Fou.base}
  e_k(\tau)
  = \begin{cases}
    \cos(k\tau)    & k\geq 0 \\
    -\sin(k\tau)   & k<0
  \end{cases}\ ;
\end{equation}
we can decompose $u\in\ell^2(\RR)$ in its Fourier
components\footnote{We will always use the subscript $k$ to denote the
  Fourier index, and the subscript $j$ for the site index.}
\begin{equation}
  \label{e.Fou}
  u(\tau) = \sum_{k\in \ZZ\setminus\graff{0}} u_k e_k(\tau)\ ,
\end{equation}
and introduce the Lyapunov-Schmidt decomposition\footnote{Please
  notice the use of the sans serif font for the present decomposition
  variables: $\vvv$ and $\www$. From subsection~\ref{ss:bifeq}, the
  letters $v$ and $w$, with the usual font, will have a different
  meaning.} which splits the first harmonics from the rest of the
Fourier expansion
\begin{equation}
  \label{e.Fou.dec}
  u = \vvv+\www\ ,\qquad \vvv = u_{-1}e_{-1}(\tau) + u_{1}e_{1}(\tau)\ ;
\end{equation}
in other words $\vvv$ solves the harmonic oscillator equation
$\frac{d^2\vvv}{dt^2}+\vvv=0$. We define
\begin{equation}
  V_2:=\inter{e_1,e_{-1}} = \ker\left(\partial_\tau^2+\Id\right)\ ,
  \quad
  W_2:= V_2^\complement\ .
  \label{frm:V2}
\end{equation}
If we consider the
unperturbed reference solution $\bar u$, we have for any $j\in S$
\begin{displaymath}
  \bar u_j = \sum_{k\in\ZZ} \bar u_{j,k} e_k(\tau) \ ,
  \qquad
  \bar u_{j,k} = \begin{cases}
    a_k\cos(k\varphi_j)   &k>0\\
    0                     &k=0\\
    -a_k\sin(k\varphi_j)  &k<0
  \end{cases} \ ,
\end{displaymath}
thus we get
\begin{equation}
  \label{e.v0}
  \bar \vvv_j = a_1\cos(\varphi_j)e_1(\tau) - a_1\sin(\varphi_j) e_{-1}(\tau)\ .
\end{equation}

We rewrite the frequency as $\gamma^2 = 1-\omega \,.$ Indeed, in the
small energy regime, the frequency $\gamma$ is close to one, and its
displacement $\omega$ is of order $\mathcal{O}(\rho^2)$.  The equation
\eqref{e.scale.Kg.eq} thus reads
\begin{equation}
  \label{e.dec.eq}
  \FF(\eps,\vvv,\www) = L_{\eps}\www -\omega \vvv -\eps L\vvv + N(\vvv+\www)=0\ .
\end{equation}

When we project \eqref{e.dec.eq} on the Range $W_0\subset X_0$ of
$\partial_\tau^2+\Id$, and its complement
$V_0$, we get\footnote{For an easier notation we drop the zero
  subscript in the projectors $\Pi_V\equiv\Pi_{V_0}$ and
  $\Pi_W\equiv\Pi_{W_0}$.}
\begin{equation}
  \label{e.RK.eq}
  \begin{cases}
    \Pi_W\FF(\eps,\vvv,\www) = L_{\eps}\www + \Pi_W N(\vvv+\www)=0
    &\quad (R)\\
    \Pi_V\FF(\eps,\vvv,\www) = -\omega \vvv - \eps L\vvv + \Pi_V N(\vvv+\www)=0
    &\quad (K)
  \end{cases}\ .
\end{equation}
Proceeding as in Section 4 of \cite{BamPP10}, the Range equation (R),
written as $\www = -L_{\eps}^{-1}\Pi_W N(\vvv+\www)$, can be locally solved and
approximated by $\tilde\www(\vvv,\eps)$
\begin{displaymath}
  \tilde\www(\vvv,\eps) := -L_{\eps}^{-1}\Pi_W N(\vvv)
  = \mathcal{O}(\norm{v}_{X_2}^3)\ ,
\end{displaymath}
with
\begin{displaymath}
    \norm{\www-\tilde\www}_{X_2}\leq C\norm{\vvv}_{X_2}^5\ .
\end{displaymath}
We move now to the Kernel equation (K), i.e., $-\omega \vvv - \eps L\vvv +
\Pi_V (\vvv+\www(\vvv,\eps))^3=0$.
Since $\www(\vvv) = \mathcal{O}(\norm{\vvv}_{X_2}^3)$, we can expand
\begin{displaymath}
  \begin{aligned}
  \Pi_V (\vvv+\www(\vvv,\eps))^3 &=
  \Pi_V (\vvv)^3 + \quadr{\Pi_V (\vvv+\www(\vvv,\eps))^3-\Pi_V (\vvv)^3} \\&=
  \Pi_V (\vvv)^3 + \mathcal{O}(\norm{\vvv}_{X_2}^5).
  \end{aligned}
\end{displaymath}
We compute explicitly the Kernel projection of the leading term of
the nonlinear part. First we have, by definition
\begin{displaymath}
  \begin{aligned}
    \Pi_V (\vvv)^3 = &\tond{\frac1{2\pi}\int_0^{2\pi}
      \vvv^3(\tau)\cos(\tau)d\tau }e_1 +
    \\
    &\tond{\frac1{2\pi}\int_0^{2\pi}
      \vvv^3(\tau)\sin(\tau)d\tau }e_{-1}\ ,
  \end{aligned}
\end{displaymath}
and since, omitting the $\tau$ dependence, we have
\begin{displaymath}
\vvv^3 = u_1^3e_1^3 +
3u_1^2u_{-1}e_1^2e_{-1} + u_{-1}^3e_{-1}^3 + 3u_1u_{-1}^2e_1e_{-1}^2\ ,
\end{displaymath}
trigonometric formulas give immediately
\begin{equation}
  \label{e.v3.K}
  \begin{aligned}
  \Pi_V (\vvv)^3 &= \frac34\left(
    \tond{u_1^2+u_{-1}^2}u_1\cos(\tau) +
    \tond{u_1^2+u_{-1}^2}u_{-1}\sin(\tau)\right)\\& =
    \frac34\tond{u_1^2+u_{-1}^2}\vvv\ .
    \end{aligned}
\end{equation}
By defining the remainder as
\begin{equation}
  \label{e.R_K}
  \R(\vvv,\eps) := \Pi_V (\vvv+\www(\vvv,\eps))^3-\Pi_V (\vvv)^3\ ,
\end{equation}
we can rewrite the Kernel equation as
\begin{equation}
  \label{e.Ker.1}
  -\omega \vvv - \eps L\vvv + \frac34\tond{u_1^2+u_{-1}^2}\vvv + \R(\vvv,\eps) =
  0\ .
\end{equation}
The Kernel equation, due to its dimension ($\vvv$ is a two-dimensional
vector of sequences), is equivalent to the system
\begin{displaymath}
  \begin{aligned}
    -\omega u_1 - \eps Lu_1 + \frac34\tond{u_1^2+u_{-1}^2}u_1
    + \inter{\R(\vvv,\eps),e_1}    &= 0
    \\
    -\omega u_{-1} - \eps Lu_{-1} + \frac34\tond{u_1^2+u_{-1}^2}u_{-1}
    + \inter{\R(\vvv,\eps),e_{-1}}     &= 0
  \end{aligned}\ \ .
\end{displaymath}
Introducing the complex variable $\phi$
\begin{equation}
  \label{e.vphi.def}
  \phi := u_1 + \im u_{-1}\ ,
\end{equation}
equation \eqref{e.Ker.1} takes the form
\begin{equation}
  \label{e.Ker.vphi}
  -\omega\phi - \eps L\phi +\frac34\phi|\phi|^2 + \R(\phi,\eps) = 0\ ,
\end{equation}
using again the letter $\R$ to denote the corresponding term
of~\eqref{e.Ker.1}.  It turns out that in the small energy regime
(i.e., for $\rho$ small enough) \eqref{e.Ker.vphi} looks as a
$\rho^2$-perturbation of the NL-dNLS stationary problem
\begin{equation}
  \label{e.Ker.app.0}
  -\omega\phi - \eps L\phi +\frac34\phi|\phi|^2 = 0\ ;
\end{equation}
in other words, the term $\R(\phi,\eps)$ can be treated
as a perturbation. Moreover, since also the remainder $\R$ is equivariant under
the rotational symmetry and conjugation, the whole \eqref{e.Ker.vphi} actually
represents the stationary equation for a non-local dNLS model. We are now ready
to give more detailed statements; Theorem~\ref{t.nonexistence} can be seen as
their corollary.

\subsection{Reformulation of the main results}
The first statement allows to derive an existence and approximation
result for a solution of~\eqref{e.scale.Kg.eq},
  $u(\rho,\eps,\tau)$, from the existence of a non-degenerate NL-dNLS solution
$\vvv(\eps,\tau')$, precisely
\begin{theorem}
  \label{T.2}
  Let $\phi(\eps)$ be a non-degenerate $\eps$-family of solutions for
  \eqref{e.Ker.app.0} and let $\vvv(\eps,\tau')$ be the corresponding
  real solution in $V_2$, see~\eqref{frm:V2}. Then, there exist $E^*$
  and $\eps^*$ and a constant $C_1$, such that, for $E<E^*$ and $\eps<
  E\,\eps^*$, there exists a non-degenerate two parameter family
  $u(\rho,\eps,\tau)$, solutions of \eqref{e.po}, which fulfills
  \begin{equation}
    \label{e.est.1}
    \norm{u(\rho,\eps,\cdot) - \vvv(\eps,\cdot)}_{X_2}< C_1 \rho^3\ .
  \end{equation}
\end{theorem}

Several remarks are in order:
\begin{itemize}
  \item Theorem \ref{T.2} applies to any discrete soliton solution of
    the dNLS model \eqref{e.NLdNLS}, which is obtained as isolated
    solution of the corresponding persistence condition via implicit
    function theorem. In particular, it applies to those standard
    phase-difference solitons of the model $H_{110}$ (but also in the
    more degenerate case $H_{101}$) which do not belong to the
    1-parameter families of solutions of {\eqref{e.per4_KG.zz},
      provided \eqref{e.M_phi.DNLS} holds}.

  \item The true solution and its approximation are of order
    $\mathcal{O}(\rho)\sim\mathcal{O}(\sqrt{E})$, thus the bound \eqref{e.est.1}
    on their difference, being $\mathcal{O}(\rho^3)$, is meaningful.
  \item The non-degeneracy assumption in Theorem \ref{T.2} for the
    NL-dNLS solution is related to the constrained Hessian $D^2
    E_1(\phi(0))$, being $E_1$ the $\eps$-depending part of the
    Hamiltonian~\eqref{e.NLdNLS}
    (see~\cite{Kap01,Kev09,PelKF05}). This condition is equivalent
    (through the variational formulation of \eqref{e.Ker.app.0}) to
    the non-degeneracy of the linearized bifurcation equation we will
    use in Proposition~\ref{t.EX}.
  \item The distinct time variables, i.e., $\tau$ and $\tau'$,
    reflect the different frequencies of the two unperturbed reference
    solutions.  Indeed, since the $\eps$-continuation is performed at fixed
    frequency, the two solutions keep this frequency difference.  The different
    time variables permit to normalize the period to $2\pi$.
\end{itemize}

The second statement claims the nonexistence of four-site 
vortices in the zigzag case
KG model \eqref{e.KG.zz}, at least in the regime of
small enough energy $E$:
\begin{theorem}
  \label{T.1}
  For any $\varphi\in(0,2\pi)$, $\varphi\neq \pi$, there exists
  $E^*(\varphi)$ such that, for $E<E^*$, the solutions
  \eqref{e.fam.dnls.zz} of \eqref{e.per4_KG.zz} cannot be continued at
  $\eps\not=0$.
\end{theorem}

Here we also have to stress that:
\begin{itemize}
  \item The nonexistence statement is based on the analogous result
    for the zigzag-dNLS model \eqref{e.NLdNLS.zz}, where the
    impossibility to solve the linearized bifurcation equation is
    sufficient to conclude the proof. The
    same holds if the linearized bifurcation equation is slightly
    perturbed, which is exactly what happens in the model
    \eqref{e.KG.zz} if the energy $E$ is taken small
    enough. This is discussed in
    Section~\ref{sbs:proofT1}.
  \item It turns out that $E^*(\varphi)\searrow 0$ as $\varphi\to
    0,\pi$, in agreement with the fact
    that for $\varphi=\{0,\pi\}$ the four-sites
    MBs exist.
  \item Theorem \ref{T.1} does not exclude that four-sites
    asymmetric vortices appear for $\eps>\eps^*(\rho,\varphi)$. It
    only claims that it does not exist a continuous
    (in $\epsilon$) branch which locally arises at $\eps=0$.
\end{itemize}

\subsection{Approximation of the Kernel equation}
\label{ss:approax}

Let us first remark the following;

\medskip
\noindent
{\bf Notation.} {\it Since there are two small parameters, the
  coupling $\eps$ and the amplitude $\rho$ (introduced
  in~\eqref{e.nonlin.osc} in order to maintain a notation
  as similar as possible with the paper~\cite{PenSPKK16}), in what
  follows we will indicate the dependence with respect to $\eps$ as an
  argument of the related quantities, and the dependence on $\rho$
  (absent in~\cite{PenSPKK16}) as a subscript. We also stress that,
  where it will be clear from the context, the absence of the
  subscript $\rho$ will mean $\rho=0$, i.e., for a generic quantity
  $Z_\rho$ we will set $Z\equiv Z_0$.}
\medskip

At $\eps=0$, we denote by $v_\rho$ the unperturbed solution of
\eqref{e.Ker.vphi}, corresponding to the Kernel projection $\bar \vvv$
in \eqref{e.v0}
\begin{equation}
  \label{e.phi0}
  v_j = a_1\cos(\varphi_j) - \im a_1\sin(\varphi_j) =
              a_1 e^{-\im\varphi_j} \ ,
  \qquad
  j\in S \ .
\end{equation}
Recalling that $\R(\vvv,\eps)=\mathcal{O}(||\vvv||^5_{X_2})$,
introducing the vector field $X_N(\phi) := \frac34\phi|\phi|^2$ and
the following $\rho$-scaling
\begin{equation}
  \label{e.scale}
  \phi =: \rho\tilde\phi         \ ,
  \qquad
  \eps =: \rho^2\tilde\eps       \ ,
  \qquad
  \omega =: \rho^2\tilde\omega \ ,
\end{equation}
and immediately {\sl dropping the tildes}, equation~\eqref{e.Ker.vphi}
reads
\begin{equation}
  \label{e.Ker.rho}
  \rho^3 \Big[ -\omega\phi - \eps L\phi + X_N(\phi)
     + \rho^2 \R_\rho(\phi,\rho^2\eps) \Big] = 0  \ .
\end{equation}
Thus we have shown that, in
the small energy regime, i.e., for $\rho$ small enough,
equation~\eqref{e.Ker.rho} (equivalent to~\eqref{e.Ker.vphi}) looks as
a $\rho^2$-perturbation of the NL-dNLS problem~\eqref{e.Ker.app.0}. We
remark that also the first three terms in the square brackets depend
on $\rho$, so that in the following we will systematically add the
corresponding subscript; it has been omitted here to help the
comparison with formula~\eqref{e.Ker.app.0} and to emphasize that
those terms, though depending on $\rho$, do not vanish with $\rho$, so
that the last term in~\eqref{e.Ker.rho} is really a small correction.

\begin{remark}
  Starting from~\eqref{e.Ker.rho}, $\eps$ won't be anymore exactly the
  KG coupling (remember we are dropping the tildes of the
  scaling~\eqref{e.scale}), but it will represent the coupling of the
  (perturbed) dNLS associated to the original KG.
\end{remark}

We introduce\footnote{We will follow again paper~\cite{PenSPKK16}, decomposing
  $\phi$ in a reference solution $v$ and a correction $w$.} the scaled
unperturbed solution of~\eqref{e.Ker.rho}, $v_\rho$, (again
  {\sl dropping the tildes}), that has amplitude ${a_1}/{\rho}=\mathcal{O}(1)$
  and uniquely defines the frequency detuning $\omega_\rho$ from the harmonic
  frequency, namely
\begin{displaymath}
  \omega_\rho = \frac34|v_\rho|^2 +
  \rho^2\frac{\R_\rho(v_\rho,0)}{v_\rho} \ .
\end{displaymath}
By definition, $v$ has to solve the uncoupled NL-dNLS problem,
i.e.,~\eqref{e.Ker.rho} with $\eps=0$ and $\rho=0$,
\begin{displaymath}
  -\omega v + \frac34v|v|^2 = 0
  \quad\Rightarrow\quad
  \omega = \frac34R^2  \ ,
  \qquad
  R:=\lim_{\rho\to0}\frac{a_1}{\rho} \neq 0 \ .
\end{displaymath}
Analyzing the leading order expansion of $a_1(\rho)$, we
  provide an estimate for the distance between the two unperturbed solutions,
  $v_\rho$ and $v$ in
\begin{lemma}
  \label{l.phi0*}
  There exists $\rho^*<1$ and two constants $C_0$ and $c_0$ such that,
  for $\rho<\rho^*$, one has
  \begin{equation}
    \label{e.est.phi0*}
    \norm{v_\rho - v}_{\ell^2(\CC)} < C_0 \rho^2 \ ,
    \qquad
    |\omega_\rho - \omega| < c_0 \rho^2\ .
  \end{equation}
\end{lemma}

Once we focus on a particular solution $v_\rho$ of the uncoupled problem, we ask
for its continuation for $\eps\neq0$; we thus look for a
  correction $w_\rho(\eps)$ around $v_\rho$, that is continuous in $\eps$,
  namely
\begin{equation*}
  w_\rho(v_\rho,\eps) := \phi_\rho(\eps) - v_\rho \ ,
  \qquad\qquad\text{with}\qquad
  w_\rho(v_\rho,0) = 0 \ ,
\end{equation*}
so that $\phi_\rho(\eps)$ solves~\eqref{e.Ker.rho}.

Inserting the above definition, and exploting that $v_\rho$ is a
solution for $\eps=0$, the Kernel equation~\eqref{e.Ker.rho} takes the
form
\begin{equation}
  \label{e.Ker.per}
  0 = \FFF(v_\rho;w_\rho,\rho,\eps) :=
      F(v;w,\eps) + \RRR(v_\rho;w_\rho,\rho,\eps) \ ,
\end{equation}
where
\begin{equation}
  \label{e.G.01}
  \begin{aligned}
    F_\rho(v_\rho;w_\rho,\eps) &:=
    -\omega_\rho w_\rho - \eps L(v_\rho + w_\rho) \\&\qquad+
    \quadr{X_N(v_\rho + w_\rho)-X_N(v_\rho)}
    \\
    \RRR(v_\rho;w_\rho,\rho,\eps) &:=
    \quadr{ F_\rho(v_\rho;w_\rho,\eps) - F(v;w,\eps) } \\&\qquad+ 
    \rho^2\quadr{\R_\rho(v_\rho + w_\rho,\eps)-\R_\rho(v_\rho,0)}
\end{aligned} \ .
\end{equation}

In contrast with ~\eqref{e.Ker.rho}, in~\eqref{e.Ker.per} we
have a term completely independent of $\rho$,
i.e., $\FFF(v_\rho;w_\rho,\rho,\eps)|_{\rho=0} \equiv F(v;w,\eps)$,
which is indeed the $\mathcal{O}(1)$ leading term in $\rho$
of~\eqref{e.Ker.app.0}, and is the dNLS model analyzed in
Section~\ref{s:break}.

The usual strategy to solve the kernel equation is to probe the
applicability of the implicit function theorem. Thus we consider the
linear operator
\begin{equation}
  \label{e.Lambda}
  \Lambda_\rho := \left( D_w \FFF \right) (v_\rho;0,\rho,0) \ .
\end{equation}
Following the same arguments shown in~\cite{PenSPKK16} it is not
difficult to check that $\Lambda_\rho$ has a four-dimensional kernel
which inhibits the application of the implicit function theorem.

\subsection{The second Lyapunov-Schmidt decomposition}
\label{ss:bifeq}

Given the above comment on the non-applicability of the implicit function
theorem in the case of $\Lambda_\rho$, we have to proceed (as in the NL-dNLS
case developed in \cite{PenSPKK16}) with a Lyapunov-Schmidt decomposition of
\begin{equation*}
  w_\rho = k_\rho + h_\rho      \ ,
  \quad
  k_\rho\in \text{Ker}(\Lambda_\rho) \ , \quad h_\rho\in \text{Range}(\Lambda_\rho)  \ .
\end{equation*}
The equation \eqref{e.Ker.per} then becomes
\begin{equation*}
  \begin{cases}
    \FFF_H(v_\rho;k_\rho+h_\rho,\rho,\eps)=0 \\
    \FFF_K(v_\rho;k_\rho+h_\rho,\rho,\eps)=0 \\
\end{cases}\ ,
\end{equation*}
where the subscripts $H$ and $K$ denote the corresponding projections over
$\text{Range}(\Lambda_\rho)$ and
  $\text{Ker}(\Lambda_\rho)$, respectively. The Range equation
$\FFF_H=0$ can be solved locally by the implicit function theorem and provides
\begin{equation}
\label{e.h.GH}
 h_\rho = h_\rho(v_\rho;k_\rho,\eps)\ ;
\end{equation}
inserting~\eqref{e.h.GH} into $\FFF_K=0$ we get the {\bf bifurcation
  equation}, redefining $\FFF_K$ as
\begin{equation}
\label{e.GK.1}
0 = \FFF_K(v_\rho;k_\rho,\rho,\eps) :=
\FFF_K(v_\rho;k_\rho+h(v_\rho,k_\rho,\eps),\rho,\eps)\ ,
\end{equation}
where now
\begin{equation*}
\FFF_K : \RR^4\times \RR\times \RR\to \RR^4\ ,
\end{equation*}
is defined once given the unperturbed reference solution
$v_\rho$. The following lemmas allow us to
then properly treat \eqref{e.GK.1}
as a $\rho$-perturbation of the corresponding bifurcation equation for
its normal form \eqref{e.NLdNLS}.

\begin{lemma}
\label{e.GK.smooth}
The function $\FFF_K(v_\rho;k,\rho,\eps)$ is smooth in $\rho$ and
\begin{equation*}
\FFF_K(v;k,0,\eps) = F_K(v;k,\eps)\ ,
\end{equation*}
where $F_K$ is the corresponding bifurcation equation for
\eqref{e.NLdNLS} defined in \eqref{e.G.01}.
\end{lemma}

\proof For the proof it is simply necessary to show that the
Lyapunov-Schmidt decomposition commutes with the limit $\rho\to0$. As
already observed, if $\rho=0$, then equation \eqref{e.Ker.per} reduces
to $F(v;k,\eps)=0$. For such an equation, the Lyapunov-Schmidt
decomposition is performed with respect to the linear operator
$\Lambda_0$. The smoothness in $\rho$ of all the involved functions
(including the $\rho$-family of isomorphism between the spaces of the
Lyapunov-Schmidt decompositions) concludes the proof.
\qed

A further important characterization of the Kernel projection of both
$\FFF$ and $F$ is that they vanish with $\eps$, so that it is possible
to introduce
  \begin{equation}
    \label{e.BE.0}
    \begin{aligned}
      \FFF_K(v_\rho^*;k_\rho,\rho,\eps) &=: \eps P_\rho(v_\rho^*;k_\rho,\rho,\eps)\ ,
      \\
      F_K(v^*;k,\eps) &=: \eps P(v^*;k,\eps) \ .
    \end{aligned}
  \end{equation}
In~\cite{PenSPKK16} we checked the
  corresponding property by a direct calculation. Here we limit to
  remark that it has to be true since, when $\eps=0$, it corresponds
  to the existence of the ``coordinates'' $(k_\rho,h_\rho)$ describing
  the four-dimensional torus around the chosen $v_\rho^*$. The
  additional property we have here is the continuity with
  respect to $\rho$, i.e.,
  \begin{equation}
    \label{e.BE.cont}
    \rho\to 0
    \qquad\Rightarrow\qquad
    P_\rho(v_\rho^*;k_\rho,\rho,\eps) \to P(v^*;k,\eps) \,
  \end{equation}
which reduces to
\begin{equation}
  \label{e.GK.2}
  P_\rho(v_\rho^*;k_\rho,\rho,\eps) = 0 \,
\end{equation}
for which we look for local solutions
$\norm{k_\rho(\eps)}\ll 1$, with $|\eps|\ll 1$ and $|\rho|\ll 1$.

\subsection{Continuation from the persistence conditions}

We will now concentrate on those particular solutions of the uncoupled system
which satisfy the persistence conditions, that we connote with a
  star superscript.  In particular let
$\bar u^*$ be an unperturbed solution given by \eqref{e.u0} whose phases
$\theta_j$ satisfy the persistence conditions~\eqref{e.per4_KG.zz} and~\eqref{e.M_phi}; then we denote by
\begin{equation}
\label{e.phi0*}
v_\rho^* := \frac1\rho\tond{\bar u_1^* + \im \bar u_{-1}^*}
\end{equation}
its unique rescaled projection on $V_2$, according to
\eqref{e.Fou.dec},\eqref{e.vphi.def} and \eqref{e.scale}; as already
noted in general, $v_\rho^*$ solves~\eqref{e.Ker.rho} with
$\eps=0$. Since the whole previous construction is continuous in
$\rho$, we also have that $v^* = \lim_{\rho\to 0} v_\rho^*$, and
\begin{equation*}
  v_j^* = \begin{cases} Re^{-\im\theta_j} \ , &j\in S \\ 0 \ ,
    &j\not\in S
  \end{cases}\ ,
\end{equation*}
where the phase-differences $\varphi_j=\theta_{j+1}-\theta_j$, introduced in
\eqref{e.phi}, satisfy the corresponding NL-dNLS persistence conditions given
by~\eqref{e.per4_KG.zz} and~\eqref{e.M_phi.DNLS}
Indeed, in the limit of vanishing amplitude, i.e., for $\rho\to 0$,
the KG persistence condition \eqref{e.per4_KG.zz} converges to the
NL-dNLS persistence condition, (see \eqref{e.shifts}), in view of the
exponential decay of the Fourier components.

At the present stage of our construction it is worth
  recalling that the persistence conditions take the form
  \begin{equation*}
    P_\rho(v_\rho^*;0,\rho,0)=0 \quad 
    \text{and}\quad
    P(v^*;0,0)=0 \ ,
  \end{equation*}
  respectively for the KG and dNLS cases, since by continuity
  the ``correction'' $k$ has to vanish
  with $\eps$.

\subsubsection{Proof of Theorem \ref{T.2}}

The first part of Theorem~\ref{T.2} follows from
\begin{proposition}
\label{t.EX}
Let $v_\rho^*$ be as in \eqref{e.phi0*}. If the corresponding $v^*$ is
linearly non-degenerate, which means that the Linearized Bifurcation
Equation
\begin{equation*}
\eps \partial_\eps P(v^*;0,0) + D_k P(v^*;0,0)[k] = 0 \ ,
\end{equation*}
can be uniquely solved (apart from the Gauge direction), then there
exists $\rho^*$ such that, for $|\rho|<\rho^*$ the same holds true for
the $\rho$-perturbed Linearized Bifurcation Equation
\begin{equation*}
  \eps \partial_\eps P_\rho(v_\rho^*;0,\rho,0) +
  D_k P_\rho(v_\rho^*;0,\rho,0)[k] = 0 \ .
\end{equation*}
Hence, there exists $\eps^*(\rho)$ such that, for $|\eps|<\eps^*$ the
bifurcation equation \eqref{e.GK.2} can be locally solved  and
\begin{equation}
  \label{e.est.k}
  \norm{k_\rho(\eps) - k(\eps)} < C \rho^2\ .
\end{equation}
\end{proposition}

\proof the proof is based on the same ideas of Theorem C.1 of \cite{BamPP10}.
From the definitions~\eqref{e.Ker.per},~\eqref{e.G.01} and~\eqref{e.BE.0}, and
by exploiting~\eqref{e.est.phi0*} and the Lipschitz-continuity of $P_\rho$, it
is possible to show that
\begin{equation*}
  P_\rho - P = \mathcal{O}(\rho^2) \ .
\end{equation*}

The non-degeneracy of the Linearized Bifurcation Equation for the
NL-dNLS model, which is equivariant under the action of the Gauge
symmetry, can be translated into the condition that the Kernel of the
four-dimensional squared matrix
\begin{equation*}
  D_k P_\rho(v^*,0,\rho,0)
\end{equation*}
is given only by the Gauge direction, being invertible in the three-dimensional
orthogonal complement\footnote{This is a delicate point and involves the
  preservation of a symmetry under the Lyapunov-Schmidt reduction. The
  equivariance of equation \eqref{e.Ker.vphi} reflects the Gauge invariance of
  the corresponding Hamiltonian: this is a common variational interpretation of
  the Kernel equation in the first LS reduction (see \cite{Ber07}). At a second
  stage, if the Kernel and Range projections commute with the symmetry, then
  also \eqref{e.GK.1} is equivariant and it is enough to restrict to the
  transversal directions.}. As we remarked already, the whole bifurcation
equation $P_\rho(v_\rho^*,k,\rho,\eps)=0$ is still Gauge equivariant, hence
invertibility isn't lost under a continuous, and small enough,
$\rho^2$-perturbation. Hence also $D_k P_\rho(v_\rho^*,0,\rho,0)$ is invertible
in the Gauge-orthogonal subspace and estimate \eqref{e.est.k} is a standard
by-product of the implicit function theorem. \qed

In order to conclude the proof of Theorem \ref{T.2}, we still have to
show that estimate \eqref{e.est.1} holds true. Let now
$w_\rho^*(v_\rho^*;\eps)$ be the solution of
\begin{equation*}
  \FFF(v_\rho^*;w_\rho^*,\rho,\eps)=0 \ ,
\end{equation*}
and, in a similar way, let $w^*(v^*;\eps)$ be the solution of
\begin{equation*}
  \FFF(v^*;w^*,0,\eps)=0 \ .
\end{equation*}

\begin{lemma}
\label{l.est.2}
There exists $\rho^*$ and $\eps^*$ and a constant $C_2$ such that, for
$|\rho|<\rho^*$ and $\eps< \rho^2 \eps^*$ one has
\begin{equation}
  \label{e.est.vphi}
  \norm{w_\rho^*(v_\rho^*;\eps) - w^*(v^*;\eps)}_{\ell^2} < C_2 \rho^2 \ .
\end{equation}
\end{lemma}

\proof As in the proof of the previous Lemma, it is possible to show that
\begin{equation*}
  \FFF_H(v_\rho^*;h+k,\rho,\eps) - F_H(v^*;h+k,\eps) = \mathcal{O}(\rho^2) \ ;
\end{equation*}
then, again from \eqref{e.est.phi0*} one can deduce
\begin{equation}
  \label{e.est.h}
  \norm{h(v_\rho^*;k,\rho,\eps) - h(v^*;k,\eps)} < C \rho^2 \ ,
\end{equation}
which combined with \eqref{e.est.k} gives the desired estimate.
\qed

Going back to \eqref{e.Fou.dec}, let $\vvv^*(\rho,\eps,\tau)$ and
$\vvv^*(0,\eps,\tau')$ be the scaled real solutions (belonging to the
Kernel $V_2$) built respectively with
$\phi_\rho^*(\eps)=v_\rho^*+w_\rho^*(v_\rho^*;\eps)$ and
$\phi^*(\eps)=v^*+w^*(v^*;\eps)$, and
\begin{equation}
  \label{e.u*.sol}
  u^*(\rho,\eps,\tau) = \vvv^*(\rho,\eps,\tau)
  + \www(\vvv^*(\rho,\eps,\tau),\eps)\ ,
\end{equation}
the reconstructed solution of the original perturbed problem
\eqref{e.problem}. Following the same steps developed in
\cite{BamPP10} one gets \eqref{e.est.1}.

\subsubsection{Proof of Theorem \ref{T.1}}\label{sbs:proofT1}

The proof of Theorem~\ref{T.1} is essentially based on a necessary
condition for the solvability of the bifurcation equation which is
shown to be violated. {Precisely, as before, we first show that the
  same property is violated in the dNLS model \eqref{e.NLdNLS.zz} and
  then we extend the result to the system under investigation.  Let
  $v^*$ represent an element of the families \eqref{e.fam.dnls.zz}
  with $\varphi\not\in\{0,\pi\}$. The first step --- deferred to
  Section~\ref{s:zz-ne-LS} --- consists in showing that the linearized
  bifurcation equation of the dNLS system ($\rho=0$)
\begin{equation*}
  \eps\partial_\eps P(v^*;0,0) + D_k P(v^*;0,0)[k] = 0
\end{equation*}
cannot be solved, because the necessary condition
\begin{equation*}
  \partial_\eps
  P(v^*;0,0)\in \text{Range}\tond{D_k P(v^*;0,0)}
\end{equation*}
does not hold.  Once the above is proven, as a consequence, the whole
nonlinear equation cannot be solved for $(k,\eps)$ close to the
origin, thus $v^*$ cannot represent a bifurcation point. The last
implication, namely the relationship between the linearized equation
and the nonlinear equation, can be understood again in terms of
bifurcation theory, and is included in the more general statement of
Proposition 2.10 of~\cite{PelKF05b} (remark that, using the notation
as in \cite{PelKF05b}, in the zigzag case
$g^{(2)}(\theta^*)\not=0$). In qualitative terms, the main idea is
that if $\partial_\eps P(v^*;0,0)\neq 0$ and the linearized equation
cannot be solved, then close enough to the origin $P(v^*;k,\eps)\neq
0$, since higher order corrections are negligible.

To add some details, one can follow Lemma 4.4 and Remark 4.4
of~\cite{PenSPKK16}, where a similar condition on the second order
term $\partial^2_\eps P(\vphi^*,k_g,0,0)$ can be derived for
$\partial_\eps P(\vphi^*,k_g,0,0)=0$. In brief, one can implement a
further Lyapunov-Schmidt decomposition, by splitting again the (four
dimensional) space into the subspace $\text{Ker}(D_k P(v^*;0,0))$,
given by the tangent directions to the $\vphi$-family and the
Gauge-symmetry, and the remaining $\text{Range}(D_k P(v^*;0,0))$. In
terms of variables, one simply introduces $k_{\K}$ and $k_{\R}$, the
set of coordinates of $\text{Ker}(D_k P(v^*;0,0))$ and $\text{Range}(D_k
P(v^*;0,0))$, respectively, such that $k=k_{\K}+k_{\R}$. After
Taylor-expanding and projecting the equation
$P(v^*;k_{\K},k_{\R},\eps)=0$ onto the Range, one immediately realizes
that $k_{\R}=\cal{O}(\eps)$. Thus, at leading order in the Kernel equation one has
\begin{equation*}
  \Pi_{\K}\quadr{\partial_\eps
    P(v^*,0,0)}=0
\end{equation*}
which is equivalent to
\begin{equation*}
\partial_\eps P(v^*;0,0)\in \text{Range}\tond{D_k P(v^*;0,0)}\ .
\end{equation*}
By continuity in $\rho$, the same conclusions can be derived in the
regime of small $\rho$ via the equation
\begin{equation*}
  \eps \partial_\eps P_\rho(v_\rho^*;0,\rho,0) + D_k
  P_\rho(v_\rho^*;0,\rho,0)[k] = 0
\end{equation*}
due to the following}

\begin{proposition}
\label{t.NON.EX.corr}
Let $v_\rho^*$ as in \eqref{e.phi0*}. If the corresponding $v^*$ is
such that 
\begin{equation}
  \partial_\eps P(v^*;0,0)\not=0 \ \wedge\ 
  \partial_\eps P(v^*;0,0) \not\perp \text{Ker}\tond{D_k P(v^*;0,0)} \ ,
  \label{frm:prop4.2}
\end{equation}
then there exists $\rho^*$ such that, for $|\rho|<\rho^*$ one has
\begin{equation*}
  \partial_\eps P_\rho(v_\rho^*,0,\rho,0)\not=0
 \ \wedge\ 
  \partial_\eps P_\rho(v_\rho^*,0,\rho,0) \not\perp
  \text{Ker}\tond{D_k P_\rho(v_\rho^*,0,\rho,0)}\ .
\end{equation*}
\end{proposition}

As previously said, this Proposition shows that, also for the
Klein-Gordon model, the nonlinear equation cannot be solved for
$(k,\epsilon)$ close to the origin, therefore $v^*$ cannot represent a
bifurcation point.

\subsection{A note on the more degenerate model: $\HH_{101}$}\label{sbs:H101}

The technique developed in this Section is not sufficient to deal with
the more degenerate model examined in Section~\ref{s:101}, i.e.,
$\HH_{101}$ in~\eqref{e.KG.deg}.  Actually this kind of degeneracy in
a dNLS model was already examined systematically in~\cite{PenSPKK16},
where we were able to prove the nonexistence of any four-sites
phase-shift discrete soliton for $\eps$ small enough.  The crucial
point is that the higher non-degeneracy required the analysis of
higher order expansions of the Bifurcation Equation: this is exactly
the reason that prevents the application of the tecniques used in the
present paper.  Indeed the small perturbation due to the energy, which
``measures'' the distance among the model~\eqref{e.KG.deg} and its
{dNLS-type} normal form~\eqref{e.NLdNLS.deg}, could be enough to
introduce small linear terms in the bifurcation equation allowing for
non-trivial solutions, which otherwise would not exist. This, however,
depends on the magnitude of the linear term in $\eps$ introduced by
the perturbation.  Since the obstruction to nonexistence comes out
from the $\eps^2$ term in the Kernel equation, the corrections of
order $\rho^2$ would not be relevant for $\rho^2\ll \eps$.  However,
as we are considering the regime $\eps\lesssim\rho^2$ (due to our
initial scaling \eqref{e.scale}), we cannot exclude the existence of
solutions for the perturbed problem.

\section{Nonexistence results for the zigzag-dNLS model}
\label{s:break}

In the present Section we give the nonexistence results for the
corresponding dNLS model upon which are based the proofs of the
previous Section. Since we will closely follow the scheme
of~\cite{PenSPKK16}, many details will be omitted.

Let us rewrite explicitly the model we consider here, i.e.,
\begin{equation}
  \label{e.ZZdNLS.phi}
  \omega\phi_j =
    -\frac\eps2\Bigl[ (\Delta_1+\Delta_2)\phi \Bigr]_j +
    \frac34\phi_j|\phi_j|^2\ ,
\quad
  \text{where}
\ \omega := \lambda-1\ ,
\end{equation}
and consider the unperturbed solutions
\begin{equation}
  \label{e.4Dtorus}
  \phi_j^{(0)} = \begin{cases}
    R e^{\im \theta_j} \, , &j\in S\ ,       \\
    0               \, , &j\not\in S\ ,
  \end{cases}
\end{equation}
where $S=\graff{1,2,3,4}$ and $R>0$.

\subsection{$C^1$ nonexistence result}
\label{s:zz-ne-current}

We first state a finite
regularity nonexistence result.  For this purpose, we assume to deal
with a continuation $\{\phi_j(\eps)\}_{j\in\ZZ}$ which is at least
$\C^1$ in $\eps$. Hence we expand the solution variables $\phi_j$ in
$\eps$
\begin{equation}
\label{e.exp.phi}
\phi_j = \phi^{(0)}_j + \eps\phi^{(1)}_j + o(\eps)\ ,
\end{equation}
where $o(\eps)$ is a continuous function. The continuation is assumed
to be performed at fixed period (frequency). With the perturbative
approach, we are able to prove
\begin{theorem}
\label{t.notsmooth}
For $\eps$ small enough, the only unperturbed solutions
\eqref{e.4Dtorus} that can be continued to $\C^1$ solutions
$\phi(\eps)$ of \eqref{e.ZZdNLS.phi}, (with $\eps\not=0$), correspond to
$\vphi_{j}\in\graff{0,\pi}$.
\end{theorem}

In the proof of the above Theorem, a key point is the fact that the
discrete map \eqref{e.ZZdNLS.phi} preserves
\begin{equation}
  \label{e.current}
  J_j := \Im\Bigl(  
    \phi_{j-1}\overline\phi_j + \phi_{j-2}\overline\phi_j +
    \phi_{j-1}\overline\phi_{j+1}
    \Bigr)\ .
\end{equation}
The conservation of this quantity, the so-called {\it
    current}, $J_j\equiv J$, together with the
hypothesis $\graff{\phi_j}_{j\in\ZZ}\in\ell^2(\CC)$, imply
\begin{equation}
\label{e.current.zero}
J_j = 0\ ,\qquad \forall j\in\ZZ\ .
\end{equation}

As in our previous paper, in what follows we take a look at the
general structure of the expansion, in the present case up to order
one, of both the equations and the conserved quantity; moreover, from
the zero order expansion, we determine the candidate solutions.

The strategy is then to investigate directly such equations evaluated
into the candidate solutions; and to exclude all the solutions
prohibited by Theorem~\ref{t.notsmooth} looking for the
incompatibility of the conserved quantity with the equations.


\subsubsection{Zero-order expansion and candidate solutions}
\label{s:zero.order}

The stationary equation~\eqref{e.ZZdNLS.phi} at order zero gives the
uncoupled system
\begin{equation}
\label{e.order.0}
\omega\phi_j^{(0)} = \frac34\phi_j^{(0)}\left|\phi_j^{(0)}\right|^2\ ,
\end{equation}
which is  invariant under the action of $e^{\im\tau}$. By using
\eqref{e.4Dtorus}, it provides the frequency $\lambda$ of the orbit,
and its detuning $\omega$ from the linear frequency 1, namely
\begin{equation}
\label{e.om_lam}
\omega = \frac34 R^2\quad\text{and}\quad \lambda =1+\frac34 R^2\ .
\end{equation}

The conservation law \eqref{e.current.zero} at order zero gives
\begin{equation}
  \label{e.current.o0}
  J_j^{(0)} := \Im\Bigl(
    \phi_{j-1}^{(0)}\overline\phi_j^{(0)} +
    \phi_{j-2}^{(0)}\overline\phi_j^{(0)} +
    \phi_{j-1}^{(0)}\overline\phi_{j+1}^{(0)}
  \Bigr) = 0 \ .
\end{equation}
If we take only 4 oscillators not at rest (as in our ansatz
\eqref{e.4Dtorus}), then \eqref{e.current.o0} is identically satisfied
for $j\leq 0$ and $j\geq 5$. For the remaining variables site $j\in
S$, by recalling the definition $ \varphi_j :=
\theta_{j+1}-\theta_j\ $ of the phase-differences as in \eqref{e.phi},
equations \eqref{e.current.o0} give
\begin{equation}
\label{e.shifts}
\begin{aligned}
\sin\tond{\varphi_1} &= -\sin\tond{\varphi_1+\varphi_2}\ ,\\
\sin\tond{\varphi_2} &= \sin\tond{\varphi_1} + \sin\tond{\varphi_3}\ ,\\
\sin\tond{\varphi_3} &= -\sin\tond{\varphi_2 + \varphi_3}\ .\\
\end{aligned}
\end{equation}

\begin{remark}
  The above system of equations for the phase-differences coincides
  with~\eqref{e.per4_KG.zz} and~\eqref{e.M_phi.DNLS}.
\end{remark}

As already anticipated in Section~\ref{s:NBA}, the solutions of the
system~\eqref{e.shifts} provide the two families
${F}_1:{\pmb{\vphi}}=(\ph,\pi,-\ph)$ and $F_2: {\pmb
  \vphi}=(\ph,\pi,\ph+\pi)$, respectively (see~\eqref{e.fam.dnls.zz}).
Their intersections give the two phase-shift solutions $F_1\cap F_2 =
{\Phi^{({\rm sv})}}= \pm\tond{\frac\pi2,\pi,-\frac\pi2}$, while they
include some in/out-of-phase solutions, i.e., $\big\{ (0,\pi,0),
(\pi,\pi,\pi) \big\}\in {F}_1$ and $ \big\{ (0,\pi,\pi), (\pi,\pi,0)
\big\}\in {F}_2$.  The remaining possible
in/out-of-phase solutions (those with $\ph_2=0$) are not included in
the above families, i.e., $F_{\text{iso}}: {\pmb{\vphi}}= \big\{
(0,0,0), (0,0,\pi), (\pi,0,0), (\pi,0,\pi) \big\}$.

\subsubsection{First order expansions}

The first order expansions of both the stationary equation
\eqref{e.ZZdNLS.phi} and the density current \eqref{e.current} are
easily deduced and take the form

\begin{equation}
  \label{e.order.1}
  \begin{aligned}
  \omega\phi_j^{(1)} = &-\frac12\quadr{
    \phi^{(0)}_{j+2} + \phi^{(0)}_{j+1} + \phi^{(0)}_{j-1} + \phi^{(0)}_{j-2}} +
  2\phi^{(0)}_j \\&+ \frac34\quadr{
    2\phi_j^{(1)}|\phi_j^{(0)}|^2 + \tond{\phi_j^{(0)}}^2\overline{\phi_j^{(1)}}}\ ;
  \end{aligned}
\end{equation}

\begin{equation}
  \label{e.current.o1}
  \begin{aligned}
  0 = \Im\Bigl( 
    &\phi_{j-1}^{(0)}\overline\phi_j^{(1)} +
    \phi_{j-2}^{(0)}\overline\phi_j^{(1)} +
    \phi_{j-1}^{(0)}\overline\phi_{j+1}^{(1)} \\&\quad+
    \phi_{j-1}^{(1)}\overline\phi_j^{(0)} +
    \phi_{j-2}^{(1)}\overline\phi_j^{(0)} +
    \phi_{j-1}^{(1)}\overline\phi_{j+1}^{(0)}
    \Bigr) \ .
  \end{aligned}
\end{equation}

\subsubsection{Second order expansion and conclusion}
\label{ss:asym3}
\label{ss:vs}

To get the nonexistence result, the solutions of the equations are
inserted into the conserved current.

Starting with the two families of asymmetric vortex solutions, ${F}_1$
and ${F}_2$, with the exclusion of the $F_{\text{iso}}$ and
$\Phi^{({\rm sv})}$ solutions, we end up, for the first family,
with the following set of linear equations
\begin{equation}
  \begin{aligned}
    B+C &= 2\sin\varphi
    \\
    (1+\cos\varphi)(B+C) + \cos^2\varphi(A+D) &= 0
    \\
    B+C &= -2\sin\varphi    
  \end{aligned}
\end{equation}
where $A,B,C,D$ are left free at previous order. The system is clearly
impossible once we exclude $\varphi=0,\pi$. The
second family is treated in the same way.

Concerning the symmetric vortex solutions $\Phi^{({\rm sv})}$, by similar
calculations we have again that the conservation law at order zero is
identically satisfied, and at order one is equivalent to
\begin{equation}
    4\im = 0\ , \quad
    B+C = 0\ , \quad
    -4\im = 0\ ,
\end{equation}
which is impossible independently of the four free parameters left
from the equation.


\subsection{$C^0$ nonexistence result}
\label{s:zz-ne-LS}


Following again~\cite{PenSPKK16}, we want to complete the nonexistence
result to $C^0$ solutions. The strategy is based on a Lyapunov-Schmidt
decomposition, where suitable expansions are performed mainly at the
level of the (regular) equations, without assumptions on the
regularity of the solutions. {We recall that this stronger result is
  technically needed, as already mentioned in the proof of
  Theorem~\ref{T.1}, in order to obtain the similar result for the
  Klein-Gordon model in the small energy regime: in fact, we are going
  to show here that condition~\eqref{frm:prop4.2} assumed in
  Proposition~\ref{t.NON.EX.corr} holds true. However, since the
  scheme is exactly the same of \cite{PenSPKK16}}, we only sketch the
key points. The main difference is that in such a paper we considered
the case $ H_{101}$ while here we are dealing with the case $
H_{110}$, so that
\begin{displaymath}
  L := \Delta_1+\Delta_2\ .
\end{displaymath}

The key part of the analysis is in the bifurcation (kernel) equation,
and in the application of Lemma 4.4 and Remark 4.4
of~\cite{PenSPKK16}; to this purpose we check the projection of
\begin{displaymath}
  \partial_\eps P(v^*,0,0) = \Pi_KL h^{(1,0)}(v^*,0,0) \ ,
\end{displaymath}
where
\begin{displaymath}
  h^{(1,0)}(v^*,0,0) :=  -\Lambda^{-1}\Pi_HLv^*
\end{displaymath}
onto the Kernel of the differential operator
\begin{displaymath}
  D_kP(v^*,0,0)[k] =
  \Pi_KLk - \frac32 \Re\tond{v^* \overline{h^{(1,0)}}}k \ ,
\end{displaymath}
where $v^*$ is in one of the families ${F}_1$ and ${F}_2$,
and $\Pi_K$ and $\Pi_H$ are the projectors onto respectively the
Kernel and the Range of $\Lambda$.  We deal explicitly with one
family only; taking $F_1:\bm{\ph} = \tond{\ph,\pi,-\ph}$, and setting
$\theta_0=0$, we have $\bm{\theta} = \tond{0,\ph,\pi+\ph,\pi}$, which
gives the following representation of $v^*$ in complex variables
\begin{displaymath}
  v^*\Big|_S(\ph) = R\tond{1,e^{\im\ph},-e^{\im\ph},-1}\ .
\end{displaymath}
As a consequence, the Kernel's basis reads
\begin{displaymath}
  \bm{e_j}\Big|_S =
  \im R\graff{(1,0,0,0),(0,e^{\im\ph},0,0),(0,0,-e^{\im\ph},0),(0,0,0,-1)}\ .
\end{displaymath}
An easy computation gives $Lv^*$, precisely
\begin{equation*}
\quadr{\ldots,0,1,1+e^{\im\ph}\Big|-4,
  -5e^{\im\ph},5e^{\im\ph},4\Big|-(1+e^{\im\ph}),-1,0,\ldots}\ ;
\end{equation*}
using
the scalar product $\Pi_KLv^* =
\sum_{j=1}^4\Re\tond{Lv^*\overline{\bm{e_j}}}\bm{e_j}$, one gets $\Pi_KLv^*=0$, since $\Re\tond{Lv^*\overline{\bm{e_j}}}=0$ for
all $j=1,\ldots,4$. Hence $\Pi_H Lv^* = Lv^*$.
Since
\begin{displaymath}
  \Lambda h = 
  \begin{cases}
             -2 \omega h\ ,    &j\in S\\
    \phantom{-2}\omega h\ ,    &j\not\in S\\
  \end{cases} \ ,
\end{displaymath}
then $ -\Lambda^{-1}\Pi_HLv^*$ takes the form
\begin{equation*}
\frac1\omega\quadr{\ldots,0,-1,-(1+e^{\im\ph})\Big|-2,
  -\frac52e^{\im\ph},\frac52e^{\im\ph},2\Big|(1+e^{\im\ph}),1,0,\ldots}\ .
\end{equation*}
Given that our last operation is a projection onto the Kernel, we
limit the next computation on the core sites,
precisely
$-\tond{L\Lambda^{-1}\Pi_HLv^*}\Big|_s$ reads
$$
\frac1\omega\quadr{6-e^{\im\ph},\frac{23}2e^{\im\ph}-1,
  -\frac{23}2e^{\im\ph}+1,-6+e^{\im\ph}}\ .
$$
We finally get
\begin{displaymath}
  \partial_\eps P(v^*,0,0) =
  -\im\sin(\ph)\quadr{0\Big|1,-e^{\im\ph},e^{\im\ph},-1 \Big|0}\ .
\end{displaymath}
Upon verifying that the four-dimensional matrix representing the linear
operator $D_kP(v^*,0,0)[k]$ has rank 2, we know for free the Kernel
generators, since the Gauge direction and the direction tangent to the
family for sure belong to it; these are respectively
\begin{displaymath}
  \partial_\ph v^*(\ph) =
  \quadr{0\Big|0,\im e^{\im\ph},-\im e^{\im\ph},0 \Big|0} \ ,
  \quad
  \partial_{\theta_0} e^{\im\theta_0} v^*(\ph) = \im e^{\im\theta_0} v^*(\ph) \ .
\end{displaymath}

Since the Gauge symmetry is preserved along the whole construction, we
have $\partial_\eps P\perp \partial_{\theta_0} e^{\im\theta_0} v^*$,
which can even be checked by the direct computation of
$\partial_{\theta_0} e^{\im\theta_0} v^*(\ph)\cdot\partial_\eps
P(v^*,0,0)=0$.  The other scalar product gives instead
\begin{displaymath}
  \begin{aligned}
  \partial_\ph v^*(\ph)\cdot\partial_\eps P(v^*,0,0) &=
  -\sin(\ph) \Bigl[\Re\tond{\im e^{\im\ph}(\im e^{-\im\ph})} \\&\qquad\qquad\qquad+
    \Re\tond{-\im e^{\im\ph}(-\im e^{-\im\ph})}\Bigr] \\&=
  2\sin(\ph) \ ,
  \end{aligned}
\end{displaymath}
which is different from zero, apart from the cases $\ph=0,\pi$. Thus,
we can conclude that the projection of $\partial_\eps P(v^*(\ph),0,0)$
onto the Kernel of $D_kP(v^*,0,0)$ is different from zero on any
phase-shift discrete soliton considered in the
family. This represents a sufficient condition for nonexistence of
the continuation.

\section{Conclusions - Future Directions}
\label{s:concl}

The present paper represents a natural follow up of \cite{PenSPKK16},
where we studied the related problem of the nonexistence of
degenerate phase-shift discrete solitons in a non-local dNLS
lattice. We recall that in \cite{PenSPKK16} the nonexistence of
phase-shift discrete solitons, which was not easily achievable
by means of averaging
methods due to the degeneracy of the problem, was obtained in
an efficient way exploting the rotational symmetry of the model and
the density current conservation along the spatial profile of any
candidate soliton. The absence of these ingredients in Klein-Gordon
models represents an additional layer of
difficulty to the degeneracy that one
has to face in the continuation problem that we here address.

Keeping in mind the connections among these two classes of Hamiltonian
models (KG and dNLS), a natural (although indirect) way to proceed is to
transfer the results which are accessible in the dNLS context to
similar results which are expected to be valid in the KG context,
keeping track of the relevant correction terms.
In this work we examined mainly KG systems with interactions beyond
nearest neighbors interactions (bearing also in mind
connections with higher dimensional lattices),
focusing principally on the zigzag model for our analytical considerations. In
this model, by means of Lyapunov-Schmidt techniques, we showed
that that this approach actually works
provided some smallness assumptions are made on the main physical
parameters of the models: the energy $E$ and the coupling strength
$\eps$.

However, the strategy presented here, is based on a first order normal
form approximation of the KG model, and thus it has some limitations
in cases where higher order degeneracies occur. In order to showcase
this fact we shortly examine a model that exhibits next-to-next
nearest neighbors interactions namely the $\HH_{101}$ model. Although
the previously described methodology cannot be applied, the numerical
exploration performed in the Section~\ref{s:101} shows elements which
strongly overlap with those that one can obtain in the corresponding
dNLS normal form $H_{101}$, for which a rigorous answer has been given
already in \cite{PenSPKK16}. This naturally leads us to conjecture that a
corresponding nonexistence statement of phase-shift four-sites
multibreathers holds true also for $\HH_{101}$.

In order to prove such a conjecture, one could still follow this
indirect approach by increasing the accuracy of the normal form
approximation by adding further non-local linear and nonlinear terms
to the dNLS $H_{101}$, in the spirit of a more general dNLS
approximation (see \cite{PalP14, PalP16}). Alternatively, one can use
a more direct approach and perform a local normal form technique
around the low-dimensional resonant torus, with the advantage of
working directly in the original KG model without passing from the
dNLS approximation (see \cite{PenSD17} for the maximal tori
case). With this scheme we expect to derive a normal form which
naturally extends the effective Hamiltonian method introduced in
\cite{AhnMS02}. In any case, and whatever the perturbation method one
prefers to apply may be, it appears natural that the accuracy required in
the approximation is directly related to the order of the degeneracy
of the problem: hence, for highly degenerate problems the help of a
computer assisted manipulation may be unavoidable and the choice of
the method can become extremely relevant.

A related comment is that in the present work we have limited
our considerations to one-dimensional settings with long-range
interactions. Extending relevant ideas to genuinely
higher-dimensional KG settings, where again the understanding
built on the basis of the dNLS~\cite{PelKF05b,Kev_book09}
may be useful, is another natural avenue for future work.

 \section*{Acknowledgements}
The authors, V.K., P.G.K.,  
acknowledge that this work made possible by NPRP grant {\#} [9-329-1-067] 
from Qatar National Research Fund (a member of Qatar Foundation). 
The findings achieved herein are solely the responsibility of the authors.


\def\cprime{$'$} \def\i{\ii}\def\cprime{$'$} \def\cprime{$'$}

\end{document}